\documentclass[twocolumn]{aastex7}
\usepackage{graphicx}	
\usepackage{amsmath}	
\usepackage{amssymb}
\usepackage{tikz}
\usepackage{placeins}
\usepackage{afterpage}
\usepackage{soul}
\usepackage{hyperref}
\usepackage[dvipsnames]{xcolor}
\hypersetup{
    colorlinks=true,
    linkcolor=blue,
    citecolor=blue,
    filecolor=orange,      
    urlcolor=cyan,
    }
\usepackage{multirow}

\begin{document}

\title{25 Years of Mrk~421 with \textit{XMM-Newton}: Unveiling Structured Jets and Energy-Dependent Escape\\}

\author[0000-0003-4586-0744]{Tek P. Adhikari}
\affiliation{CAS Key Laboratory for Research in Galaxies and Cosmology, Department of Astronomy,
University of Science and Technology of China, Hefei, Anhui 230026, People’s Republic of China}
\email[]{}
\affiliation{School of Astronomy and Space Science, University of Science and Technology of China,
Hefei, Anhui 230026, People’s Republic of China}
\email[show]{tek@ustc.edu.cn}

\author[0000-0002-0705-6619]{Gopal Bhatta}
\affiliation{Janusz Gil Institute of Astronomy, University of Zielona Góra,
ul. Szafrana 2, 65-516 Zielona Góra, Poland}
\email[show]{g.bhatta@ia.uz.zgora.pl}

\author[0009-0000-4001-2012]{Sangeetha Kizhakkekalam}
\affiliation{Janusz Gil Institute of Astronomy, University of Zielona Góra,
ul. Szafrana 2, 65-516 Zielona Góra, Poland}
\affiliation{Doctoral School of Exact and Technical Sciences, University of Zielona Góra,
al. Wojska Polskiego 69, 65-762 Zielona Góra, Poland}
\email[]{s.kizhakkekalam@stud.uz.zgora.pl}

\author[0000-0001-9092-2903]{Navaneeth P K}
\affiliation{Janusz Gil Institute of Astronomy, University of Zielona Góra,
ul. Szafrana 2, 65-516 Zielona Góra, Poland}
\affiliation{Doctoral School of Exact and Technical Sciences, University of Zielona Góra,
al. Wojska Polskiego 69, 65-762 Zielona Góra, Poland}
\email[]{g.bhatta@ia.uz.zgora.pl}

\author[0000-0002-2173-0673]{Alex Markowitz}
\affiliation{Nicolaus Copernicus Astronomical Center, Polish Academy of Sciences, ul. Bartycka 18, 00-716 Warsaw, Poland}
\email[]{almarkowitz@camk.edu.pl}

\author[0000-0003-4586-0744]{Zhicheng He}
\affiliation{CAS Key Laboratory for Research in Galaxies and Cosmology, Department of Astronomy,
University of Science and Technology of China, Hefei, Anhui 230026, People’s Republic of China}
\affiliation{School of Astronomy and Space Science, University of Science and Technology of China,
Hefei, Anhui 230026, People’s Republic of China}
\email[show]{zcho@ustc.edu.cn}

\author[0000-0003-0793-6066]{Santanu Mondal}
\affiliation{Department of Physics and Electronics, Christ University, Bangalore 560029, India}
\email[]{santanuicsp@gmail.com}

\begin{abstract}
We present a long-term X-ray spectral and timing analysis of the TeV blazar Mrk~421 based on nearly a quarter century of archival \textit{XMM-Newton}/pn observations obtained between 2000 and 2025. Our sample comprises 76 epochs obtained with the pn instrument in both IMAGING and TIMING modes, allowing us to trace the source's variability and spectral evolution. Mrk~421 exhibits flux variations by approximately an order of magnitude, with prominent flaring periods interspersed with low-activity states. A clear harder-when-brighter trend is observed, accompanied by scatter in the hardness ratio (HR), suggesting flux alone does not uniquely determine the spectral state. Fractional variability in the hard band ($2-10$ keV) correlates tightly with that in the soft band ($0.3-2$ keV), with a steeper slope indicating pronounced variability at higher energies. The flux distribution is non-Gaussian and lognormal-like, dominated by moderate flux levels with a tail of bright flares. Spectral analysis shows that power-law model is not sufficient; 59 epochs are best described by a broken power-law (BPL) and 17 by a log-parabola (LP). This fluctuation between BPL and LP models suggests the emission region transitions between a distinct acceleration shockfront and downstream stochastic turbulence. The spectral break energy of BPL remains predominantly clustered around $2$ keV across all flux states, while the average spectral steepening above the break is mild ($\Delta \Gamma \approx 0.19$), inconsistent with standard one-zone cooling models. Our results support structured jet scenarios or energy-dependent electron escape, providing key constraints on the long-term behavior of relativistic jets in high-synchrotron-peaked blazars.
\end{abstract}


\section{Introduction}\label{sec:intro}
Blazars are powerful extragalactic sources capable of accelerating particles to extremely high energies and producing non-thermal radiation across the entire electromagnetic spectrum. They constitute a subclass of radio-loud active galactic nuclei characterized by relativistic jets closely aligned with the observer's line of sight \citep{urry1995unified}. Because of strong Doppler boosting, blazars appear exceptionally bright and highly variable spanning timescales from minutes to years, making them ideal laboratories for exploring the physical mechanisms governing particle acceleration and radiation in relativistic jets \citep{bottcher2003coordinated,bottcher2019,Acciari2011, bhatta2020nature}. The broadband spectral energy distribution of blazars displays two prominent emission peaks \citep[e.g.,][]{Fossati1998}. The low-energy component is attributed to synchrotron radiation from relativistic electrons, while the origin of the high-energy component remains a topic of active discussion. Leptonic models explain the gamma-ray emission through inverse-Compton scattering of either internally generated synchrotron photons (Synchrotron Self-Compton, SSC) \citep{Maraschi1992,Mastichiadis2002} or photons originating from external structures such as the accretion disk, broad-line region, and dusty torus (External Compton, EC) \citep{Dermer1993, Sikora1994, Blazejowski2000}. The blazar population is broadly classified into two subclasses, according to their optical spectral characteristics: Flat-Spectrum Radio Quasars (FSRQs) and BL Lacertae objects (BL Lacs). FSRQs are distinguished by strong broad emission lines with equivalent widths greater than 5 \AA{} and are generally associated with higher luminosities and more powerful jets. Conversely, BL Lac objects exhibit weak or absent emission lines with equivalent widths smaller than 5 \AA{}, which is indicative of their comparatively less powerful accretion and jet systems.
Moreover, BL Lac objects generally exhibit synchrotron peaks at relatively high frequencies, ranging from the infrared to the X-ray band, and produce strong high-energy emission extending up to TeV energies \citep{falomo2014optical, ghisellini1998, urry1995unified}. In contrast, FSRQs are characterized by synchrotron peaks located at lower frequencies and by a high-energy component that is predominantly Compton dominated \citep{Fossati1998,ghisellini1998}. Furthermore, based on the synchrotron peak frequency ($\nu_{\rm p}$), blazars are classified into low-synchrotron-peaked (LSP; $\nu_{\rm p}<10^{14}$ Hz), intermediate-synchrotron-peaked (ISP; $10^{14}<\nu_{\rm p}<10^{15}$ Hz), and high-synchrotron-peaked (HSP; $\nu_{\rm p}>10^{15}$ Hz) sources \citep[see][]{abdo2010spectral}.

High-synchrotron-peaked BL Lac objects (HSPs) are characterized by synchrotron emission extending into the X-ray band, making X-ray observations particularly important for studying their spectral and temporal properties and probing characteristics of the highest-energy electrons. Among the HSP population, Mrk~421 is one of the nearest and most extensively studied blazars. Located at a redshift of $z=0.031$, it is a bright X-ray and TeV gamma-ray source that exhibits pronounced variability across a broad range of timescales \citep{Aharonian2006, punch1992detection, falomo2014optical}. The exceptional brightness of Mrk~421 has enabled extensive multiwavelength investigations spanning from radio frequencies to TeV gamma rays using both space- and ground-based observatories \citep{Aleksic2015, Balokovi2016,Markowitz2022}. Recent X-ray polarimetric observations with Imaging X-ray Polarimetry Explorer (\textit{IXPE}) have demonstrated that the X-ray emission of Mrk~421 is highly polarized and exhibits temporal variations in the polarization angle, providing new insight to the magnetic field geometry and particle acceleration within the jet \citep{kim2024}. Early X-ray work with \textit{EXOSAT} identified increased spectral softening towards relatively lower-flux states, 
interpreted as evidence for SSC emission \citep{Warwick1986}. Subsequent analyses 
using \textit{RXTE} revealed spectral variability behaviors that varied between different flares 
on hour-to-month 
timescales \citep{Cui2004}. 
Mrk~421 shows a distinctly curved X-ray spectrum, characteristic of HSP
blazars \citep{fossati2000x}, with flux variability commonly accompanied by
marked spectral evolution \citep{Tramacere2007,Massaro2004}.  
The stochastic and temporal properties of the source have been 
investigated across a wide range of timescales \citep{Sobolewska2014,bhattacharyya2020blazar,Panis2025}. 
Time-resolved X-ray spectral studies employing different spectral models
enable a systematic examination of parameter correlations and spectral
evolution, as demonstrated by recent works \citep[e.g.,][]{Hota2021,Kizhakkekalam2026}.
In a study investigating intra-day and long-term spectral and flux variability in the X-ray bands using \textit{XMM-Newton} observations, several blazars exhibited significant flux and spectral variability. The analysis revealed a multi-polar flux distribution, linear trends in the root mean square (RMS)-flux relation, and a ``harder-when-brighter'' trend, indicating that the X-ray spectra harden as the flux increases.
\citep[see][]{2025ApJ...981..118B,2023ApJ...955..121D,2022MNRAS.510.5280M,Mondal2022}. More recently, a timing analysis of $\sim$ two decades of \textit{XMM-Newton} observations of Mrk~421, employing deterministic (RQA) and stochastic (ARMA/ARIMA) methods, reveals distinct temporal scales. It is shown that deterministic patterns dominate at small scales, while stochastic processes with memory propagate from large to small scales \citep{Panis2025}.
 \textit{Swift} observations further uncovered intrinsic 
spectral curvature, pointing to curved electron energy distributions and 
energy-dependent acceleration mechanisms \citep{Tramacere2009}. Recent studies have deepened our understanding of Mrk~421’s variability. 
\citet{Acciari2011} reported a harder-when-brighter trend linked to TeV 
activity. \citet{Yan2018}, in their detailed statistical analysis of 
Mrk~421 X-ray flares, argued that X-ray flares likely 
originate from magnetic reconnection. Multi-band observations 
confirmed rapid sub-hour variability in both X-ray and TeV bands, supporting 
magnetic reconnection as a plausible driver \citep{Acciari2020}. 
\textit{IXPE} X-ray polarization measurements favor shock-accelerated synchrotron 
emission in ordered jet magnetic fields \citep{DiGesu2022}. Polarization 
studies further revealed intermittent angle swings and stable high polarization 
fractions ($\sim$14\%) \citep{kim2024}. Several other multi-wavelength campaigns
show Mrk~421 is persistently bright from optical to TeV energies, 
with prominent variability on $\lesssim$1--2 days \citep{Blazejowski2005,Aleksic2015,Balokovi2016,Markowitz2022}.

In this work, we present the first systematic 25-year 
\textit{XMM-Newton}/pn spectro-timing analysis
leveraging 10 years of new archival data to resolve long-term jet evolution, constrain acceleration/cooling models, and link spectral parameters to flux states.
We study the time evolution of the light curves and fit the spectra using
several emission models: power-law (PL), broken power-law (BPL), 
and log-parabola (LP), identifying the best-fitting model for each epoch.
We investigate the time evolution of the hardness ratio (HR), 
spectral index, break energy, and spectral curvature. In particular,
we explore how spectral parameters vary across different flux states and
construct hardness-intensity diagrams to probe the interplay between
particle acceleration and cooling over decade-long timescales. 
The present study complements prior works by providing the longest temporal
baseline for understanding the evolution of jet emission processes 
in Mrk~421.

The paper is organized as follows: Section~\ref{sec:obs} describes the \textit{XMM-Newton} observations and data reduction procedures. Sections~\ref{sec:light-curve} and~\ref{sec:spec} present the timing and spectral analyses, respectively. The discussion of the results and the derived conclusions are presented in Section~\ref{sec:discuss_conclude}.
\section{25 Years of {\it XMM-Newton} observations of the TeV blazar Mrk~421}
\label{sec:obs}
Markarian 421 is one of the brightest extragalactic X-ray sources and has been extensively monitored by several X-ray observatories, including \textit{XMM-Newton}, which has been observing the source over the past 25 years. 
\citet{Brinkmann2001} presented the first \textit{XMM--Newton} observations of Mrk~421, revealing rapid X-ray variability and complex spectral evolution. Subsequently, \citet{Brinkmann2003,Brinkmann2005} performed detailed spectral and timing analyses of the early \textit{XMM--Newton} observations, investigating spectral variability, energy-dependent time lags, and the flux--spectral index relationship. \citet{Yan2018} analyzed 15 years of {\it XMM-Newton} data, limiting their study to the statistical properties of the source's X-ray flares. Using an extended 17-year \textit{XMM--Newton} data set, \citet{Noel2022} investigated the intraday X-ray variability of Mrk~421. More recently, \citet{Panis2025} employed the 17-year \textit{XMM--Newton} archive to investigate the deterministic and stochastic characteristics of the X-ray variability.
We systematically search for the publicly available  observational records of 
Mrk~421 in the {\it XMM-Newton} Science Archive\footnote{\url{https://nxsa.esac.esa.int/nxsa-web/search}}, and downloaded the data with valid pn exposures, both in IMAGING and TIMING mode of observations. Only the pn data were used for the analysis, since this detector provides significantly higher effective area and signal-to-noise ratio than the MOS detectors, enabling more reliable determination of spectral parameters. 
The observations spanning 2000 to 2025 were selected, with the observation log presented in Table~\ref{tab:observation_log1}. It is worth noting that \textit{XMM-Newton} observations provide targeted snapshots of Mrk~421 across different activity states rather than continuous monitoring. The source was observed both as a planned target and in response to enhanced activity or flare alerts, with several observations also obtained as part of coordinated multi-wavelength campaigns. Together, these observations provide a diverse sampling of the source over its 25-year \textit{XMM-Newton} history, although the observed flux distribution reflects the activity states covered by the available observations. 

This work extends the available \textit{XMM--Newton} observational baseline to nearly 25 years by incorporating an additional decade of high-quality observations, enabling a comprehensive investigation of the long-term timing and spectral evolution of Mrk~421.

\begin{figure*}
    \includegraphics[width=0.99\linewidth]{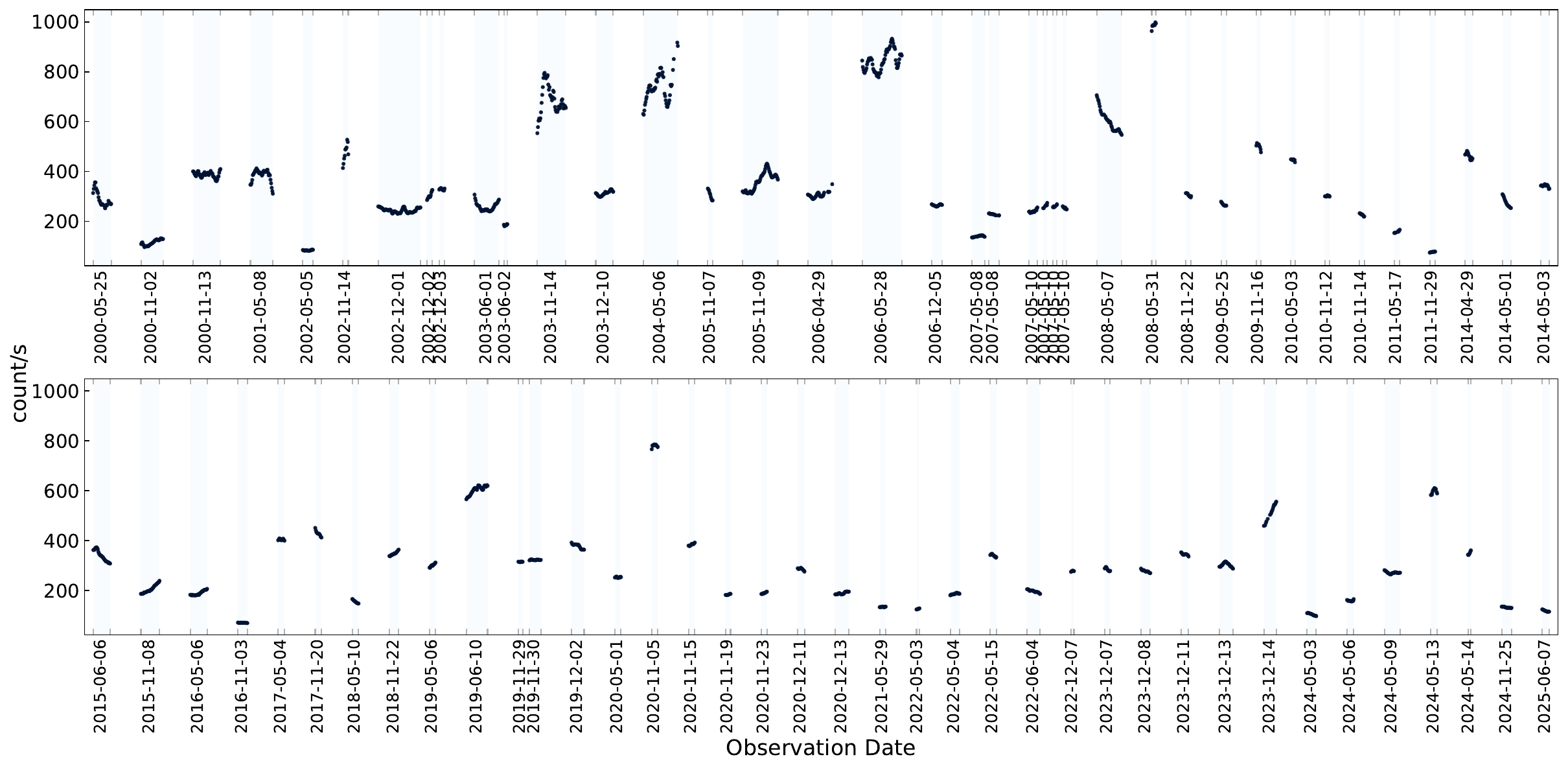}
    \caption{X-ray (0.3–10 keV) light curve of the TeV blazar Mrk~421, rebinned to 1000 s, obtained from approximately 2.5 decades of \textit{XMM-Newton}/pn observations. The y-axis represents the count rate (counts s$^{-1}$), and the x-axis shows the observation dates in chronological order, with the intervals between successive observations not drawn to scale.}
    \label{fig:LC-whole}
\end{figure*}

\subsection{Data processing using {\it XMM}-\texttt{SAS}}
We followed the standard data reduction procedures outlined in the \texttt{SAS}\footnote{\url{https://www.cosmos.esa.int/web/xmm-newton/sas-threads}} data analysis threads
to reprocess the downloaded observational data files (ODFs) for obtaining the calibrated and concatenated event lists using \texttt{SAS} \texttt{version 21.0} employing the task \texttt{epproc} for the pn observations. Background particle flares were then checked by extracting the high-energy (10 -- 12 keV) light curve from the single-pattern event lists, and good time intervals (GTIs) were generated by visually inspecting the high-energy light curve manually. In the majority of cases, a background count rate threshold of $0.4 \text{ counts/s}$ was applied for the pn instrument; intervals exceeding this rate were excluded.
The standard \texttt{PATTERN} and \texttt{FLAG} expressions recommended in the official manual of 
{\it XMM-Newton} data reduction were followed. The expression \texttt{"FLAG==0 \&\& PATTERN<=4"} was set for all pn data.
Finally, the clean event lists, free of background flares were generated by applying the GTI sets to the original event lists.

In the next step, we systematically checked each observation for possible pile-up effects because of the high X-ray flux of Mrk~421.
Using the \texttt {SAS} task \texttt {epatplot}, applied to the unfiltered events, it was found that majority of the observations were affected by the pileup, and, this effect was corrected by iteratively excising the central source core in \texttt{ ds9}\footnote{\url{https://sites.google.com/cfa.harvard.edu/saoimageds9/home?authuser=0}} until the \texttt{epatplot} output confirmed negligible pileup, 
retaining an annular (rectangular for TIMING mode) source region for spectral/light curve extraction. The light curves and the spectra were extracted from the annular region for the observations in IMAGING mode, and the rectangular region for the TIMING mode using \texttt{ds9}. The background regions were selected from a source free region in the same CCD where the source is present.  

The light curves in three energy bands--$0.3-10$ keV (full band), $0.3-2$ keV (soft band), and $2-10$ keV (hard band)--were generated for each observational epoch with a time-bin size of 100 s, 
and the task \texttt{epiclccor} was applied to
each to obtain the final, background corrected source light curves. 
Response matrix files (RMF) and ancillary response files (ARF) were generated for the extracted spectra using the \texttt{rmfgen} and
\texttt{arfgen} tasks. Finally, each spectrum is grouped and binned to have at least 25 counts for each background-subtracted spectral channel by using the task \texttt{specgroup}.

\section{Light curves of Mrk~421}
\label{sec:light-curve}
The long-term $0.3-10$ keV light curve of Mrk~421, obtained from the 76 epochs of XMM-Newton/pn observations, is depicted in Fig. 1. The data are ordered sequentially by observation date along the X-axis and displayed across a two-panel plot to accommodate the 25-year baseline.
As the figure shows, across timescales ranging from hours to decades, the source exhibits multiple
epochs of enhanced X-ray activity, including several strong flaring states during the early 2000s and 
around $2019-2020$, where the count rate exceeds $\sim$700-900 count/s, interspersed with comparatively low-flux
states below $\sim$ 100 count/s. 

\begin{figure*}
\includegraphics[scale=0.59]{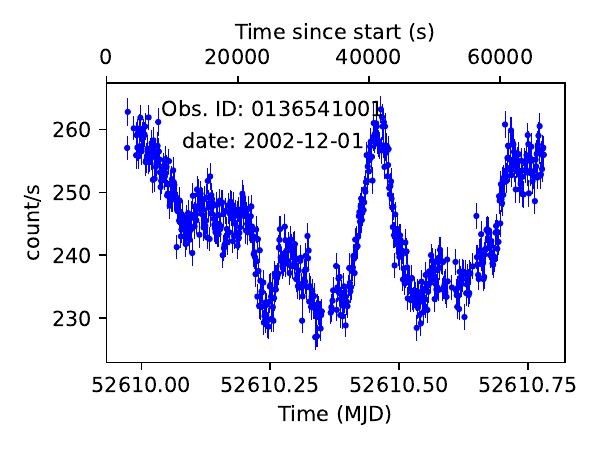}\includegraphics[scale=0.59]{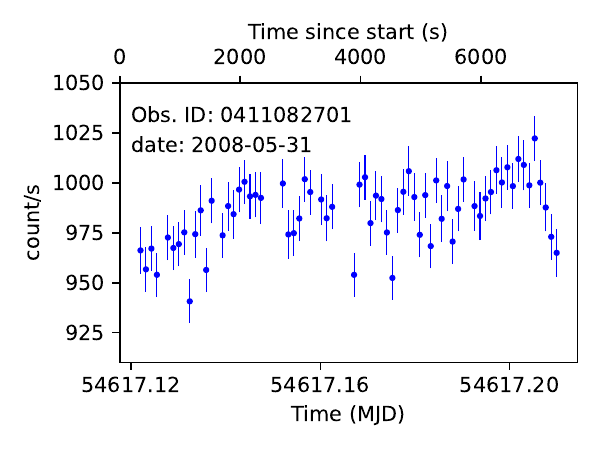}\includegraphics[scale=0.59]{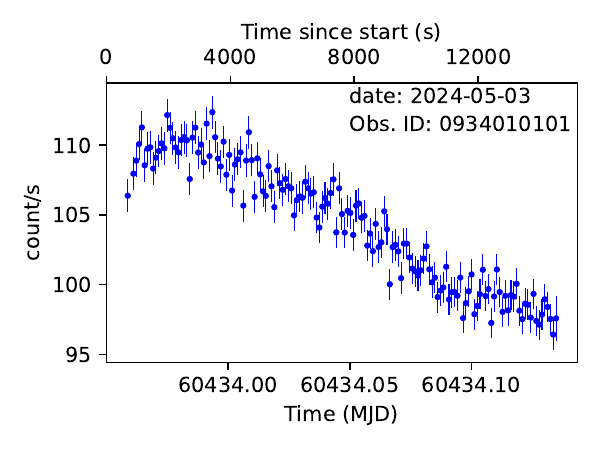}
    \caption{Selected light-curves showing the temporal variability properties of Mrk~421 across different observational epochs depicting the various timescales of variability that the source exhibits. 
    }
    \label{fig:Individual-LC}
\end{figure*}

The selected individual light curves, illustrating the intra-observational X-ray variability of Mrk~421 highlighting a 
diversity of temporal behavior is depicted in Fig.~\ref{fig:Individual-LC}. The left panel (Obs. ID: 0136541001) shows pronounced short-timescale variability, with multiple flux fluctuations on timescales of a few tens of kilo-seconds (ks). The middle panel (Obs. ID: 0411082701) corresponds to the highest-flux state among the 76 epochs and displays a comparatively smoother variability, with modest fluctuations around the mean count rate, suggesting a stable emission phase over a timescale of a $\sim$few ks. In contrast, the right panel (Obs. ID: 0934010101) reveals a gradual, monotonic decline in the count rate over the course of the observation, consistent with decaying
emission. Altogether, these light curves demonstrate that Mrk~421 exhibits variability over a wide range of timescales - covering the intra-day variability (IDV), from rapid fluctuations to slower, systematic trends, reflecting changes in the physical conditions within the emitting region of the relativistic jet.
Furthermore, we employed various timing techniques for assessing and 
understanding the flux variability seen in the long-term light-curve and the individual light curves of Mrk~421. The methods used and the results are described in the following subsections.

\startlongtable
\begin{deluxetable*}{llccccl}
\tablecaption{Archival \textit{XMM-Newton}/pn observation log of Mrk~421.
Columns list observation IDs (Obs. ID), observation dates (Obs. Date),  the exposure ID (Expo. ID), the mode of observation, the total exposure time, the mean count rate for the light curve ($0.3-10$ keV) of each observation and the fractional variability.\label{tab:observation_log1}}
\tablehead{
\colhead{Obs. ID} &
\colhead{Obs. Date} &
\colhead{Expo. ID} &
\colhead{Mode} &
\colhead{Expo. Time (ks)} &
\colhead{Mean Count Rate (count/s)} &
\colhead{$F_{\rm var}$ (\%)}
}

\startdata
0099280101 & 2000-05-25 & S010 & IMAGING & 12.39 & $300.69 \pm 0.41$ & $10.87 \pm 0.14$ \\
0099280201 & 2000-11-02 & S010 & IMAGING & 24.24 & $115.55 \pm 0.09$ & $9.62 \pm 0.08$ \\
0099280301 & 2000-11-13 & S010 & IMAGING & 25.64 & $388.30 \pm 0.29$ & $3.84 \pm 0.08$ \\
0136540101 & 2001-05-08 & S008 & IMAGING & 25.73 & $384.77 \pm 0.28$ & $6.27 \pm 0.07$ \\
0153950701 & 2002-05-05 & S005 & IMAGING & 15.85 & $83.80 \pm 0.18$ & $1.21 \pm 0.42$ \\
0136540801 & 2002-11-14 & S008 & IMAGING & 5.49 & $475.29 \pm 0.92$ & $7.52 \pm 0.20$ \\
0136541001 & 2002-12-01 & S008 & TIMING & 54.33 & $244.02 \pm 0.08$ & $3.46 \pm 0.03$ \\
0136541101 & 2002-12-02 & S008 & IMAGING & 7.24 & $302.74 \pm 0.65$ & $4.13 \pm 0.23$ \\
0136541201 & 2002-12-03 & S008 & IMAGING & 7.11 & $328.72 \pm 0.64$ & $0.95 \pm 0.34$ \\
0158970101 & 2003-06-01 & U002 & IMAGING & 22.53 & $256.87 \pm 0.22$ & $6.43 \pm 0.09$ \\
0158970201 & 2003-06-02 & S009 & IMAGING & 14.67 & $184.77 \pm 2.92$ & $-- \pm --$ \\
0150498701 & 2003-11-14 & S003 & TIMING & 17.76 & $682.16 \pm 0.34$ & $8.35 \pm 0.05$ \\
0162960101 & 2003-12-10 & S007 & IMAGING & 13.48 & $310.86 \pm 0.30$ & $3.03 \pm 0.10$ \\
0158971201 & 2004-05-06 & S003 & TIMING & 6.37 & $715.17 \pm 0.26$ & $6.86 \pm 0.04$ \\
0153951201 & 2005-11-07 & S005 & TIMING & 3.78 & $307.20 \pm 0.38$ & $5.80 \pm 0.13$ \\
0158971301 & 2005-11-09 & S003 & TIMING & 30.75 & $361.24 \pm 0.12$ & $10.12 \pm 0.03$ \\
0302180101 & 2006-04-29 & S003 & TIMING & 24.98 & $302.19 \pm 0.16$ & $2.57 \pm 0.05$ \\
0411080301 & 2006-05-28 & S003 & IMAGING & 29.61 & $843.64 \pm 0.37$ & $4.86 \pm 0.04$ \\
0411080701 & 2006-12-05 & S003 & TIMING & 17.40 & $264.38 \pm 0.13$ & $1.22 \pm 0.05$ \\
0510610201 & 2007-05-08 & S003 & TIMING & 16.71 & $140.04 \pm 0.11$ & $2.03 \pm 0.08$ \\
0510610101 & 2007-05-08 & S003 & TIMING & 11.04 & $228.01 \pm 0.16$ & $1.11 \pm 0.08$ \\
0411081301 & 2007-05-10 & S003 & IMAGING & 9.28 & $238.06 \pm 0.51$ & $1.16 \pm 0.37$ \\
0411081401 & 2007-05-10 & S003 & IMAGING & 4.46 & $262.84 \pm 6.12$ & $-- \pm --$ \\
0411081501 & 2007-05-10 & S003 & IMAGING & 5.35 & $262.59 \pm 6.77$ & $-- \pm --$ \\
0411081601 & 2007-05-10 & S003 & IMAGING & 3.17 & $255.04 \pm 0.92$ & $1.16 \pm 0.61$ \\
0502030101 & 2008-05-07 & S003 & TIMING & 27.64 & $596.99 \pm 0.16$ & $7.21 \pm 0.03$ \\
0411082701 & 2008-05-31 & U002 & IMAGING & 4.50 & $985.60 \pm 1.42$ & $1.18 \pm 0.18$ \\
0560980101 & 2008-11-22 & S600 & IMAGING & 8.48 & $306.04 \pm 0.50$ & $1.97 \pm 0.19$ \\
0560983301 & 2009-05-25 & S600 & IMAGING & 8.47 & $268.09 \pm 0.46$ & $2.10 \pm 0.20$ \\
0411083201 & 2009-11-16 & S600 & IMAGING & 6.60 & $503.60 \pm 4.07$ & $-- \pm --$ \\
0656380101 & 2010-05-03 & S600 & IMAGING & 6.36 & $446.89 \pm 0.84$ & $-- \pm --$ \\
0656380801 & 2010-11-12 & S600 & IMAGING & 7.63 & $301.56 \pm 0.52$ & $-- \pm --$ \\
0656381301 & 2010-11-14 & S600 & IMAGING & 7.64 & $226.88 \pm 0.41$ & $1.78 \pm 0.22$ \\
0658800101 & 2011-05-17 & S600 & IMAGING & 4.86 & $158.91 \pm 0.39$ & $2.91 \pm 0.27$ \\
0658800801 & 2011-11-29 & S600 & TIMING & 8.53 & $76.94 \pm 0.10$ & $1.73 \pm 0.15$ \\
0670920301 & 2014-04-29 & S003 & TIMING & 8.59 & $463.06 \pm 0.54$ & $2.77 \pm 0.12$ \\
0670920401 & 2014-05-01 & S003 & TIMING & 13.48 & $275.78 \pm 0.16$ & $6.70 \pm 0.06$ \\
0670920501 & 2014-05-03 & S003 & TIMING & 11.29 & $342.45 \pm 0.21$ & $1.63 \pm 0.06$ \\
0658801301 & 2015-06-06 & S003 & IMAGING & 19.27 & $339.11 \pm 0.23$ & $6.44 \pm 0.07$ \\
0658801801 & 2015-11-08 & S003 & IMAGING & 21.20 & $205.51 \pm 0.17$ & $7.55 \pm 0.08$ \\
0658802301 & 2016-05-06 & S003 & IMAGING & 19.54 & $189.28 \pm 0.17$ & $4.76 \pm 0.09$ \\
0791780101 & 2016-11-03 & S001 & IMAGING & 11.21 & $70.37 \pm 0.10$ & $0.64 \pm 0.32$ \\
0791780601 & 2017-05-04 & S001 & IMAGING & 7.71 & $404.01 \pm 0.39$ & $0.57 \pm 0.16$ \\
0791781401 & 2017-11-20 & S601 & IMAGING & 7.71 & $428.34 \pm 0.43$ & $2.48 \pm 0.11$ \\
0791782001 & 2018-05-10 & S604 & IMAGING & 7.70 & $155.36 \pm 0.30$ & $3.72 \pm 0.21$ \\
0810860201 & 2018-11-22 & S601 & IMAGING & 10.96 & $348.02 \pm 0.33$ & $2.18 \pm 0.10$ \\
0810860701 & 2019-05-06 & S601 & IMAGING & 7.11 & $300.60 \pm 0.38$ & $1.97 \pm 0.14$ \\
0845000901 & 2019-06-10 & S015 & TIMING & 10.19 & $600.31 \pm 0.20$ & $2.93 \pm 0.03$ \\
0810862501 & 2019-11-29 & S601 & TIMING & 7.47 & $314.86 \pm 0.22$ & $0.32 \pm 0.12$ \\
0850380301 & 2019-11-30 & S001 & TIMING & 18.70 & $322.61 \pm 0.14$ & $0.28 \pm 0.08$ \\
0850380401 & 2019-12-02 & S001 & TIMING & 20.68 & $377.49 \pm 0.17$ & $2.48 \pm 0.04$ \\
0810863001 & 2020-05-01 & S602 & IMAGING & 7.11 & $252.92 \pm 0.41$ & $0.49 \pm 0.42$ \\
0871590201 & 2020-11-05 & S604 & IMAGING & 5.28 & $780.59 \pm 1.02$ & $0.26 \pm 0.49$ \\
0870190301 & 2020-11-15 & S001 & TIMING & 10.82 & $384.26 \pm 0.23$ & $1.15 \pm 0.06$ \\
0853600201 & 2020-11-19 & U002 & TIMING & 8.42 & $183.51 \pm 0.17$ & $1.02 \pm 0.11$ \\
0853600301 & 2020-11-23 & S003 & TIMING & 8.79 & $189.85 \pm 0.22$ & $1.69 \pm 0.13$ \\
0870190501 & 2020-12-11 & S001 & TIMING & 11.22 & $285.93 \pm 0.20$ & $1.33 \pm 0.07$ \\
0870190601 & 2020-12-13 & S001 & TIMING & 21.86 & $188.91 \pm 0.10$ & $2.33 \pm 0.06$ \\
0871590701 & 2021-05-29 & S602 & IMAGING & 7.07 & $134.06 \pm 0.27$ & $-- \pm --$ \\
0902110401 & 2022-05-03 & S001 & TIMING & 5.21 & $125.85 \pm 0.19$ & $1.09 \pm 0.18$ \\
0910790101 & 2022-05-04 & S001 & TIMING & 14.58 & $186.25 \pm 0.12$ & $1.59 \pm 0.07$ \\
0871591801 & 2022-05-15 & S602 & IMAGING & 7.11 & $340.61 \pm 0.42$ & $1.33 \pm 0.15$ \\
0891200101 & 2022-06-04 & S001 & TIMING & 21.32 & $196.84 \pm 0.18$ & $2.41 \pm 0.10$ \\
0902111701 & 2022-12-07 & S001 & TIMING & 5.31 & $277.63 \pm 0.25$ & $0.61 \pm 0.11$ \\
0902112401 & 2023-12-07 & S001 & TIMING & 9.18 & $285.02 \pm 0.19$ & $2.22 \pm 0.07$ \\
0902112501 & 2023-12-08 & S001 & TIMING & 15.00 & $277.61 \pm 0.17$ & $1.77 \pm 0.06$ \\
0920900201 & 2023-12-11 & S004 & TIMING & 9.73 & $345.56 \pm 0.20$ & $1.06 \pm 0.06$ \\
0891200201 & 2023-12-13 & S001 & TIMING & 16.93 & $303.02 \pm 0.18$ & $2.71 \pm 0.06$ \\
0920901301 & 2023-12-14 & S004 & TIMING & 9.37 & $515.97 \pm 0.33$ & $6.04 \pm 0.06$ \\
0934010101 & 2024-05-03 & U002 & TIMING & 14.96 & $104.42 \pm 0.09$ & $4.10 \pm 0.09$ \\
0914191801 & 2024-05-06 & S604 & IMAGING & 7.74 & $158.49 \pm 0.28$ & $0.90 \pm 0.32$ \\
0934010201 & 2024-05-09 & U002 & TIMING & 25.20 & $271.22 \pm 0.13$ & $1.50 \pm 0.05$ \\
0920901401 & 2024-05-13 & S004 & TIMING & 7.69 & $599.30 \pm 0.50$ & $1.51 \pm 0.09$ \\
0920902101 & 2024-05-14 & S004 & TIMING & 3.95 & $348.93 \pm 0.46$ & $1.76 \pm 0.14$ \\
0914192901 & 2024-11-25 & S602 & IMAGING & 11.24 & $132.16 \pm 0.21$ & $1.63 \pm 0.21$ \\
0914193701 & 2025-06-07 & S602 & IMAGING & 8.42 & $118.32 \pm 0.20$ & $2.79 \pm 0.19$ \\
\enddata
\end{deluxetable*}
\noindent\textit{Note.} A dashed entry (``--'') indicates that the total variance of the observation is smaller than the mean square of the measurement uncertainties.

\subsection{Fractional variability ($F_{\rm var}$)}
The estimation of the fractional variability (FV) allows us to measure
the extent and nature of the intrinsic variability present in the light curve of any astrophysical source. It is defined as 
\begin{equation}
F_{\mathrm{var}} = \sqrt{\frac{S^2 - \bar{\sigma}_{\mathrm{err}}^2}{\bar{X}^2}}
\end{equation}
\citep{vaughan2003characterizing,edelson1990broad}. 
Here, \( S^2 \) is the total variance of the light curve, $\bar{\sigma}_{\mathrm{err}}^2$ is the mean square of the measurement uncertainties, and $\bar{X}$ is the mean flux.

The associated error in fractional variability is obtained using the following equation:
\begin{equation}
\sigma_{F_{\mathrm{var}}} = \sqrt{
\left( \frac{1}{\sqrt{2N}} \cdot \frac{\bar{\sigma}_{\mathrm{err}}^2}{F_{\mathrm{var}} \bar{X}^2} \right)^2
+ \left( \sqrt{\frac{\bar{\sigma}_{\mathrm{err}}^2}{N}} \cdot \frac{1}{\bar{X}^2} \right)^2 }
\end{equation}
\citep{vaughan2003characterizing, Aleksic2015,bhatta2018microvariability}.
The fractional variability quantifies the intrinsic variability amplitude of the source after accounting for measurement uncertainties. A larger value of $F_{\mathrm{var}}$ indicates stronger flux variations and hence a higher degree of intrinsic variability, whereas smaller values correspond to a relatively stable emission state.

The fractional variability for the full $0.3-10$ keV band across all 76 epochs is presented in Table 1, demonstrating substantial long-term changes with $F_{\text{var}}$ ranging from a minimum of $\sim 0.26\%$ during quiescent states to a maximum of $\sim 10.87\%$ during intense flaring episodes.
This signifies substantial long-term changes in the underlying particle acceleration and cooling processes within the
relativistic jet. As the observations are distributed irregularly over $2.5$ decades and correspond to 
different observing campaigns and source states, the light curve primarily traces the long-term evolution
of the X-ray activity rather than continuous temporal behavior. Nevertheless, the pn monitoring provides 
a valuable overview of the broadband X-ray variability properties of the source across a wide range of 
accretion and jet activity conditions.
 \begin{figure}[!b]
    \centering
        \includegraphics[scale=0.65]{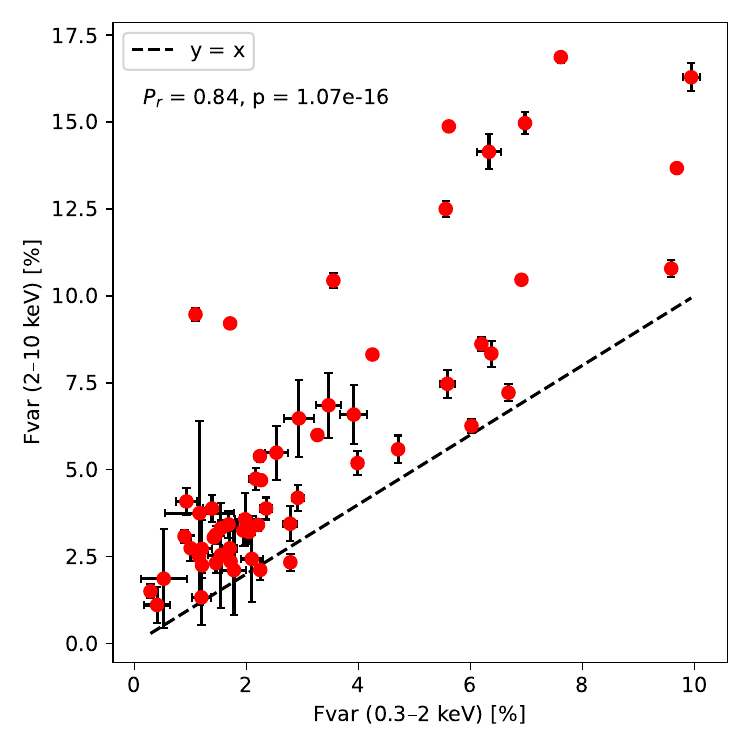}
    \caption{Fractional variability of hard band ($2-10$ keV) of all the observational epochs plotted against the soft band ($0.3-2$ keV). A line $y=x$ which would pass through the origin is shown by the black dashed line.}
    \label{fig:Fvar-Hard-soft}
\end{figure}

The correlation between the fractional variability measured between the soft X-ray band and hard X-ray band for all the light curves of Mrk~421 is presented in Fig.~\ref{fig:Fvar-Hard-soft}. 
The figure demonstrates that the variability amplitude in the hard X-ray band grows systematically with the variability measured in the soft band. A line $y=x$ is also overplotted. The Pearson correlation coefficient $P_r=0.84$ with an associated probability of $p=1.07\times 10^{-16}$ demonstrates that the correlation is highly significant. 
The best fitting slope is $1.41\pm0.12$; it's being larger than unity indicates that the value of $F_{\rm var}$ in the hard X-rays tend to be larger than the soft X-ray
for the most variable epochs further suggesting that the higher energy electrons responsible for the emission of the hard X-ray emission undergo more rapid changes in their energy distribution compared to the low energy electrons producing soft X-rays. This trend is consistent with synchrotron emission models in which the high energy tail of the electron population is more sensitive to fluctuations in particle acceleration efficiency. Overall, the tight correlation and the steep slope support a scenario in which the soft and hard X-ray bands originate from closely related electron populations within the same emission region, but with the higher-energy electrons exhibiting intrinsically stronger variability.

\subsection{Flux distributions}
A flux distribution is a measure of frequency of the source appearance at 
different brightness level. This is usually 
obtained by plotting a histogram of flux values observed over time in the light curve of the source, and it is a useful tool to constrain the origin of the variability in the blazar light curve.
\begin{figure}
   \centering
  \includegraphics[width=0.9\linewidth]{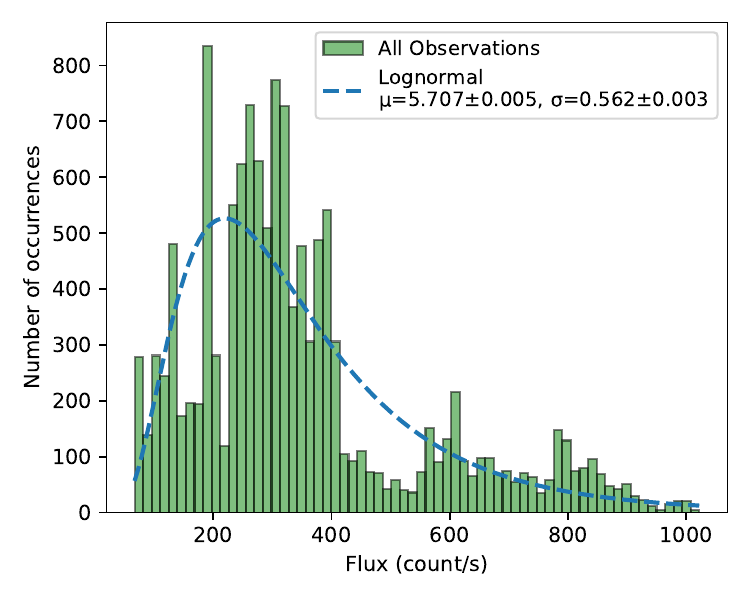}
    \caption{Flux distribution histogram of Mrk~421 obtained by combining all the observations of \textit{XMM-Newton}/pn.}
      \label{fig:flux-dist}
\end{figure}

The flux distribution constructed from all combined \textit{XMM-Newton}/pn observations is shown by the histogram in Fig.~\ref{fig:flux-dist}. The distribution is clearly non-Gaussian and asymmetric, reflecting the strong long-term variability of the source. Flux values span up to $1000$ counts s$^{-1}$, with the majority of occurrences concentrated below $400$ counts s$^{-1}$. The distribution exhibits a prominent peak at low-to-intermediate flux levels (with a maximum occurrence count of $\sim 830$), followed by an asymmetric tail extending toward higher fluxes. This extended high-flux tail, reaching values beyond $\sim 1000$ counts s$^{-1}$, indicates the presence of rare but intense flaring episodes, which are also visible in the full light curve (Fig.~\ref{fig:LC-whole}, notably Obs. ID 0411082701 observed on 2008-05-31). \textit{XMM-Newton} observations provide targeted sampling of Mrk~421 across a range of activity states, with the resulting flux distribution reflecting the observed activity states.

To investigate the nature of the flux variability, the flux distribution was fitted with a lognormal function of the form

\begin{equation}
P(x)=\frac{A}{x\sigma\sqrt{2\pi}}
\exp\left[-\frac{(\ln x-\mu)^2}{2\sigma^2}\right],
\end{equation}
where $x$ denotes the observed count rate, $A$ is a normalization constant, and $\mu$ and $\sigma$ represent the mean and standard deviation of ln $x$, respectively. The best-fitting lognormal model, with $\mu = 5.707\pm0.005$ and $\sigma = 0.562\pm0.003$, is overlaid on the histogram.

The strong skewness and long-tail behavior characteristic of this distribution are consistent with multiplicative variability processes. Such a departure from a Gaussian shape suggests that Mrk~421 does not vary around a single characteristic flux level, but rather occupies multiple distinct X-ray emission states over long timescales. Overall, the combined histogram demonstrates that the source spends most of its time in low-to-intermediate flux states, while intermittently transitioning to the significantly bright episodes. 
This highlights the highly variable nature of the X-ray emission and underscores the importance of long-term monitoring to study its non-linear variability dynamics. Recently, \citet{Kizhakkekalam2026}, from their analysis of 45 \textit{NICER} observations of Mrk~421 spanning 2 years of observations, identified a complex, multi-modal flux distribution with a primary peak corresponding to an intermediate flux state and a tail extending toward higher flux levels.

\begin{figure}
   \centering
  \includegraphics[width=0.9\linewidth]{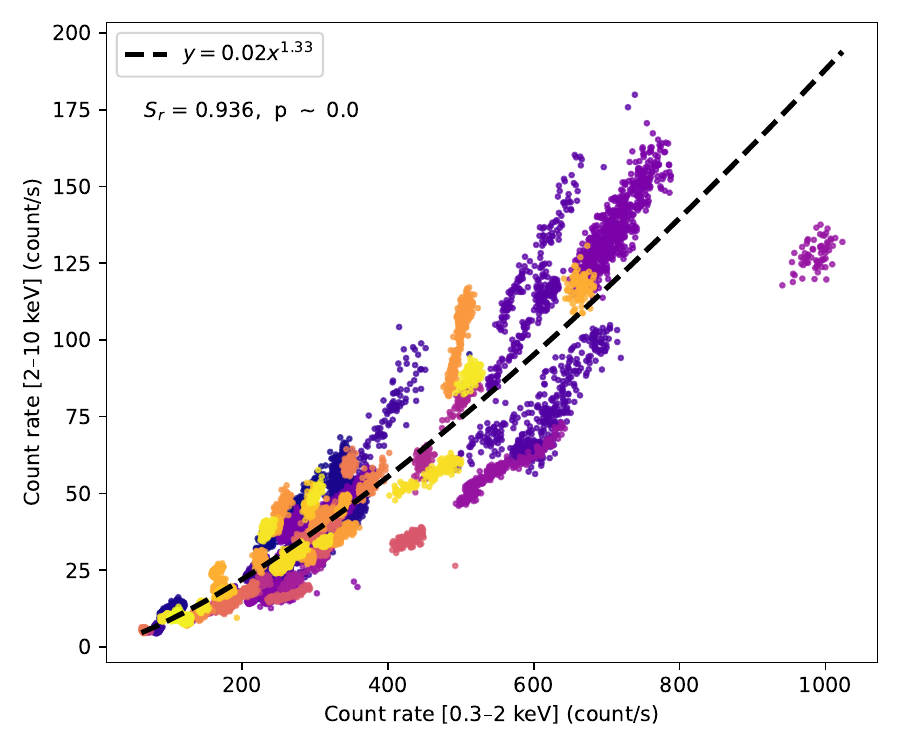}
    \caption{Count rates of hard X-ray ($2-10$ keV) light curve plotted against the count rates of soft X-ray ($0.3-2$ keV) light curve. The power-law representation of the correlation among the hard and soft count rates is shown by the dashed line.}
      \label{fig:soft-hard-counts}
\end{figure}

We further examine the relation between soft‑band (0.3–2 keV) and hard‑band (2–10 keV) X‑ray count rates as demonstrated in Fig~\ref{fig:soft-hard-counts}.
A strong positive non‑linear correlation is observed, well described by a power‑law of the form 
$y =(0.019 \pm 0.001)x^{1.33 \pm 0.005}$ with Spearman correlation coefficient $S_r=0.936$,
indicating that hard and soft bands are coupled. The power-law index being greater than unity demonstrates that hard‑band flux increases more rapidly than soft‑band flux with rising source brightness, independently confirming the harder‑when‑brighter spectral behaviour seen in other analyses. Despite the overall trend, several branches and clustered tracks are visible, suggesting that the spectral evolution is not fully identical across different activity states.

\section{Spectral Analysis}
\label{sec:spec}
We investigated the X-ray spectral evolution of Mrk~421 using two complementary approaches. The first is a model-independent analysis based on the X-ray HR, which provides a direct probe of spectral variability and enables the characterization of spectral hardening and softening independent of any assumed spectral model. The second approach consists of quantitative spectral fitting using the \texttt{XSPEC} package, where the observed spectra are modeled with appropriate spectral functions to derive the underlying spectral parameters and examine their temporal evolution. The combination of these two methods provides both a robust empirical characterization of the spectral variability and a physically interpretable description of the source's X-ray emission.

\subsection{Hardness ratio (HR)} 
The hardness ratio (HR) is calculated by using the equation
\begin{equation}
    HR=\frac{H}{S},
\end{equation}
where H and S are the count rates in the soft (0.3 -- 2 keV), and hard (2 -- 10 keV) energy bands, respectively. The HR error is calculated by using the expression
\begin{equation}
\sigma_{\rm HR}
= \frac{H}{S}\,
\sqrt{
\left( \frac{\sigma_H}{H} \right)^{2}
+
\left( \frac{\sigma_S}{S} \right)^{2}
}
\end{equation}
where $\sigma_H$ and $\sigma_S$ are the uncertainties in $H$ and $S$, respectively. The HR plot tracks the changes in the X-ray spectrum with flux and reveals whether the spectral changes are driven by particle accleration or cooling. 
It can also provide information on changes in the spectral shape during different flux states.

\begin{figure}[!t]
   \centering
  \includegraphics[scale=0.6]{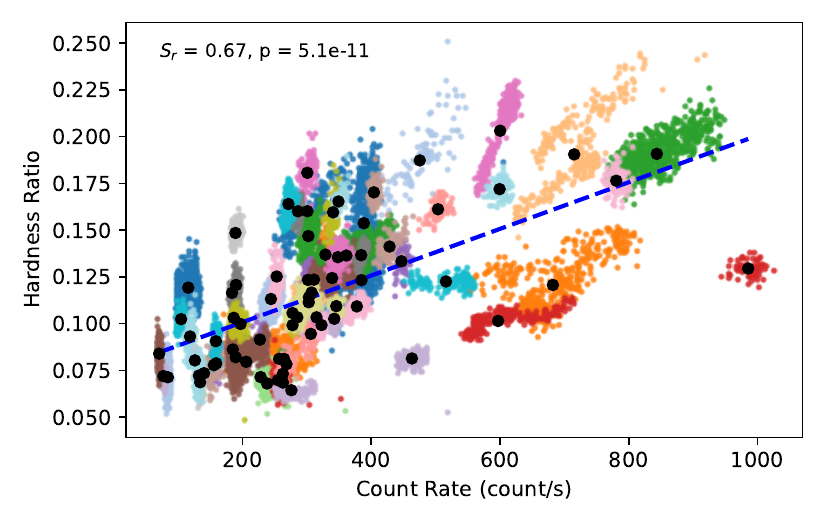}
     \caption{ Hardness Ratio (HR) evolution over 25 years of observations of Mrk~421 where the 76 epochs of observations (represented by different colors) are plotted together. The average HR for each epoch is shown by a black solid circle.}
      \label{fig:hardness-ratio}
\end{figure}

A distribution of spectral HR as a function of the X-ray ($0.3-10$ keV) count rates
for all 76 \textit{XMM-Newton}/pn epochs, is presented in 
Fig.~\ref{fig:hardness-ratio}, where a pronounced long-term spectral 
variability is observed while the value of HR varies between $\sim$~$0.05-0.25$. The black solid circles are the average HR value of each epoch of observation. The Spearman correlation analysis between the average HR and the average mean count rate shows a significant harder-when-brighter trend,
indicating that the X-ray spectrum generally hardens as the source brightens. 
In particular, the low-flux states ($<200$ count/s) are predominantly associated 
with softer spectra with HR values around $\sim~0.06-0.12$, whereas the brighter 
states ($>~600$ count/s) preferentially occupy harder spectral regions with HR 
values reaching $\sim$~$0.18-0.25$. However, the HR distribution exhibits a 
visible scatter and several partially separated spectral tracks, 
implying that similar flux levels can correspond to different spectral states. 
Such behavior suggests that the spectral evolution of Mrk~421 is not solely 
determined by the instantaneous flux level, but also depends on the temporal 
state of the source and the physical conditions within the emitting region. 
A similar trend for HR but for a few number of observations has also 
been reported from {\it Swift}-XRT observations of this source 
by \citet{Tramacere2009}.
The distribution further reveals the presence of multiple spectral branches with 
different HR evolution patterns across different observational epochs. Several tracks 
display a strong positive HR - flux correlation, while others remain comparatively 
flat over a narrower hardness range. 
In particular, the observations corresponding to the highest flux states ($\geq$700 count/s)
show substantial spectral hardening, whereas some intermediate-flux states 
($\sim$300-500 count/s) exhibit a broader dispersion in HR values ranging 
from $\sim$ 0.10 to $\sim$ 0.18. 

Spectral hysteresis has been widely used as a diagnostic of time-dependent particle acceleration and synchrotron cooling in X-ray blazars. Loop-like patterns in hardness--flux or photon index--flux diagrams have been reported for several HSP blazars, including Mrk~421, PKS~2155--304, and H~1426+428, where the X-ray spectrum generally hardens during the rising phase of a flare and softens during its decay \citep{Takahashi1996,Kataoka2000,Falcone2004,Tramacere2009,Abeysekara2017}.

\begin{figure}[!t]
   \centering
   \includegraphics[scale=0.91]{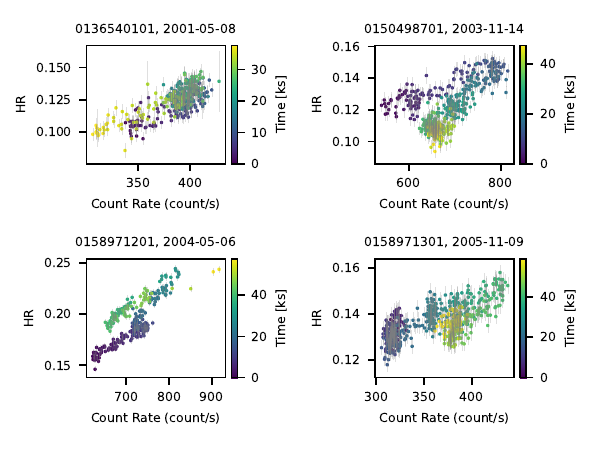}
    \caption{Representative examples of the HR variations during four selected observational epochs with Obs. IDs: 0136540101, 0150498701, 0158971201 and 0158971301 displaying pronounced rising and falling flux states within individual observations. 
    The temporal evolution is color coded according to the elapsed time since the start of the observation.}
      \label{fig:hyst}
\end{figure}
To further investigate the complex HR variability patterns observed in the
overall HR distribution, we selected four representative epochs 
(Obs. IDs: 0136540101, 0150498701, 0158971201, and 0158971301) and 
examined their intra-observation HR evolution, as shown in Fig.~\ref{fig:hyst}. 
In general, the source exhibits a 
harder-when-brighter behavior, consistent with the trend observed from 
the full sample of 76 observations, however, the temporal evolution within
individual epochs reveals a more complex behavior than a simple monotonic 
HR-flux correlation. In particular, the source frequently occupies two
distinct HR states at the similar flux level during different 
stages of the observation forming a hysteresis loop.

From our selected single‑epoch observations, the 2001‑05‑08 and 2004‑05-06 datasets show clear clockwise HR hysteresis. As time progresses, flux rises with harder spectra and decays with softer spectra. In contrast, the 2003‑11‑14 and 2005‑11‑09 epochs display complex two‑stage HR evolution: an initial mild counter-clockwise tendency, where flux rise accompanies weak spectral softening, is followed by a harder‑when‑brighter phase with increasing HR up to peak flux, after which both flux and HR decrease.

\subsection{Spectral Modeling}
Spectral analysis of all XMM-Newton observations was performed using \texttt{XSPEC} version 12.13.1 \citep{Arnaud1996}, and the
$\chi^{2}$ statistic was used to assess the quality of the spectral fits.

The X-ray emission of HSP  blazars is widely attributed to synchrotron radiation from relativistic electrons accelerated within the jet. In the framework of diffusive shock acceleration, the injected electron population is expected to follow a power-law energy distribution, resulting in a power-law synchrotron spectrum over the X-ray band. However, synchrotron cooling is inherently energy dependent, with the radiative cooling timescale scaling as $t_{\rm syn}\propto\gamma^{-1}$, where $\gamma$ is the electron Lorentz factor. Consequently, higher-energy electrons cool more rapidly than lower-energy electrons, leading to spectral steepening and curvature in the X-ray spectrum.

Motivated by these physical considerations, we employed three commonly adopted models to characterize the X-ray spectra. We first considered a simple power-law (PL) model to represent the synchrotron spectrum produced by a power-law distribution of relativistic electrons.
A broken power-law (BPL) model was then employed to account for spectral steepening associated with energy-dependent synchrotron cooling, which may produce a distinct spectral break in the X-ray band. Finally, a log-parabola (LP) model was fitted to the spectra to describe smoothly curved continua that may arise from gradual energy-dependent particle acceleration and/or radiative cooling.
Here, the PL model is defined as
\begin{equation}
    A(E)=N_{\rm f}E^{-\Gamma},
\end{equation}
where $\Gamma$ is the spectral index and $N_{\rm f}$ is the normalization factor. The broken power-law (BPL) model is given by

\begin{equation}
A(E) = 
\begin{cases}
N_{\rm f} E^{-\Gamma_1} & \text{if } E \leq E_{\text{b}} \\
N_{\rm f} E_{\text{b}}^{\Gamma_2 - \Gamma_1} \left( \frac{E}{1\,\text{keV}} \right)^{-\Gamma_2}, & \text{if } E > E_{\text{b}}
\end{cases}
\end{equation}
where the parameters $E_{\rm b}$, $\Gamma_1$, $\Gamma_2$ and $N_{\rm f}$ are the break energy, photon indices (low and high energy) and the normalization constant respectively. Finally, the log-parabola (LP) model is expressed as,
\begin{equation}
A(E) = N_{\rm f} \left( \frac{E}{E_1} \right)^{ -\alpha - \beta \log\left( \frac{E}{E_1} \right) }, 
\end{equation}
where the parameters $E_{\rm 1}$, $\alpha$, $\beta$ and $N_{\rm f}$ are the pivot energy fixed at 1 keV, slope at the pivot energy, spectral curvature term, and the normalization constant respectively. 

To account for the Galactic absorption, we multiplied all the \texttt{XSPEC} models by the Tuebingen--Boulder interstellar medium absorption model, \texttt {TBabs} \citep{Wilms2000} and fitted the X-ray spectra sequentially using (i) \texttt {TBabs$\times$pow}, (ii) \texttt {TBabs$\times$bknpower}, and (iii) \texttt {TBabs$\times$logpar}. Throughout the analysis, the absorption column density was fixed at $N_{\rm H} = 1.92\times10^{20}$ cm$^{-2}$ following the results from the Leiden/Argentine/Bonn (LAB) survey \citep{Kalberla2005}. For all the spectral fittings, we considered the $0.6-10$ keV X-ray energy range in order to minimize the effects of calibration uncertainties at lower energies.

To determine the statistically preferred spectral representation for each observation, we performed comparative fitting using the PL, BPL, and LP models. The significance of introducing additional spectral complexity in the BPL and LP models relative to the simple PL model was initially evaluated using the F-test \citep{Bevington1992}, and the corresponding values are presented in Col.~8 of Table~\ref{tab:fitting-results}. To complement the applicability of the F-test,
we further employed the corrected Akaike Information Criterion (AICc) \citep{Akaike1974, Sugiura1978, Liddle2007} and the Bayesian Information Criterion (BIC) \citep{1978AnSta...6..461S, Liddle2007} for the best fit model selection. The models favored by the AICc and BIC tests are listed in Col.~9 of Table~\ref{tab:fitting-results}. The final preferred model for each observation was selected by jointly considering the F-test significance together with the AICc and BIC results. In cases where the same spectral model was favored by two or more statistical tests, that model was adopted as the best-fit representation of the X-ray spectrum. This combined approach provides a more robust model-selection framework by balancing improvements in the goodness-of-fit against the increase in model complexity.

The results of the spectral fitting using the three different models are presented in Table~\ref{tab:fitting-results}. 
For a majority of the cases, an unbroken PL model is not an adequate fit, as reported previously for X-ray spectra of many blazars, including Mrk~421 \citep{Garson2010,Brinkmann2005,Ravasio2004}.

In the next step, we fitted the spectra with the BPL model, which allows the spectral slope to change at a break energy, $E_{\rm b}$, thereby naturally accounting for variations in the spectral shape between the soft and hard X-ray bands. We found that 59 X-ray spectra of Mrk~421 are well described by the BPL model, as indicated by the significant improvement in the reduced $\chi^2$ values presented in Col.~7 of Table~\ref{tab:fitting-results}. Most of the derived $E_{\rm b}$ values are distributed around $2$ keV, with some values scattered within the $\sim1.3-5.1$ keV range, while the spectra typically steepen above the break energy. 

Several previous X-ray studies of Mrk~421 have suggested the presence of intrinsic spectral curvature in its X-ray spectrum \citep{Massaro2004,Tramacere2009,Kizhakkekalam2026}, which may trace gradual changes in the underlying electron energy distribution. Motivated by these findings, we also fitted the spectra using the LP model to investigate the possibility of smooth spectral curvature. In 17 observations, the spectra are better represented by the LP model, yielding improved fit statistics compared to the BPL model. The spectral curvature parameter $\beta$ spans a broad range from weak to strong curvature, indicating substantial spectral variability among different epochs. All the fitting results are summarized in Table~\ref{tab:fitting-results}. 

Based on the combined statistical tests described above, we find that, out of the total 76 observations, 59 X-ray spectra are best represented by the BPL model, while the remaining 17 spectra are better described by the LP model. None of the observations are statistically favored by the simple PL model, indicating that the X-ray spectra of Mrk~421 are generally characterized either by a spectral break or by intrinsic spectral curvature within the $0.6-10$ keV energy range. 
We also identified that, in many instances, the difference in the statistical significance between the BPL and LP models is weak.

\subsection{Study of spectral parameter trends}
\begin{figure}[!t]
    \includegraphics[scale=0.5]{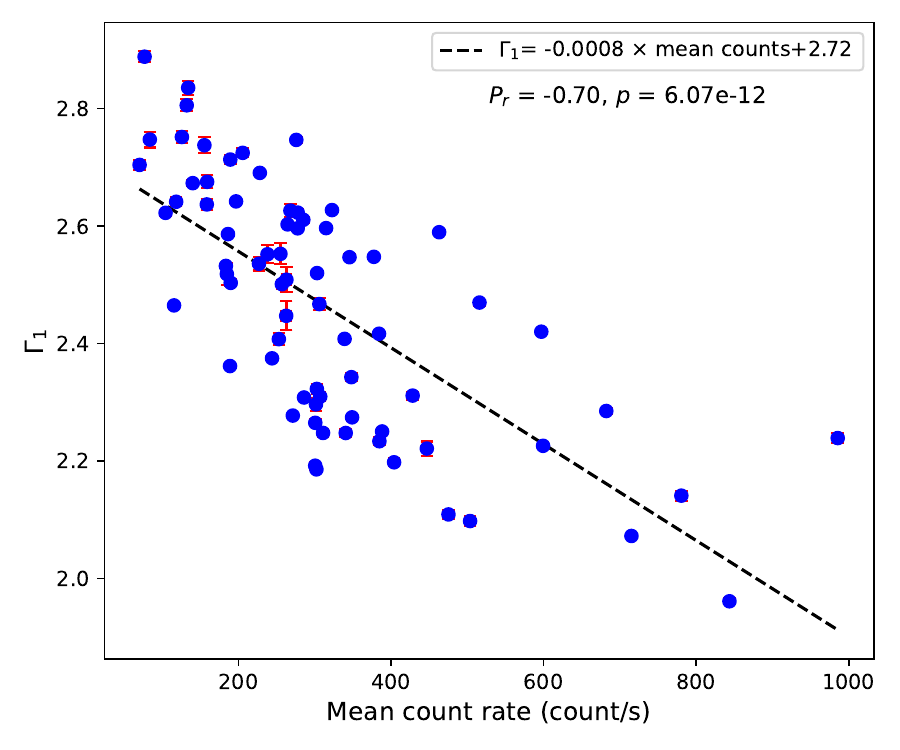}
    \includegraphics[scale=0.33]{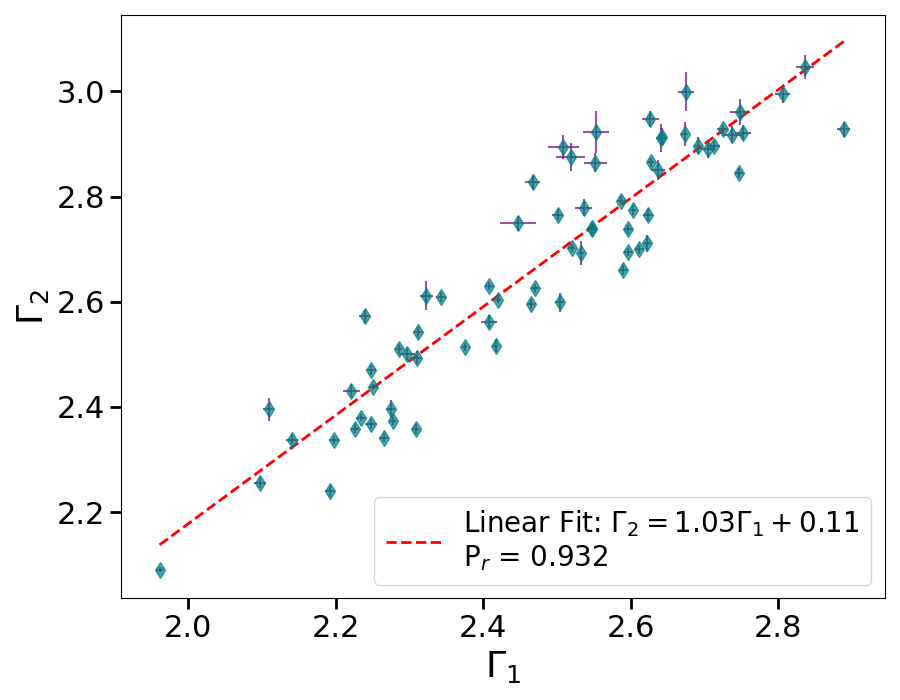}
    \caption{\textit{Top panel:} Pre-break spectral index $\Gamma_1$ plotted against the mean count rate across the observational epochs. \textit{Bottom panel:} correlation between the  pre-break and post-break BPL slopes $\Gamma_{1}$ and $\Gamma_{2}$.}
    \label{fig:Gamma1-mean-counts}
\end{figure}

We studied the correlations among the best-fit spectral parameters obtained using the BPL model for all
observational epochs. The top panel of Fig.~\ref{fig:Gamma1-mean-counts} shows a clear anti-correlation
between the low-energy photon index, $\Gamma_1$, and the mean count rate, confirming the 
well-known harder-when-brighter behavior previously inferred from the temporal analysis presented in
Section~\ref{sec:light-curve}. The Pearson correlation coefficient of $P_r =-0.70$ (with probability $p=6.07\times10^{-12}$) indicates that the 
trend is statistically significant. Similar spectral variability
has been widely reported in 
Mrk~421 and other HSP sources \citep{Brinkmann2005,Kizhakkekalam2026,Tramacere2009}.

The bottom panel of Fig.~\ref{fig:Gamma1-mean-counts} shows that
the two spectral indices,  $\Gamma_1$ and  $\Gamma_2$, are
strongly linearly correlated, with a best-fit slope of
approximately $1.03\pm$0.05 and a Pearson correlation coefficient of $P_r = 0.93$ (with the probability $p=2.88\times10^{-13}$). 
A similar correlation between the pre-break and post-break spectral slopes was
also reported by \citet{Garson2010} using \textit{Suzaku} observations of Mrk~421. 
The near one-to-one correspondence between  $\Gamma_1$ and  $\Gamma_2$ 
indicates that both the low and high energy segments of the X-ray spectrum 
evolve coherently during different flux states. The best-fit intercept further suggests an 
approximate relation of $\Gamma_2 \approx \Gamma_1 + 0.11$, implying
the presence of an approximately constant spectral steepening across the
observations. 

\begin{figure}
    \includegraphics[scale=0.60]{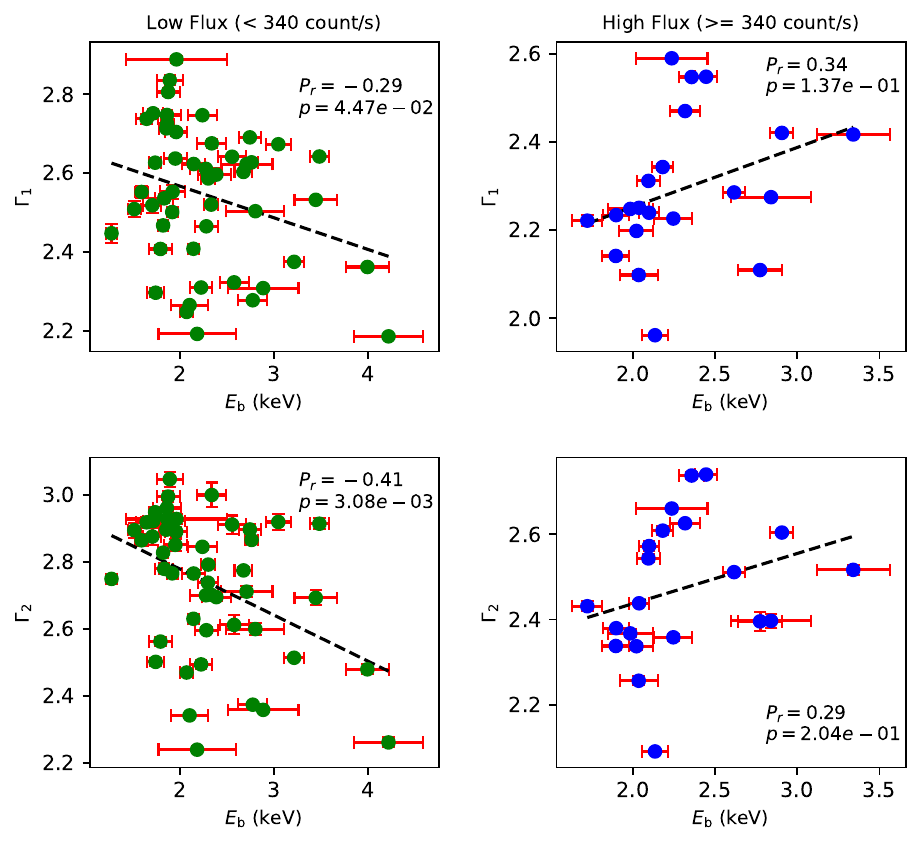}
    \caption{Flux-resolved correlations between spectral indices ($\Gamma_1$, $\Gamma_2$) and break energy $E_{\rm b}$ for Mrk~421. Data are divided into low-flux (mean counts $<340\;{\rm count\ s^{-1}}$) and high-flux (mean counts $\geq340\;{\rm count\ s^{-1}}$) subsets. A  reversal in correlation polarity is observed between these two source states.}
    \label{fig:Gamma_vs_BreakEnergy_FluxSplit}
\end{figure}

We further investigated how the low-energy ($\Gamma_1$) and high-energy ($\Gamma_2$) spectral indices correlate with the spectral break energy $E_{\rm b}$, splitting the analysis into two states defined by mean count rates: low flux ($<340~{\rm count\ s^{-1}}$) and high flux ($\geq340~{\rm count\ s^{-1}}$). The trends measured for the low- and high-flux states of Mrk~421 are presented in Fig.~\ref{fig:Gamma_vs_BreakEnergy_FluxSplit}. 
We find clear state-dependent
correlation behaviour between the spectral indices and $E_{\rm b}$. 
In the low-flux regime, both indices exhibit negative correlations 
with $E_{\rm b}$. Specifically, $\Gamma_1$ shows a marginally significant
anti-correlation ($P_r=-0.29$, $p=4.47\times10^{-2}$), while $\Gamma_2$
displays a much stronger, statistically robust 
anti-correlation ($P_r=-0.41$, $p=3.08\times10^{-3}$). 
This trend indicates that higher $E_{\rm b}$ correspond to steeper 
X-ray spectral slopes during faint source states \citep[see also][]{2020ApJS..247...27K,2016ApJ...831..102K}.
In contrast, the high-flux regime shows weak positive trends for both $\Gamma_1$ ($P_r=0.34, ~p=0.137$) and $\Gamma_2$ ($P_r=0.29, ~p=0.204$), although neither correlation is statistically significant.
Similar positive correlations of $\Gamma_1,\Gamma_2$ with $E_{\rm b}$ were reported by \citet{Garson2010} during the X-ray flare of Mrk~421 observed by the {\it Suzaku} X-ray observatory. 
This pronounced flux-driven
reversal in correlation sign demonstrates fundamentally different spectral
evolution behaviour between low- and high-brightness epochs. 
Such discrete, state-dependent variations stand in contrast to the smooth,
continuous spectral curvature described by the LP model adopted in earlier
studies by \citet{Tramacere2009,Tramacere2007}, implying that distinct
electron acceleration and radiative cooling physical regimes govern 
the jet emission across different flux levels.

\begin{figure}[!t]
    \includegraphics[scale=0.40]{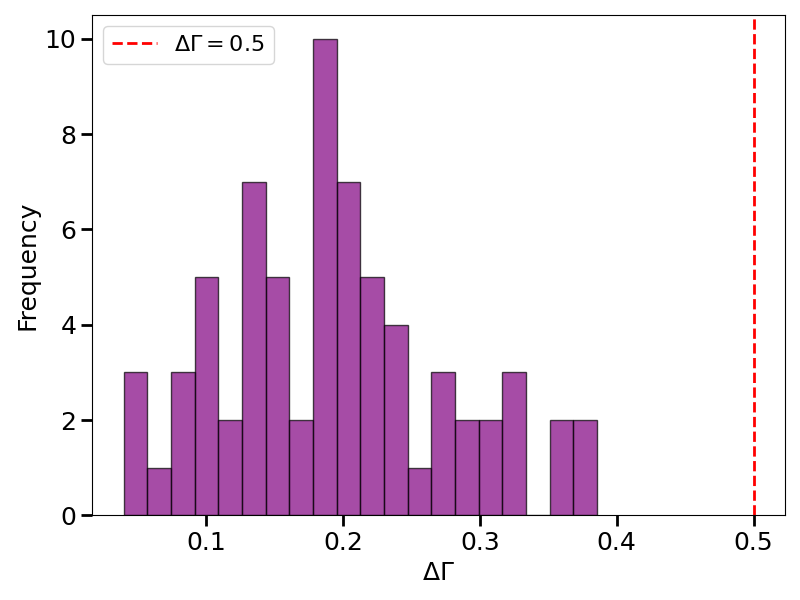}
    \includegraphics[scale=0.40]{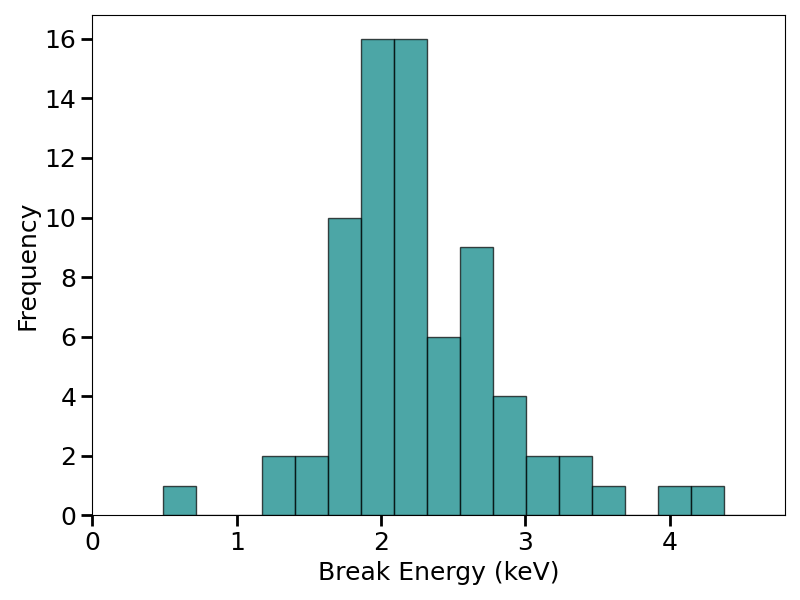}
    \caption{{\it Top panel:} Histogram of the difference between the pre-break ($\Gamma_1$) and post-break ($\Gamma_2$) photon indices, illustrating the degree of spectral steepening. The dashed red vertical line denotes the theoretical expectation for a standard single-shock synchrotron cooling scenario. {\it Bottom panel:}  Histogram displaying the distribution of the spectral break energy in keV.}
    \label{fig:breakE-mean-counts}
\end{figure}

The top panel of Fig.~\ref{fig:breakE-mean-counts} presents the
distribution of the spectral
steepening parameter, $\Delta \Gamma = \Gamma_2 - \Gamma_1$. 
The distribution is
strongly concentrated around $\Delta \Gamma \approx0.19$, with all
observations showing positive values, indicating that the spectrum consistently
becomes steeper above $E_{\rm b}$. 

The bottom panel of Fig.~\ref{fig:breakE-mean-counts} shows the relatively narrow distribution of $E_{\rm b}$ which suggests the presence of a characteristic energy scale in the emitting electron population. In blazar emission models, such a characteristic electron energy can arise from the balance between particle acceleration and radiative cooling or between radiative cooling and particle escape \citep{2002A&A...386..833G,1998ApJ...509..608T}. For Mrk~421, \citet{2022MNRAS.514.3074B} investigated the physical origin of the characteristic X-ray photon energy and found that its observed behavior is not consistent with a simple radiative-loss interpretation. Energy-dependent particle diffusion has also been proposed as a mechanism shaping the X-ray spectral properties of Mrk~421 \citep{2018MNRAS.480.2046G}. The persistence of $E_{\rm b}$ around ($\sim2$) keV over the long observational baseline considered here may therefore point to a relatively stable physical scale in the electron population or in the processes governing particle acceleration, cooling, and escape. 
 In HSP objects such as Mrk~421, 
where the synchrotron component peaks within the X-ray band, 
the $E_{\rm b}$ traces an intrinsic feature of the underlying electron
energy distribution. The observed stability of $E_{\rm b}$ may further indicate 
that the fundamental physical conditions within the emitting region, such as the
magnetic field strength or Doppler boosting factor, do not vary dramatically during
the observed flux changes. Consequently, the observed variability may primarily
arise from changes in the number or injection rate of relativistic particles 
rather than from major structural changes in the jet.

\section{Discussion and Conclusions}
\label{sec:discuss_conclude}
The long-term X-ray variability properties of Mrk~421 derived from the 2.5 decades 
of \textit{XMM-Newton}/pn observations reveal a dynamic synchrotron-emitting jet 
characterized by strong spectral and flux variability across timescales ranging from 
intra-day variations to decades. The source exhibits pronounced harder-when-brighter 
behavior in both the HR evolution and the spectral fitting results, indicating that 
the X-ray emission is governed by changes in the high-energy tail of 
the relativistic electron population.
The harder-when-brighter trend is commonly
interpreted within the standard blazar jet scenario, where enhanced particle
acceleration during flaring states increases the population of high-energy electrons,
producing a flatter X-ray spectrum at higher flux levels, while comparatively softer
states may arise during cooling-dominated phases following flaring activity.
The strong correlation between the soft and 
hard X-ray bands, together with the leading variability amplitude observed in the hard band, 
suggests that both energy ranges originate from closely connected electron 
populations within the same emission region, while the highest-energy electrons 
respond more rapidly to changes in the acceleration process.
This coupled soft-hard variability supports a scenario in which the electron population
responsible for the synchrotron emission evolves in a coupled manner over the observed
energy range, rather than through independent variations of separate spectral components.
Similar behavior has been widely observed in HSP blazars and is commonly 
interpreted in the framework of shock acceleration and synchrotron cooling 
operating within relativistic jets\citep[e. g.,][]{Kirk1998,Brinkmann2005,Tramacere2009}.
The presence of clockwise and counter-clockwise hysteresis patterns during different 
epochs further supports a time-dependent interplay between particle acceleration and
cooling processes, where the relative timescales of electron injection, 
acceleration, and radiative losses vary between different activity states
\citep[e.g., ][]{Takahashi1996,Zhang2002,Garson2010}.
Such behavior is generally
interpreted as the result of the competition between particle acceleration,
synchrotron cooling, and particle escape, with the loop direction reflecting the
relative ordering of these characteristic timescales \citep{Kirk1998,Li2000,Bottcher2002}.

Moreover, the spectral analysis demonstrates that the $0.6-10$ keV X-ray spectra of Mrk~421 
are generally inconsistent with a simple PL description and instead require either a 
BPL or a LP representation. We found that the observed spectral steepening remains systematically
smaller than the canonical value expected from the standard one-zone synchrotron cooling
scenario, with the distribution strongly concentrated around $\Delta\Gamma \sim 0.19$.
In the simplest synchrotron cooling picture, relativistic electrons cool
radiatively above a characteristic energy, producing a spectral steepening of
$\Delta\Gamma = 0.5$ above the cooling break \citep{Kardashev1962,Rybicki1979,Takahashi1996,Longair2011}.
The significantly smaller value of $\Delta\Gamma$ observed here therefore suggests that the simple
one-zone cooling interpretation is insufficient to explain the X-ray spectral properties of the source.
Such weak spectral breaks suggest that the X-ray emission region is likely more complex than a
simple homogeneous cooling zone. Continuous or stochastic particle acceleration, incomplete
synchrotron cooling, and energy-dependent particle transport may all contribute to softening the
expected cooling break \citep{Kirk1998,Massaro2004Mrk421,Tramacere2007}.
Alternatively, the weak spectral break may also arise from multi-zone emission scenarios within the jet
\citep{Tavecchio1998,Ghisellini2005} or from particle transport effects such as energy-dependent escape and
diffusion that can soften the canonical synchrotron cooling break \citep[e.g.,][]{Zheng2019}, naturally
producing the weaker spectral steepening observed in our data.
In addition, the break energy is predominantly clustered around $\sim2$ keV over 2.5 decades of observations. Here, the BPL model should therefore be regarded primarily as a phenomenological representation of the spectral steepening, rather than as evidence for a sharp break in the underlying electron energy distribution. The concentration of the fitted break energies may nevertheless indicate a characteristic energy scale in the observed X-ray spectra, although its physical interpretation should be treated with caution.
The flux-resolved analysis further reveals that the correlation between
the spectral indices ($\Gamma_1,\Gamma_2$) and break energy $E_{\rm b}$ changes sign between 
low and high flux states, implying that distinct physical regimes may dominate the spectral
evolution during quiescent and flaring phases. Such state-dependent behavior favors a 
scenario in which the jet emission is governed by evolving acceleration conditions 
and possibly multiple emitting zones rather than by a single stationary electron population.

To physically interpret the persistent spectral steepening ($\Delta\Gamma \approx 0.19$) observed across the 25-year dataset, we invoke a steady-state analytical solution to the electron continuity equation. We assume the X-ray emitting region is supplied by a continuous injection of energetic electrons from a primary acceleration zone (e.g., a shock front) following a power-law distribution, $Q(\gamma) = Q_0 \gamma^{-p}$, where $\gamma$ is the electron Lorentz factor. The steady-state spectrum is determined by the competition between two primary depletion mechanisms: radiative cooling (dominated by synchrotron emission in the X-ray band) and spatial escape from the active region. The synchrotron cooling timescale scales inversely with energy, $t_{\text{cool}} \propto \gamma^{-1}$. However, rather than assuming an energy-independent advective escape time, we adopt an energy-dependent spatial diffusion model, $t_{\text{esc}} \propto \gamma^{-k}$, where $k$ is the diffusion index \citep[see also][]{2018MNRAS.480.2046G,2018MNRAS.478L.105J}.
This energy-dependent escape provides a natural explanation for the weaker-than-canonical spectral steepening discussed above, by allowing the relative importance of escape and radiative cooling to vary with electron energy.
The resulting steady-state electron distribution, $N(\gamma) \approx Q(\gamma) \left( t_{\text{cool}}^{-1} + t_{\text{esc}}^{-1} \right)^{-1}$, produces a broken power-law shape. At lower energies, where $t_{\text{esc}} \ll t_{\text{cool}}$, the distribution is escape-dominated with $N(\gamma) \propto \gamma^{-(p+k)}$. At higher energies, where $t_{\text{cool}} \ll t_{\text{esc}}$, the distribution becomes cooling-dominated with $N(\gamma) \propto \gamma^{-(p+1)}$. The difference in these indices yields an electron spectral break of $\Delta p = 1 - k$, corresponding to an observable photon spectral steepening of $\Delta\Gamma = (1 - k) / 2$. Matching our observed median steepening of $\Delta\Gamma = 0.19$ requires an escape index of $k \approx 0.62$, indicating that the characteristic escape timescale decreases 
with increasing electron energy.

The physical consistency of this acceleration–escape scenario can be examined using the temporal and spectral constraints provided by our XMM-Newton observations. The characteristic intra-day variability timescale of Mrk~421, \(t_{\rm var}\sim10^{4}\) s (\(\sim2.8\) hr), constrains the comoving size of the emitting region through the causality condition \(R\lesssim c\,t_{\rm var}\delta/(1+z)\), yielding \(R\lesssim10^{16}\) cm for a representative Doppler factor of \(\delta\sim20\). Comparable or even shorter X-ray variability timescales have previously been reported for Mrk~421 \citep[e. g.,][]{Brinkmann2005,Aggrawal2018,Gokus2024}. Furthermore, adopting representative jet parameters of \(B\sim0.1\) G and \(\delta\sim20\), electrons producing synchrotron emission near the observed break energy, \(E_{\rm b}\simeq2\) keV, have Lorentz factors of order \(\gamma_{\rm b}\sim10^{5}\)–\(10^{6}\). The corresponding observer-frame synchrotron cooling timescale, $t_{\mathrm{syn, obs}} \approx 7.7 \times 10^8 [(1+z)/\delta] (1\mathrm{\,G}/B)^2 (1/\gamma_{\mathrm{b}})$, is of order \(10^{4}\) s, comparable to the observed intra-day variability timescale. If the spectral break represents the transition at which radiative cooling becomes comparable to particle escape, \(t_{\rm cool}\sim t_{\rm esc}\), this correspondence implies a characteristic escape timescale of the same order, i.e. several hours.

This temporal consistency further supports the proposed balance between energy-dependent particle escape and synchrotron cooling.
The $0.6$–$10$ keV XMM-Newton bandpass is well positioned to probe the key energetic region of the spectrum where the electrons transition from a diffusion-dominated escape regime into an efficient radiative cooling regime within a highly compact, localized emitting region.

Based on comprehensive temporal and spectral study of Mrk~421 using 76 archival
\textit{XMM-Newton}/pn observations spanning $\sim$ 2.5 decades, we conclude 
our findings as follows:
\begin{enumerate}
\item The source exhibits strong, multi-timescale X-ray variability with a log-normal flux distribution dominated by low to intermediate states and sporadic intense flares during which the flux changes by $\sim$ an order of magnitude.

\item The HR analysis and spectral fitting results reveal a 
significant harder-when-brighter behavior, indicating that the X-ray spectral
evolution is linked to enhanced particle acceleration during high flux
states. The hard X-ray band also shows systematically larger variability 
compared to the soft band, suggesting that the high energy electrons 
are more sensitive to changes in the acceleration process.

\item The X-ray spectra are predominantly described by BPL or LP, 
while a simple PL model fails to adequately reproduce the observed spectral shape. 
Out of 76 observations, 59 are statistically favored by the BPL model and 17 by 
the LP model, indicating the widespread presence of spectral breaks or intrinsic 
curvature in the X-ray spectrum of Mrk~421 across the observational epochs.

\item The spectral indices, $\Gamma_1, \Gamma_2$ below and above $E_{\rm b}$ are 
strongly correlated, implying that both spectral components evolve coherently 
during different activity states, with the fitted 
break energies remaining predominantly clustered around $\sim 2$ keV.

\item The flux resolved study shows that the spectral evolution is 
strongly state-dependent: low flux states show negative correlations 
between spectral indices $\Gamma_1,\Gamma_2$ and $E_{\rm b}$, 
while high flux flares produce weak positive correlations. 
This polarity reversal demonstrates distinct 
acceleration and cooling regimes governing faint and bright 
jet emission states.

\item The observed spectral steepening, $\Delta\Gamma \sim 0.19$, is 
smaller than the canonical synchrotron cooling value of $0.5$ expected from a 
simple one-zone homogeneous emission scenario. This result suggests that the 
X-ray emission of Mrk~421 is likely governed by more complex physical processes 
such as continuous or stochastic particle acceleration, incomplete cooling, 
energy-dependent particle transport, or multi-zone jet structure.

\item The comparable statistical performance of the BPL and LP models indicates that both broken and smoothly curved spectral representations can describe the observed X-ray spectra, suggesting that the spectral curvature may arise from a combination of particle acceleration, cooling, and transport processes.

\end{enumerate}

Finally, our 25-years of systematic analysis demonstrates that while the standard leptonic framework broadly applies to Mrk~421, the simplistic one-zone cooling approximations historically favored in the literature are insufficient. The observed spectral steepening of $\Delta\Gamma \approx 0.19$, rather than the canonical 0.5, supports the inclusion of energy-dependent spatial diffusion ($k \approx 0.62$) in future blazar jet models. Furthermore, the fitted break energies remain predominantly clustered around $E_b\sim2$ keV across the observations, although the physical origin of this clustering requires further investigation.  
The combined spectral and timing results therefore suggest that the X-ray spectral curvature of Mrk~421 may reflect the interplay between particle acceleration, radiative cooling, and energy-dependent particle transport within the jet. 

\begin{longrotatetable}
\begin{deluxetable}{ccccccccc}
\footnotesize
\startlongtable
\tablecaption{Results of the spectral fitting in \texttt{XSPEC}. Each epoch of observations was fitted with the models: power-law (PL), broken-power-law (BPL) and log-parabola (LP), all modified by Galactic absorption model (\texttt{TBabs}) with $N_{\rm H} = 1.92 \times 10^{20}$ cm$^{-2}$ fixed. For the LP model, $E_{\rm 1}\,({\rm scale}) = 1.0\; {\rm keV}$ is fixed.
Cols. 1--8 represent the serial number, observation ID, spectral model, photon index/first photon index/log-parabolic slope parameter ($\Gamma/\Gamma_1/\alpha$), second photon index/log-parabolic curvature parameter ($\Gamma_2/\beta$), break energy ($E_{\rm b}$), reduced $\chi^2$, and F-statistic, respectively. Here, $\Gamma$ is the photon index of the PL model, $\Gamma_1$ and $\Gamma_2$ are the photon indices below and above the break energy for the BPL model, and $\alpha$ and $\beta$ are the slope and curvature parameters of the LP model, respectively. The best selected model based on the AICc and BIC tests are shown in the column 9. The best fit model chosen by comparing the F-statistic, AICc and BIC tests is highlighted in bold faced character in the column 2.}
\label{tab:fitting-results}
\tablehead{
\colhead{No.} &
\colhead{Obs. ID} &
\colhead{Model} &
\colhead{$\Gamma/\Gamma_1/\alpha$} &
\colhead{$\Gamma_2/\beta$} &
\colhead{$E_{\rm b}$ (keV)} &
\colhead{$\chi^2$/dof (red. $\chi^2$)} &
\colhead{F-value (prob.)} &
\colhead{AICc/BIC}
}
\startdata
1 & 0099280101 & PL & 2.299 $\pm$ 0.003 & -- & -- & 215.72/166 (1.300) & -- & BPL/BPL \\
  & & \textbf{BPL} & 2.265 $\pm$ 0.006 & 2.342 $\pm$ 0.008 & 2.101 $\pm$ 0.198 & \textbf{154.85/164 (0.944)} & 32.23 (1.56e-12) & \\
  & & LP & 2.262 $\pm$ 0.006 & 0.061 $\pm$ 0.009 & -- & 166.93/165 (1.012) & 48.23 (8.32e-11) & \\
\hline
2 & 0099280201 & PL & 2.516 $\pm$ 0.003 & -- & -- & 338.83/164 (2.066) & -- & BPL/BPL \\
  & & \textbf{BPL} & 2.465 $\pm$ 0.005 & 2.596 $\pm$ 0.009 & 2.279 $\pm$ 0.120 & \textbf{157.64/162 (0.973)} & 93.10 (1.21e-27) & \\
  & & LP & 2.452 $\pm$ 0.006 & 0.107 $\pm$ 0.009 & -- & 180.42/163 (1.107) & 143.12 (4.48e-24) & \\
\hline
3 & 0099280301 & PL & 2.347 $\pm$ 0.002 & -- & -- & 750.95/167 (4.497) & -- & LP/LP \\
  & & BPL & 2.250 $\pm$ 0.006 & 2.438 $\pm$ 0.006 & 2.039 $\pm$ 0.062 & 187.35/165 (1.135) & 248.18 (1.80e-50) & \\
  & & \textbf{LP} & 2.236 $\pm$ 0.005 & 0.170 $\pm$ 0.007 & -- & \textbf{174.48/166 (1.051)} & 548.45 (1.73e-54) & \\
\hline
4 & 0136540101 & PL & 2.312 $\pm$ 0.002 & -- & -- & 451.11/166 (2.718) & -- & BPL/LP \\
  & & \textbf{BPL} & 2.234 $\pm$ 0.007 & 2.380 $\pm$ 0.006 & 1.899 $\pm$ 0.080 & \textbf{174.80/164 (1.066)} & 129.62 (1.73e-34) & \\
  & & LP & 2.230 $\pm$ 0.006 & 0.131 $\pm$ 0.008 & -- & 179.46/165 (1.088) & 249.76 (7.50e-35) & \\
\hline
5 & 0153950701 & PL & 2.822 $\pm$ 0.007 & -- & -- & 171.31/125 (1.370) & -- & LP/LP \\
  & & BPL & 2.747 $\pm$ 0.014 & 2.961 $\pm$ 0.025 & 1.860 $\pm$ 0.149 & 102.69/123 (0.835) & 41.10 (2.14e-14) & \\
  & & \textbf{LP} & 2.736 $\pm$ 0.012 & 0.211 $\pm$ 0.025 & -- & \textbf{96.32/124 (0.777)} & 96.54 (3.35e-17) & \\
\hline
6 & 0136540801 & PL & 2.187 $\pm$ 0.004 & -- & -- & 415.42/154 (2.698) & -- & BPL/LP \\
  & & \textbf{BPL} & 2.109 $\pm$ 0.008 & 2.396 $\pm$ 0.022 & 2.775 $\pm$ 0.132 & \textbf{186.74/152 (1.229)} & 93.07 (4.06e-27) & \\
  & & LP & 2.061 $\pm$ 0.009 & 0.215 $\pm$ 0.015 & -- & 189.32/153 (1.237) & 182.72 (6.76e-28) & \\
\hline
7 & 0136541001 & PL & 2.417 $\pm$ 0.002 & -- & -- & 538.42/168 (3.205) & -- & BPL/BPL \\
  & & \textbf{BPL} & 2.375 $\pm$ 0.003 & 2.514 $\pm$ 0.007 & 3.212 $\pm$ 0.109 & \textbf{147.25/166 (0.887)} & 220.49 (1.84e-47) & \\
  & & LP & 2.340 $\pm$ 0.004 & 0.105 $\pm$ 0.006 & -- & 173.00/167 (1.036) & 352.75 (5.02e-43) & \\
\hline
8 & 0136541101 & PL & 2.396 $\pm$ 0.006 & -- & -- & 320.44/143 (2.241) & -- & LP/LP \\
  & & BPL & 2.323 $\pm$ 0.009 & 2.612 $\pm$ 0.028 & 2.574 $\pm$ 0.154 & 172.56/141 (1.224) & 60.42 (1.12e-19) & \\
  & & \textbf{LP} & 2.281 $\pm$ 0.011 & 0.228 $\pm$ 0.019 & -- & \textbf{161.98/142 (1.141)} & 138.91 (8.68e-23) & \\
\hline
9 & 0136541201 & PL & 2.256 $\pm$ 0.005 & -- & -- & 315.28/145 (2.174) & -- & LP/LP \\
  & & BPL & 2.256 $\pm$ 0.005 & 2.353 $\pm$ -1.000 & -- & 315.28/143 (2.205) & -- (--) & \\
  & & \textbf{LP} & 2.171 $\pm$ 0.011 & 0.161 $\pm$ 0.018 & -- & \textbf{226.03/144 (1.570)} & 56.86 (4.82e-12) & \\
\hline
10 & 0158970101 & PL & 2.616 $\pm$ 0.003 & -- & -- & 651.06/161 (4.044) & -- & BPL/BPL \\
  & & \textbf{BPL} & 2.501 $\pm$ 0.007 & 2.766 $\pm$ 0.010 & 1.916 $\pm$ 0.056 & \textbf{145.50/159 (0.915)} & 276.23 (1.84e-52) & \\
  & & LP & 2.485 $\pm$ 0.007 & 0.252 $\pm$ 0.012 & -- & 171.07/160 (1.069) & 448.93 (2.68e-48) & \\
\hline
11 & 0158970201 & PL & 2.673 $\pm$ 0.009 & -- & -- & 245.78/120 (2.048) & -- & BPL/LP \\
  & & \textbf{BPL} & 2.518 $\pm$ 0.019 & 2.876 $\pm$ 0.026 & 1.704 $\pm$ 0.095 & \textbf{109.60/118 (0.929)} & 73.31 (2.03e-21) & \\
  & & LP & 2.523 $\pm$ 0.016 & 0.342 $\pm$ 0.031 & -- & 113.61/119 (0.955) & 138.44 (1.13e-21) & \\
\hline
12 & 0150498701 & PL & 2.366 $\pm$ 0.002 & -- & -- & 970.16/168 (5.775) & -- & LP/LP \\
  & & BPL & 2.285 $\pm$ 0.004 & 2.511 $\pm$ 0.008 & 2.616 $\pm$ 0.066 & 213.85/166 (1.288) & 293.54 (3.10e-55) & \\
  & & \textbf{LP} & 2.243 $\pm$ 0.005 & 0.190 $\pm$ 0.007 & -- & \textbf{213.50/167 (1.278)} & 591.86 (8.84e-57) & \\
\hline
13 & 0162960101 & PL & 2.344 $\pm$ 0.003 & -- & -- & 585.47/163 (3.592) & -- & BPL/BPL \\
  & & \textbf{BPL} & 2.248 $\pm$ 0.007 & 2.470 $\pm$ 0.009 & 2.069 $\pm$ 0.071 & \textbf{162.86/161 (1.012)} & 208.89 (1.85e-45) & \\
  & & LP & 2.230 $\pm$ 0.006 & 0.199 $\pm$ 0.010 & -- & 178.55/162 (1.102) & 369.20 (1.26e-43) & \\
\hline
14 & 0158971201 & PL & 2.085 $\pm$ 0.002 & -- & -- & 232.72/167 (1.394) & -- & BPL/BPL \\
  & & \textbf{BPL} & 2.072 $\pm$ 0.003 & 2.243 $\pm$ 0.029 & 5.063 $\pm$ 2.279 & \textbf{154.99/165 (0.939)} & 41.38 (2.73e-15) & \\
  & & LP & 2.056 $\pm$ 0.005 & 0.044 $\pm$ 0.007 & -- & 196.75/166 (1.185) & 30.35 (1.35e-07) & \\
\hline
15 & 0153951201 & PL & 2.379 $\pm$ 0.004 & -- & -- & 451.99/158 (2.861) & -- & LP/LP \\
  & & BPL & 2.310 $\pm$ 0.007 & 2.494 $\pm$ 0.013 & 2.223 $\pm$ 0.118 & 260.62/156 (1.671) & 57.27 (2.23e-19) & \\
  & & \textbf{LP} & 2.288 $\pm$ 0.007 & 0.161 $\pm$ 0.012 & -- & \textbf{260.00/157 (1.656)} & 115.93 (1.36e-20) & \\
\hline
16 & 0158971301 & PL & 2.497 $\pm$ 0.002 & -- & -- & 1.78e3/167 (10.584) & -- & LP/LP \\
  & & BPL & 2.497 $\pm$ 0.002 & 9.218 $\pm$ -1.000 & -- & 1.77e3/165 (10.712) & -- (--) & \\
  & & \textbf{LP} & 2.330 $\pm$ 0.005 & 0.225 $\pm$ 0.006 & -- & \textbf{248.89/166 (1.499)} & 1012.87 (1.45e-72) & \\
\hline
17 & 0302180101 & PL & 2.197 $\pm$ 0.002 & -- & -- & 196.98/168 (1.172) & -- & BPL/BPL \\
  & & \textbf{BPL} & 2.186 $\pm$ 0.003 & 2.261 $\pm$ 0.015 & 4.216 $\pm$ 0.369 & \textbf{153.19/166 (0.923)} & 23.73 (8.65e-10) & \\
  & & LP & 2.175 $\pm$ 0.005 & 0.032 $\pm$ 0.006 & -- & 171.88/167 (1.029) & 24.39 (1.90e-06) & \\
\hline
18 & 0411080301 & PL & 2.034 $\pm$ 0.002 & -- & -- & 631.04/168 (3.756) & -- & BPL/LP \\
  & & \textbf{BPL} & 1.961 $\pm$ 0.005 & 2.091 $\pm$ 0.004 & 2.136 $\pm$ 0.080 & \textbf{229.37/166 (1.382)} & 145.35 (3.31e-37) & \\
  & & LP & 1.951 $\pm$ 0.004 & 0.111 $\pm$ 0.006 & -- & 233.98/167 (1.401) & 283.40 (8.14e-38) & \\
\hline
19 & 0411080701 & PL & 2.659 $\pm$ 0.002 & -- & -- & 540.93/166 (3.259) & -- & BPL/BPL \\
  & & \textbf{BPL} & 2.603 $\pm$ 0.004 & 2.774 $\pm$ 0.009 & 2.674 $\pm$ 0.095 & \textbf{172.97/164 (1.055)} & 174.44 (2.49e-41) & \\
  & & LP & 2.573 $\pm$ 0.005 & 0.136 $\pm$ 0.008 & -- & 206.63/165 (1.252) & 266.95 (2.60e-36) & \\
\hline
20 & 0510610201 & PL & 2.712 $\pm$ 0.003 & -- & -- & 369.36/163 (2.266) & -- & LP/LP \\
  & & BPL & 2.673 $\pm$ 0.005 & 2.919 $\pm$ 0.023 & 3.044 $\pm$ 0.131 & 172.21/161 (1.070) & 92.16 (2.10e-27) & \\
  & & \textbf{LP} & 2.633 $\pm$ 0.006 & 0.161 $\pm$ 0.012 & -- & \textbf{167.11/162 (1.032)} & 196.07 (1.06e-29) & \\
\hline
21 & 0510610101 & PL & 2.736 $\pm$ 0.003 & -- & -- & 377.22/164 (2.300) & -- & BPL/BPL \\
  & & \textbf{BPL} & 2.691 $\pm$ 0.005 & 2.897 $\pm$ 0.016 & 2.741 $\pm$ 0.116 & \textbf{152.21/162 (0.940)} & 119.74 (1.19e-32) & \\
  & & LP & 2.659 $\pm$ 0.006 & 0.150 $\pm$ 0.011 & -- & 167.98/163 (1.031) & 203.04 (1.94e-30) & \\
\hline
22 & 0411081301 & PL & 2.697 $\pm$ 0.006 & -- & -- & 315.56/129 (2.446) & -- & LP/LP \\
  & & BPL & 2.552 $\pm$ 0.015 & 2.864 $\pm$ 0.018 & 1.590 $\pm$ 0.071 & 115.66/127 (0.911) & 109.75 (2.09e-28) & \\
  & & \textbf{LP} & 2.568 $\pm$ 0.011 & 0.310 $\pm$ 0.023 & -- & \textbf{112.75/128 (0.881)} & 230.24 (2.17e-30) & \\
\hline
23 & 0411081401 & PL & 2.698 $\pm$ 0.008 & -- & -- & 302.49/115 (2.630) & -- & LP/LP \\
  & & BPL & 2.509 $\pm$ 0.021 & 2.894 $\pm$ 0.023 & 1.513 $\pm$ 0.071 & 124.91/113 (1.105) & 80.32 (1.98e-22) & \\
  & & \textbf{LP} & 2.534 $\pm$ 0.014 & 0.402 $\pm$ 0.031 & -- & \textbf{110.06/114 (0.965)} & 199.32 (8.73e-27) & \\
\hline
24 & 0411081501 & PL & 2.637 $\pm$ 0.007 & -- & -- & 264.45/127 (2.082) & -- & LP/LP \\
  & & BPL & 2.447 $\pm$ 0.025 & 2.749 $\pm$ 0.015 & 1.271 $\pm$ 0.062 & 126.26/125 (1.010) & 68.41 (8.56e-21) & \\
  & & \textbf{LP} & 2.515 $\pm$ 0.012 & 0.292 $\pm$ 0.025 & -- & \textbf{113.32/126 (0.899)} & 168.04 (6.08e-25) & \\
\hline
25 & 0411081601 & PL & 2.670 $\pm$ 0.010 & -- & -- & 204.60/108 (1.894) & -- & LP/LP \\
  & & BPL & 2.553 $\pm$ 0.018 & 2.923 $\pm$ 0.040 & 1.920 $\pm$ 0.135 & 106.70/106 (1.007) & 48.63 (1.03e-15) & \\
  & & \textbf{LP} & 2.540 $\pm$ 0.016 & 0.336 $\pm$ 0.036 & -- & \textbf{106.69/107 (0.997)} & 98.19 (8.19e-17) & \\
\hline
26 & 0502030101 & PL & 1.400 $\pm$ -1.000 & -- & -- & -- & -- & BPL/BPL \\
  & & \textbf{BPL} & 2.420 $\pm$ 0.003 & 2.604 $\pm$ 0.006 & 2.907 $\pm$ 0.070 & \textbf{116.18/166 (0.700)} & 4.8e5 (4.42e-313) & \\
  & & LP & 2.377 $\pm$ 0.004 & 0.150 $\pm$ 0.005 & -- & 167.85/167 (1.005) & 6.7e5 (7.87e-303) & \\
\hline
27 & 0411082701 & PL & 2.375 $\pm$ 0.004 & -- & -- & 770.63/159 (4.847) & -- & LP/LP \\
  & & BPL & 2.239 $\pm$ 0.008 & 2.572 $\pm$ 0.013 & 2.099 $\pm$ 0.062 & 175.53/157 (1.118) & 266.14 (3.67e-51) & \\
  & & \textbf{LP} & 2.208 $\pm$ 0.008 & 0.300 $\pm$ 0.013 & -- & \textbf{171.71/158 (1.087)} & 551.10 (2.21e-53) & \\
\hline
28 & 0560980101 & PL & 2.625 $\pm$ 0.004 & -- & -- & 643.41/147 (4.377) & -- & BPL/BPL \\
  & & \textbf{BPL} & 2.467 $\pm$ 0.010 & 2.827 $\pm$ 0.014 & 1.819 $\pm$ 0.056 & \textbf{137.78/145 (0.950)} & 266.06 (2.99e-49) & \\
  & & LP & 2.461 $\pm$ 0.009 & 0.337 $\pm$ 0.016 & -- & 143.01/146 (0.980) & 510.86 (1.57e-49) & \\
\hline
29 & 0560983301 & PL & 2.767 $\pm$ 0.005 & -- & -- & 444.52/142 (3.130) & -- & BPL/BPL \\
  & & \textbf{BPL} & 2.626 $\pm$ 0.012 & 2.948 $\pm$ 0.015 & 1.735 $\pm$ 0.064 & \textbf{128.71/140 (0.919)} & 171.76 (2.09e-38) & \\
  & & LP & 2.627 $\pm$ 0.009 & 0.311 $\pm$ 0.018 & -- & 133.76/141 (0.949) & 327.58 (1.36e-38) & \\
\hline
30 & 0411083201 & PL & 2.170 $\pm$ 0.003 & -- & -- & 311.13/160 (1.945) & -- & BPL/BPL \\
  & & \textbf{BPL} & 2.098 $\pm$ 0.008 & 2.256 $\pm$ 0.010 & 2.037 $\pm$ 0.117 & \textbf{148.01/158 (0.937)} & 87.06 (3.24e-26) & \\
  & & LP & 2.089 $\pm$ 0.007 & 0.137 $\pm$ 0.011 & -- & 157.56/159 (0.991) & 154.97 (2.88e-25) & \\
\hline
31 & 0656380101 & PL & 2.328 $\pm$ 0.004 & -- & -- & 311.82/155 (2.012) & -- & BPL/BPL \\
  & & \textbf{BPL} & 2.221 $\pm$ 0.012 & 2.431 $\pm$ 0.012 & 1.724 $\pm$ 0.091 & \textbf{145.49/153 (0.951)} & 87.46 (4.71e-26) & \\
  & & LP & 2.232 $\pm$ 0.009 & 0.182 $\pm$ 0.015 & -- & 165.42/154 (1.074) & 136.29 (5.88e-23) & \\
\hline
32 & 0656380801 & PL & 2.398 $\pm$ 0.004 & -- & -- & 331.72/153 (2.168) & -- & BPL/LP \\
  & & \textbf{BPL} & 2.297 $\pm$ 0.011 & 2.501 $\pm$ 0.012 & 1.740 $\pm$ 0.090 & \textbf{155.21/151 (1.028)} & 85.86 (1.25e-25) & \\
  & & LP & 2.300 $\pm$ 0.009 & 0.188 $\pm$ 0.015 & -- & 158.08/152 (1.040) & 166.96 (3.05e-26) & \\
\hline
33 & 0656381301 & PL & 2.639 $\pm$ 0.005 & -- & -- & 301.18/144 (2.092) & -- & BPL/BPL \\
  & & \textbf{BPL} & 2.536 $\pm$ 0.011 & 2.779 $\pm$ 0.016 & 1.831 $\pm$ 0.094 & \textbf{127.11/142 (0.895)} & 97.23 (2.51e-27) & \\
  & & LP & 2.529 $\pm$ 0.010 & 0.229 $\pm$ 0.018 & -- & 133.18/143 (0.931) & 180.39 (4.07e-27) & \\
\hline
34 & 0658800101 & PL & 2.752 $\pm$ 0.007 & -- & -- & 245.95/125 (1.968) & -- & BPL/BPL \\
  & & \textbf{BPL} & 2.675 $\pm$ 0.011 & 3.000 $\pm$ 0.037 & 2.336 $\pm$ 0.155 & \textbf{130.07/123 (1.057)} & 54.79 (9.66e-18) & \\
  & & LP & 2.649 $\pm$ 0.012 & 0.247 $\pm$ 0.024 & -- & 135.75/124 (1.095) & 100.66 (1.05e-17) & \\
\hline
35 & 0658800801 & PL & 2.903 $\pm$ 0.004 & -- & -- & 198.97/151 (1.318) & -- & BPL/PL \\
  & & \textbf{BPL} & 2.888 $\pm$ 0.009 & 2.928 $\pm$ 0.015 & 1.961 $\pm$ 0.534 & \textbf{193.09/149 (1.296)} & 2.27 (1.07e-01) & \\
  & & LP & 2.892 $\pm$ 0.008 & 0.024 $\pm$ 0.015 & -- & 196.49/150 (1.310) & 1.89 (1.71e-01) & \\
\hline
36 & 0670920301 & PL & 2.617 $\pm$ 0.003 & -- & -- & 220.91/164 (1.347) & -- & BPL/BPL \\
  & & \textbf{BPL} & 2.590 $\pm$ 0.006 & 2.660 $\pm$ 0.009 & 2.237 $\pm$ 0.218 & \textbf{169.50/162 (1.046)} & 24.57 (4.80e-10) & \\
  & & LP & 2.584 $\pm$ 0.006 & 0.056 $\pm$ 0.009 & -- & 181.02/163 (1.111) & 35.92 (1.28e-08) & \\
\hline
37 & 0670920401 & PL & 2.783 $\pm$ 0.003 & -- & -- & 238.11/165 (1.443) & -- & BPL/BPL \\
  & & \textbf{BPL} & 2.747 $\pm$ 0.005 & 2.845 $\pm$ 0.009 & 2.237 $\pm$ 0.154 & \textbf{143.05/163 (0.878)} & 54.16 (9.22e-19) & \\
  & & LP & 2.736 $\pm$ 0.006 & 0.085 $\pm$ 0.009 & -- & 151.24/164 (0.922) & 94.20 (7.00e-18) & \\
\hline
38 & 0670920501 & PL & 2.441 $\pm$ 0.002 & -- & -- & 161.73/166 (0.974) & -- & LP/LP \\
  & & BPL & 2.356 $\pm$ 0.004 & 2.441 $\pm$ 0.003 & 0.486 $\pm$ 0.045 & 161.73/164 (0.986) & -- (--) & \\
  & & \textbf{LP} & 2.456 $\pm$ 0.006 & -0.025 $\pm$ 0.008 & -- & \textbf{152.78/165 (0.926)} & 9.67 (2.21e-03) & \\
\hline
39 & 0658801301 & PL & 2.508 $\pm$ 0.002 & -- & -- & 817.11/166 (4.922) & -- & BPL/BPL \\
  & & \textbf{BPL} & 2.408 $\pm$ 0.006 & 2.630 $\pm$ 0.007 & 2.142 $\pm$ 0.059 & \textbf{160.94/164 (0.981)} & 334.32 (1.38e-58) & \\
  & & LP & 2.392 $\pm$ 0.005 & 0.189 $\pm$ 0.008 & -- & 236.04/165 (1.431) & 406.19 (2.36e-46) & \\
\hline
40 & 0658801801 & PL & 2.820 $\pm$ 0.003 & -- & -- & 498.63/159 (3.136) & -- & BPL/BPL \\
  & & \textbf{BPL} & 2.725 $\pm$ 0.008 & 2.928 $\pm$ 0.009 & 1.863 $\pm$ 0.067 & \textbf{161.88/157 (1.031)} & 163.30 (4.43e-39) & \\
  & & LP & 2.718 $\pm$ 0.006 & 0.189 $\pm$ 0.011 & -- & 184.61/158 (1.168) & 268.76 (6.46e-36) & \\
\hline
41 & 0658802301 & PL & 2.795 $\pm$ 0.003 & -- & -- & 394.52/159 (2.481) & -- & BPL/BPL \\
  & & \textbf{BPL} & 2.713 $\pm$ 0.007 & 2.896 $\pm$ 0.009 & 1.853 $\pm$ 0.077 & \textbf{134.66/157 (0.858)} & 151.49 (2.26e-37) & \\
  & & LP & 2.711 $\pm$ 0.006 & 0.165 $\pm$ 0.011 & -- & 157.45/158 (0.997) & 237.90 (2.49e-33) & \\
\hline
42 & 0791780101 & PL & 2.772 $\pm$ 0.004 & -- & -- & 257.61/148 (1.741) & -- & BPL/BPL \\
  & & \textbf{BPL} & 2.704 $\pm$ 0.009 & 2.890 $\pm$ 0.015 & 1.958 $\pm$ 0.117 & \textbf{128.70/146 (0.882)} & 73.12 (9.98e-23) & \\
  & & LP & 2.695 $\pm$ 0.008 & 0.166 $\pm$ 0.015 & -- & 137.52/147 (0.936) & 128.37 (8.79e-22) & \\
\hline
43 & 0791780601 & PL & 2.266 $\pm$ 0.003 & -- & -- & 365.75/165 (2.217) & -- & BPL/BPL \\
  & & \textbf{BPL} & 2.198 $\pm$ 0.007 & 2.337 $\pm$ 0.008 & 2.023 $\pm$ 0.103 & \textbf{173.34/163 (1.063)} & 90.47 (3.72e-27) & \\
  & & LP & 2.193 $\pm$ 0.006 & 0.117 $\pm$ 0.009 & -- & 203.33/164 (1.240) & 131.00 (1.14e-22) & \\
\hline
44 & 0791781401 & PL & 2.413 $\pm$ 0.003 & -- & -- & 630.69/162 (3.893) & -- & BPL/BPL \\
  & & \textbf{BPL} & 2.311 $\pm$ 0.007 & 2.543 $\pm$ 0.009 & 2.096 $\pm$ 0.070 & \textbf{178.19/160 (1.114)} & 203.15 (1.22e-44) & \\
  & & LP & 2.291 $\pm$ 0.007 & 0.207 $\pm$ 0.010 & -- & 198.05/161 (1.230) & 351.70 (2.42e-42) & \\
\hline
45 & 0791782001 & PL & 2.818 $\pm$ 0.006 & -- & -- & 182.56/135 (1.352) & -- & BPL/LP \\
  & & \textbf{BPL} & 2.738 $\pm$ 0.013 & 2.918 $\pm$ 0.017 & 1.644 $\pm$ 0.117 & \textbf{103.05/133 (0.775)} & 51.31 (3.05e-17) & \\
  & & LP & 2.744 $\pm$ 0.010 & 0.174 $\pm$ 0.020 & -- & 105.74/134 (0.789) & 97.35 (1.35e-17) & \\
\hline
46 & 0810860201 & PL & 2.454 $\pm$ 0.003 & -- & -- & 763.70/162 (4.714) & -- & BPL/BPL \\
  & & \textbf{BPL} & 2.343 $\pm$ 0.006 & 2.608 $\pm$ 0.010 & 2.182 $\pm$ 0.063 & \textbf{145.18/160 (0.907)} & 340.83 (2.08e-58) & \\
  & & LP & 2.312 $\pm$ 0.006 & 0.237 $\pm$ 0.010 & -- & 158.95/161 (0.987) & 612.55 (9.42e-57) & \\
\hline
47 & 0810860701 & PL & 2.211 $\pm$ 0.003 & -- & -- & 159.18/162 (0.983) & -- & BPL/LP \\
  & & \textbf{BPL} & 2.192 $\pm$ 0.007 & 2.239 $\pm$ 0.011 & 2.181 $\pm$ 0.411 & \textbf{143.15/160 (0.895)} & 8.96 (2.05e-04) & \\
  & & LP & 2.190 $\pm$ 0.007 & 0.037 $\pm$ 0.010 & -- & 146.06/161 (0.907) & 14.46 (2.03e-04) & \\
\hline
48 & 0845000901 & PL & 1.981 $\pm$ 0.002 & -- & -- & 336.29/168 (2.002) & -- & BPL/BPL \\
  & & \textbf{BPL} & 2.073 $\pm$ 0.012 & 1.958 $\pm$ 0.003 & 1.336 $\pm$ 0.070 & \textbf{185.76/166 (1.119)} & 67.26 (4.03e-22) & \\
  & & LP & 2.020 $\pm$ 0.005 & -0.055 $\pm$ 0.006 & -- & 250.73/167 (1.501) & 56.99 (2.71e-12) & \\
\hline
49 & 0810862501 & PL & 2.650 $\pm$ 0.003 & -- & -- & 372.00/164 (2.268) & -- & BPL/BPL \\
  & & \textbf{BPL} & 2.597 $\pm$ 0.005 & 2.738 $\pm$ 0.009 & 2.295 $\pm$ 0.110 & \textbf{159.61/162 (0.985)} & 107.79 (1.71e-30) & \\
  & & LP & 2.580 $\pm$ 0.006 & 0.120 $\pm$ 0.009 & -- & 175.40/163 (1.076) & 182.70 (2.09e-28) & \\
\hline
50 & 0850380301 & PL & 2.704 $\pm$ 0.002 & -- & -- & 995.35/168 (5.925) & -- & BPL/BPL \\
  & & \textbf{BPL} & 2.627 $\pm$ 0.004 & 2.865 $\pm$ 0.008 & 2.760 $\pm$ 0.064 & \textbf{163.60/166 (0.986)} & 421.98 (8.17e-66) & \\
  & & LP & 2.579 $\pm$ 0.005 & 0.194 $\pm$ 0.007 & -- & 188.76/167 (1.130) & 713.61 (3.49e-62) & \\
\hline
51 & 0850380401 & PL & 2.626 $\pm$ 0.002 & -- & -- & 774.78/167 (4.639) & -- & BPL/BPL \\
  & & \textbf{BPL} & 2.548 $\pm$ 0.005 & 2.740 $\pm$ 0.007 & 2.446 $\pm$ 0.067 & \textbf{155.58/165 (0.943)} & 328.35 (3.01e-58) & \\
  & & LP & 2.513 $\pm$ 0.005 & 0.170 $\pm$ 0.007 & -- & 167.27/166 (1.008) & 602.90 (3.85e-57) & \\
\hline
52 & 0810863001 & PL & 2.481 $\pm$ 0.004 & -- & -- & 220.33/152 (1.450) & -- & BPL/LP \\
  & & \textbf{BPL} & 2.408 $\pm$ 0.011 & 2.562 $\pm$ 0.012 & 1.791 $\pm$ 0.123 & \textbf{119.21/150 (0.795)} & 63.62 (9.84e-21) & \\
  & & LP & 2.407 $\pm$ 0.009 & 0.143 $\pm$ 0.015 & -- & 122.19/151 (0.809) & 121.28 (4.50e-21) & \\
\hline
53 & 0871590201 & PL & 2.239 $\pm$ 0.003 & -- & -- & 447.71/161 (2.781) & -- & BPL/BPL \\
  & & \textbf{BPL} & 2.141 $\pm$ 0.008 & 2.338 $\pm$ 0.009 & 1.897 $\pm$ 0.081 & \textbf{169.23/159 (1.064)} & 130.82 (2.57e-34) & \\
  & & LP & 2.136 $\pm$ 0.007 & 0.176 $\pm$ 0.011 & -- & 175.47/160 (1.097) & 248.24 (2.30e-34) & \\
\hline
54 & 0870190301 & PL & 2.439 $\pm$ 0.002 & -- & -- & 252.56/167 (1.512) & -- & BPL/BPL \\
  & & \textbf{BPL} & 2.417 $\pm$ 0.004 & 2.516 $\pm$ 0.012 & 3.341 $\pm$ 0.223 & \textbf{160.66/165 (0.974)} & 47.19 (6.20e-17) & \\
  & & LP & 2.399 $\pm$ 0.005 & 0.062 $\pm$ 0.007 & -- & 181.10/166 (1.091) & 65.50 (1.17e-13) & \\
\hline
55 & 0853600201 & PL & 2.557 $\pm$ 0.003 & -- & -- & 239.85/164 (1.462) & -- & BPL/BPL \\
  & & \textbf{BPL} & 2.532 $\pm$ 0.004 & 2.693 $\pm$ 0.022 & 3.442 $\pm$ 0.226 & \textbf{150.81/162 (0.931)} & 47.82 (4.76e-17) & \\
  & & LP & 2.507 $\pm$ 0.006 & 0.090 $\pm$ 0.010 & -- & 160.23/163 (0.983) & 81.00 (5.64e-16) & \\
\hline
56 & 0853600301 & PL & 2.526 $\pm$ 0.003 & -- & -- & 184.68/160 (1.154) & -- & BPL/BPL \\
  & & \textbf{BPL} & 2.503 $\pm$ 0.006 & 2.599 $\pm$ 0.018 & 2.800 $\pm$ 0.308 & \textbf{147.50/158 (0.934)} & 19.91 (1.94e-08) & \\
  & & LP & 2.493 $\pm$ 0.007 & 0.060 $\pm$ 0.012 & -- & 158.40/159 (0.996) & 26.38 (8.11e-07) & \\
\hline
57 & 0870190501 & PL & 2.323 $\pm$ 0.002 & -- & -- & 147.71/167 (0.884) & -- & BPL/BPL \\
  & & \textbf{BPL} & 2.308 $\pm$ 0.004 & 2.358 $\pm$ 0.010 & 2.883 $\pm$ 0.376 & \textbf{119.74/165 (0.726)} & 19.27 (3.01e-08) & \\
  & & LP & 2.301 $\pm$ 0.005 & 0.035 $\pm$ 0.008 & -- & 126.48/166 (0.762) & 27.86 (4.03e-07) & \\
\hline
58 & 0870190601 & PL & 2.379 $\pm$ 0.002 & -- & -- & 235.77/167 (1.412) & -- & BPL/BPL \\
  & & \textbf{BPL} & 2.362 $\pm$ 0.003 & 2.479 $\pm$ 0.015 & 3.991 $\pm$ 0.228 & \textbf{134.02/165 (0.812)} & 62.64 (5.77e-21) & \\
  & & LP & 2.341 $\pm$ 0.005 & 0.057 $\pm$ 0.007 & -- & 164.64/166 (0.992) & 71.72 (1.26e-14) & \\
\hline
59 & 0871590701 & PL & 2.909 $\pm$ 0.006 & -- & -- & 200.64/131 (1.532) & -- & BPL/BPL \\
  & & \textbf{BPL} & 2.836 $\pm$ 0.012 & 3.047 $\pm$ 0.022 & 1.890 $\pm$ 0.136 & \textbf{116.96/129 (0.907)} & 46.15 (7.63e-16) & \\
  & & LP & 2.830 $\pm$ 0.011 & 0.190 $\pm$ 0.022 & -- & 125.00/130 (0.962) & 78.67 (4.93e-15) & \\
\hline
60 & 0902110401 & PL & 2.825 $\pm$ 0.005 & -- & -- & 245.60/148 (1.659) & -- & LP/LP \\
  & & BPL & 2.752 $\pm$ 0.011 & 2.922 $\pm$ 0.014 & 1.713 $\pm$ 0.107 & 142.04/146 (0.973) & 53.22 (4.36e-18) & \\
  & & \textbf{LP} & 2.749 $\pm$ 0.009 & 0.174 $\pm$ 0.017 & -- & \textbf{134.20/147 (0.913)} & 122.03 (4.94e-21) & \\
\hline
61 & 0910790101 & PL & 2.668 $\pm$ 0.003 & -- & -- & 650.30/164 (3.965) & -- & BPL/BPL \\
  & & \textbf{BPL} & 2.586 $\pm$ 0.005 & 2.792 $\pm$ 0.009 & 2.299 $\pm$ 0.073 & \textbf{171.31/162 (1.057)} & 226.48 (1.19e-47) & \\
  & & LP & 2.561 $\pm$ 0.006 & 0.176 $\pm$ 0.008 & -- & 208.01/163 (1.276) & 346.59 (3.41e-42) & \\
\hline
62 & 0871591801 & PL & 2.302 $\pm$ 0.003 & -- & -- & 282.03/161 (1.752) & -- & BPL/BPL \\
  & & \textbf{BPL} & 2.248 $\pm$ 0.008 & 2.367 $\pm$ 0.009 & 1.985 $\pm$ 0.138 & \textbf{174.59/159 (1.098)} & 48.92 (2.77e-17) & \\
  & & LP & 2.244 $\pm$ 0.007 & 0.102 $\pm$ 0.010 & -- & 186.27/160 (1.164) & 82.25 (4.14e-16) & \\
\hline
63 & 0891200101 & PL & 2.690 $\pm$ 0.002 & -- & -- & 637.66/165 (3.865) & -- & BPL/BPL \\
  & & \textbf{BPL} & 2.642 $\pm$ 0.004 & 2.914 $\pm$ 0.016 & 3.481 $\pm$ 0.098 & \textbf{176.45/163 (1.083)} & 213.03 (3.35e-46) & \\
  & & LP & 2.584 $\pm$ 0.006 & 0.172 $\pm$ 0.008 & -- & 193.37/164 (1.179) & 376.81 (2.39e-44) & \\
\hline
64 & 0902111701 & PL & 2.678 $\pm$ 0.003 & -- & -- & 295.54/161 (1.836) & -- & BPL/LP \\
  & & \textbf{BPL} & 2.623 $\pm$ 0.006 & 2.765 $\pm$ 0.011 & 2.144 $\pm$ 0.121 & \textbf{142.74/159 (0.898)} & 85.10 (7.45e-26) & \\
  & & LP & 2.608 $\pm$ 0.007 & 0.128 $\pm$ 0.011 & -- & 147.13/160 (0.920) & 161.39 (5.16e-26) & \\
\hline
65 & 0902112401 & PL & 2.646 $\pm$ 0.003 & -- & -- & 215.32/164 (1.313) & -- & BPL/BPL \\
  & & \textbf{BPL} & 2.611 $\pm$ 0.006 & 2.700 $\pm$ 0.009 & 2.274 $\pm$ 0.168 & \textbf{129.66/162 (0.800)} & 53.51 (1.44e-18) & \\
  & & LP & 2.601 $\pm$ 0.006 & 0.075 $\pm$ 0.009 & -- & 141.62/163 (0.869) & 84.83 (1.56e-16) & \\
\hline
66 & 0902112501 & PL & 2.633 $\pm$ 0.002 & -- & -- & 247.63/165 (1.501) & -- & BPL/BPL \\
  & & \textbf{BPL} & 2.596 $\pm$ 0.005 & 2.695 $\pm$ 0.009 & 2.385 $\pm$ 0.154 & \textbf{133.64/163 (0.820)} & 69.52 (1.48e-22) & \\
  & & LP & 2.581 $\pm$ 0.006 & 0.085 $\pm$ 0.008 & -- & 139.08/164 (0.848) & 128.00 (2.66e-22) & \\
\hline
67 & 0920900201 & PL & 2.621 $\pm$ 0.002 & -- & -- & 644.36/166 (3.882) & -- & BPL/BPL \\
  & & \textbf{BPL} & 2.547 $\pm$ 0.005 & 2.738 $\pm$ 0.008 & 2.358 $\pm$ 0.075 & \textbf{179.69/164 (1.096)} & 212.05 (3.33e-46) & \\
  & & LP & 2.520 $\pm$ 0.005 & 0.164 $\pm$ 0.008 & -- & 205.31/165 (1.244) & 352.85 (7.86e-43) & \\
\hline
68 & 0891200201 & PL & 2.594 $\pm$ 0.002 & -- & -- & 683.25/167 (4.091) & -- & LP/LP \\
  & & BPL & 2.520 $\pm$ 0.005 & 2.702 $\pm$ 0.008 & 2.330 $\pm$ 0.076 & 222.27/165 (1.347) & 171.10 (5.82e-41) & \\
  & & \textbf{LP} & 2.493 $\pm$ 0.005 & 0.161 $\pm$ 0.008 & -- & \textbf{217.98/166 (1.313)} & 354.32 (4.92e-43) & \\
\hline
69 & 0920901301 & PL & 2.532 $\pm$ 0.002 & -- & -- & 455.75/167 (2.729) & -- & BPL/BPL \\
  & & \textbf{BPL} & 2.470 $\pm$ 0.005 & 2.626 $\pm$ 0.008 & 2.319 $\pm$ 0.091 & \textbf{142.84/165 (0.866)} & 180.73 (2.69e-42) & \\
  & & LP & 2.449 $\pm$ 0.005 & 0.136 $\pm$ 0.008 & -- & 150.70/166 (0.908) & 336.02 (9.69e-42) & \\
\hline
70 & 0934010101 & PL & 2.644 $\pm$ 0.003 & -- & -- & 194.18/165 (1.177) & -- & LP/LP \\
  & & BPL & 2.623 $\pm$ 0.005 & 2.711 $\pm$ 0.016 & 2.706 $\pm$ 0.276 & \textbf{151.62/163 (0.930)} & 22.88 (1.75e-09) & \\
  & & \textbf{LP} & 2.609 $\pm$ 0.006 & 0.066 $\pm$ 0.010 & -- & 153.04/164 (0.933) & 44.09 (4.39e-10) & \\
\hline
71 & 0914191801 & PL & 2.715 $\pm$ 0.005 & -- & -- & 291.10/141 (2.065) & -- & BPL/BPL \\
  & & \textbf{BPL} & 2.637 $\pm$ 0.010 & 2.851 $\pm$ 0.019 & 1.949 $\pm$ 0.122 & \textbf{157.56/139 (1.134)} & 58.90 (2.96e-19) & \\
  & & LP & 2.630 $\pm$ 0.009 & 0.186 $\pm$ 0.017 & -- & 168.21/140 (1.202) & 102.28 (2.18e-18) & \\
\hline
72 & 0934010201 & PL & 2.309 $\pm$ 0.002 & -- & -- & 302.40/168 (1.800) & -- & BPL/BPL \\
  & & \textbf{BPL} & 2.277 $\pm$ 0.004 & 2.374 $\pm$ 0.008 & 2.772 $\pm$ 0.155 & \textbf{159.46/166 (0.961)} & 74.40 (8.55e-24) & \\
  & & LP & 2.258 $\pm$ 0.005 & 0.077 $\pm$ 0.007 & -- & 173.63/167 (1.040) & 123.85 (7.11e-22) & \\
\hline
73 & 0920901401 & PL & 2.280 $\pm$ 0.003 & -- & -- & 364.13/166 (2.194) & -- & BPL/BPL \\
  & & \textbf{BPL} & 2.226 $\pm$ 0.006 & 2.358 $\pm$ 0.009 & 2.247 $\pm$ 0.116 & \textbf{178.15/164 (1.086)} & 85.60 (3.48e-26) & \\
  & & LP & 2.211 $\pm$ 0.006 & 0.114 $\pm$ 0.009 & -- & 192.95/165 (1.169) & 146.38 (1.58e-24) & \\
\hline
74 & 0920902101 & PL & 2.306 $\pm$ 0.003 & -- & -- & 258.88/163 (1.588) & -- & BPL/BPL \\
  & & \textbf{BPL} & 2.274 $\pm$ 0.006 & 2.397 $\pm$ 0.017 & 2.841 $\pm$ 0.244 & \textbf{192.47/161 (1.195)} & 27.78 (4.33e-11) & \\
  & & LP & 2.257 $\pm$ 0.007 & 0.085 $\pm$ 0.011 & -- & 201.81/162 (1.246) & 45.81 (2.27e-10) & \\
\hline
75 & 0914192901 & PL & 2.875 $\pm$ 0.005 & -- & -- & 259.80/139 (1.869) & -- & BPL/LP \\
  & & \textbf{BPL} & 2.806 $\pm$ 0.010 & 2.995 $\pm$ 0.017 & 1.873 $\pm$ 0.122 & \textbf{151.74/137 (1.108)} & 48.78 (1.01e-16) & \\
  & & LP & 2.801 $\pm$ 0.009 & 0.173 $\pm$ 0.017 & -- & 156.44/138 (1.134) & 91.18 (6.71e-17) & \\
\hline
76 & 0914193701 & PL & 2.704 $\pm$ 0.005 & -- & -- & 297.65/144 (2.067) & -- & BPL/BPL \\
  & & \textbf{BPL} & 2.641 $\pm$ 0.008 & 2.911 $\pm$ 0.027 & 2.552 $\pm$ 0.147 & \textbf{147.75/142 (1.040)} & 72.03 (2.53e-22) & \\
  & & LP & 2.613 $\pm$ 0.009 & 0.193 $\pm$ 0.017 & -- & 159.93/143 (1.118) & 123.14 (4.99e-21) & \\
\enddata
\end{deluxetable}
\end{longrotatetable}

\section{Acknowledgments}
This work is based on observations obtained with XMM-Newton, an ESA
science mission with instruments and contributions directly funded by
ESA Member States and NASA.
ZH and TPA acknowledges the support of the National Natural Science 
Foundation of China (grant nos.\ 12222304, 12192220, and 12192221).
This work was partially supported by a program of the Polish Ministry of Science under the title “Regional Excellence Initiative”, project No. RID/SP/0050/2024/1.



\bibliography{refs}{}

@INPROCEEDINGS{Arnaud1996,
       author = {{Arnaud}, K.~A.},
        title = "{XSPEC: The First Ten Years}",
    booktitle = {Astronomical Data Analysis Software and Systems V},
         year = 1996,
       editor = {{Jacoby}, George H. and {Barnes}, Jeannette},
       series = {Astronomical Society of the Pacific Conference Series},
       volume = {101},
        month = jan,
        pages = {17},
       adsurl = {https://ui.adsabs.harvard.edu/abs/1996ASPC..101...17A}
}

@ARTICLE{Aggrawal2018,
       author = {{Aggrawal}, Vishi and {Pandey}, Ashwani and {Gupta}, Alok C. and {Zhang}, Zhongli and {Wiita}, Paul J. and {Yadav}, K.~K. and {Tiwari}, S.~N.},
        title = "{X-ray intraday variability of the TeV blazar Mrk 421 with Chandra}",
      journal = {\mnras},
         year = 2018,
        month = nov,
       volume = {480},
       number = {4},
        pages = {4873-4883},
          doi = {10.1093/mnras/sty2173},
archivePrefix = {arXiv},
       eprint = {1808.05158},
 primaryClass = {astro-ph.HE},
       adsurl = {https://ui.adsabs.harvard.edu/abs/2018MNRAS.480.4873A}
}

@ARTICLE{2016ApJ...831..102K,
       author = {{Kapanadze}, B. and {Dorner}, D. and {Vercellone}, S. and {Romano}, P. and {Aller}, H. and {Aller}, M. and {Hughes}, P. and {Reynolds}, M. and {Kapanadze}, S. and {Tabagari}, L.},
        title = "{X-Ray Flaring Activity of MRK 421 in the First Half of 2013}",
      journal = {\apj},
         year = 2016,
        month = nov,
       volume = {831},
       number = {1},
          eid = {102},
        pages = {102},
          doi = {10.3847/0004-637X/831/1/102},
       adsurl = {https://ui.adsabs.harvard.edu/abs/2016ApJ...831..102K}
}

@ARTICLE{2018MNRAS.478L.105J,
       author = {{Jagan}, Sitha K. and {Sahayanathan}, S. and {Misra}, R. and {Ravikumar}, C.~D. and {Jeena}, K.},
        title = "{Establishing the spectral turnover of blazar PKS 2155-304 as an outcome of radiative losses}",
      journal = {\mnras},
         year = 2018,
        month = jul,
       volume = {478},
       number = {1},
        pages = {L105-L109},
          doi = {10.1093/mnrasl/sly086},
archivePrefix = {arXiv},
       eprint = {1805.03684},
 primaryClass = {astro-ph.HE},
       adsurl = {https://ui.adsabs.harvard.edu/abs/2018MNRAS.478L.105J}
}

@ARTICLE{2022MNRAS.514.3074B,
       author = {{Baheeja}, C. and {Sahayanathan}, S. and {Rieger}, Frank M. and {Jagan}, Sitha K. and {Ravikumar}, C.~D.},
        title = "{Do radiative losses determine the characteristic emission of the blazar Mkn 421?}",
      journal = {\mnras},
         year = 2022,
        month = aug,
       volume = {514},
       number = {2},
        pages = {3074-3081},
          doi = {10.1093/mnras/stac1548},
archivePrefix = {arXiv},
       eprint = {2206.00659},
 primaryClass = {astro-ph.HE},
       adsurl = {https://ui.adsabs.harvard.edu/abs/2022MNRAS.514.3074B}
}

@ARTICLE{1998ApJ...509..608T,
       author = {{Tavecchio}, Fabrizio and {Maraschi}, Laura and {Ghisellini}, Gabriele},
        title = "{Constraints on the Physical Parameters of TeV Blazars}",
      journal = {\apj},
         year = 1998,
        month = dec,
       volume = {509},
       number = {2},
        pages = {608-619},
          doi = {10.1086/306526},
archivePrefix = {arXiv},
       eprint = {astro-ph/9809051},
 primaryClass = {astro-ph},
       adsurl = {https://ui.adsabs.harvard.edu/abs/1998ApJ...509..608T}
}

@ARTICLE{2002A&A...386..833G,
       author = {{Ghisellini}, G. and {Celotti}, A. and {Costamante}, L.},
        title = "{Low power BL Lacertae objects and the blazar sequence. Clues on the particle acceleration process}",
      journal = {\aap},
         year = 2002,
        month = may,
       volume = {386},
        pages = {833-842},
          doi = {10.1051/0004-6361:20020275},
archivePrefix = {arXiv},
       eprint = {astro-ph/0202367},
 primaryClass = {astro-ph},
       adsurl = {https://ui.adsabs.harvard.edu/abs/2002A&A...386..833G}
}

@ARTICLE{2018MNRAS.480.2046G,
       author = {{Goswami}, Pranjupriya and {Sahayanathan}, Sunder and {Sinha}, Atreyee and {Misra}, Ranjeev and {Gogoi}, Rupjyoti},
        title = "{Influence of energy-dependent particle diffusion on the X-ray spectral curvature of MKN 421}",
      journal = {\mnras},
         year = 2018,
        month = oct,
       volume = {480},
       number = {2},
        pages = {2046-2053},
          doi = {10.1093/mnras/sty2003},
archivePrefix = {arXiv},
       eprint = {1807.09541},
 primaryClass = {astro-ph.HE},
       adsurl = {https://ui.adsabs.harvard.edu/abs/2018MNRAS.480.2046G}
}

@ARTICLE{2020ApJS..247...27K,
       author = {{Kapanadze}, B. and {Gurchumelia}, A. and {Dorner}, D. and {Vercellone}, S. and {Romano}, P. and {Hughes}, P. and {Aller}, M. and {Aller}, H. and {Kharshiladze}, O.},
        title = "{Swift Observations of Mrk 421 in Selected Epochs. III. Extreme X-Ray Timing/Spectral Properties and Multiwavelength Lognormality during 2015 December-2018 April}",
      journal = {\apjs},
         year = 2020,
        month = mar,
       volume = {247},
       number = {1},
          eid = {27},
        pages = {27},
          doi = {10.3847/1538-4365/ab6322},
archivePrefix = {arXiv},
       eprint = {2004.00676},
 primaryClass = {astro-ph.GA},
       adsurl = {https://ui.adsabs.harvard.edu/abs/2020ApJS..247...27K}
}

@ARTICLE{Abeysekara2017,
       author = {{Abeysekara}, A.~U. and {Archambault}, S. and {Archer}, A. and {Benbow}, W. and {Bird}, R. and {Buchovecky}, M. and {Buckley}, J.~H. and {Bugaev}, V. and {Cardenzana}, J.~V. and {Cerruti}, M. and {Chen}, X. and {Ciupik}, L. and {Connolly}, M.~P. and {Cui}, W. and {Eisch}, J.~D. and {Falcone}, A. and {Feng}, Q. and {Finley}, J.~P. and {Fleischhack}, H. and {Flinders}, A. and {Fortson}, L. and {Furniss}, A. and {Griffin}, S. and {H{\r{a}}kansson}, N. and {Hanna}, D. and {Hervet}, O. and {Holder}, J. and {Humensky}, T.~B. and {H{\"u}tten}, M. and {Kaaret}, P. and {Kar}, P. and {Kertzman}, M. and {Kieda}, D. and {Krause}, M. and {Kumar}, S. and {Lang}, M.~J. and {Maier}, G. and {McArthur}, S. and {McCann}, A. and {Meagher}, K. and {Moriarty}, P. and {Mukherjee}, R. and {Nieto}, D. and {O'Brien}, S. and {Ong}, R.~A. and {Otte}, A.~N. and {Park}, N. and {Pelassa}, V. and {Pohl}, M. and {Popkow}, A. and {Pueschel}, E. and {Ragan}, K. and {Reynolds}, P.~T. and {Richards}, G.~T. and {Roache}, E. and {Sadeh}, I. and {Santander}, M. and {Sembroski}, G.~H. and {Shahinyan}, K. and {Staszak}, D. and {Telezhinsky}, I. and {Tucci}, J.~V. and {Tyler}, J. and {Wakely}, S.~P. and {Weinstein}, A. and {Wilhelm}, A. and {Williams}, D.~A. and {VERITAS Collaboration} and {Ahnen}, M.~L. and {Ansoldi}, S. and {Antonelli}, L.~A. and {Antoranz}, P. and {Arcaro}, C. and {Babic}, A. and {Banerjee}, B. and {Bangale}, P. and {Barres de Almeida}, U. and {Barrio}, J.~A. and {Becerra Gonz{\'a}lez}, J. and {Bednarek}, W. and {Bernardini}, E. and {Berti}, A. and {Biasuzzi}, B. and {Biland}, A. and {Blanch}, O. and {Bonnefoy}, S. and {Bonnoli}, G. and {Borracci}, F. and {Bretz}, T. and {Carosi}, R. and {Carosi}, A. and {Chatterjee}, A. and {Colin}, P. and {Colombo}, E. and {Contreras}, J.~L. and {Cortina}, J. and {Covino}, S. and {Cumani}, P. and {Da Vela}, P. and {Dazzi}, F. and {De Angelis}, A. and {De Lotto}, B. and {de O{\~n}a Wilhelmi}, E. and {Di Pierro}, F. and {Doert}, M. and {Dom{\'\i}nguez}, A. and {Dominis Prester}, D. and {Dorner}, D. and {Doro}, M. and {Einecke}, S. and {Eisenacher Glawion}, D. and {Elsaesser}, D. and {Engelkemeier}, M. and {Fallah Ramazani}, V. and {Fern{\'a}ndez-Barral}, A. and {Fidalgo}, D. and {Fonseca}, M.~V. and {Font}, L. and {Fruck}, C. and {Galindo}, D. and {Garc{\'\i}a L{\'o}pez}, R.~J. and {Garczarczyk}, M. and {Gaug}, M. and {Giammaria}, P. and {Godinovi{\'c}}, N. and {Gora}, D. and {Guberman}, D. and {Hadasch}, D. and {Hahn}, A. and {Hassan}, T. and {Hayashida}, M. and {Herrera}, J. and {Hose}, J. and {Hrupec}, D. and {Hughes}, G. and {Idec}, W. and {Kodani}, K. and {Konno}, Y. and {Kubo}, H. and {Kushida}, J. and {Lelas}, D. and {Lindfors}, E. and {Lombardi}, S. and {Longo}, F. and {L{\'o}pez}, M. and {L{\'o}pez-Coto}, R. and {Majumdar}, P. and {Makariev}, M. and {Mallot}, K. and {Maneva}, G. and {Manganaro}, M. and {Mannheim}, K. and {Maraschi}, L. and {Marcote}, B. and {Mariotti}, M. and {Mart{\'\i}nez}, M. and {Mazin}, D. and {Menzel}, U. and {Mirzoyan}, R. and {Moralejo}, A. and {Moretti}, E. and {Nakajima}, D. and {Neustroev}, V. and {Niedzwiecki}, A. and {Nievas Rosillo}, M. and {Nilsson}, K. and {Nishijima}, K. and {Noda}, K. and {Nogu{\'e}s}, L. and {N{\"o}the}, M. and {Paiano}, S. and {Palacio}, J. and {Palatiello}, M. and {Paneque}, D. and {Paoletti}, R. and {Paredes}, J.~M. and {Paredes-Fortuny}, X. and {Pedaletti}, G. and {Peresano}, M. and {Perri}, L. and {Persic}, M. and {Poutanen}, J. and {Prada Moroni}, P.~G. and {Prandini}, E. and {Puljak}, I. and {Garcia}, J.~R. and {Reichardt}, I. and {Rhode}, W. and {Rib{\'o}}, M. and {Rico}, J. and {Saito}, T. and {Satalecka}, K. and {Schroeder}, S. and {Schweizer}, T. and {Shore}, S.~N. and {Sillanp{\"a}{\"a}}, A. and {Sitarek}, J. and {Snidaric}, I. and {Sobczynska}, D. and {Stamerra}, A.},
        title = "{A Search for Spectral Hysteresis and Energy-dependent Time Lags from X-Ray and TeV Gamma-Ray Observations of Mrk 421}",
      journal = {\apj},
         year = 2017,
        month = jan,
       volume = {834},
       number = {1},
          eid = {2},
        pages = {2},
          doi = {10.3847/1538-4357/834/1/2},
archivePrefix = {arXiv},
       eprint = {1611.04626},
 primaryClass = {astro-ph.HE},
       adsurl = {https://ui.adsabs.harvard.edu/abs/2017ApJ...834....2A}
}

@ARTICLE{2022MNRAS.510.5280M,
       author = {{Mohorian}, Maksym and {Bhatta}, Gopal and {Adhikari}, Tek P. and {Dhital}, Niraj and {P{\'a}nis}, Radim and {Dinesh}, Adithiya and {Chaudhary}, Suvas C. and {Bachchan}, Rajesh K. and {Stuchl{\'\i}k}, Zden{\v{e}}k},
        title = "{X-ray timing and spectral variability properties of blazars S5 0716 + 714, OJ 287, Mrk 501, and RBS 2070}",
      journal = {\mnras},
         year = 2022,
        month = mar,
       volume = {510},
       number = {4},
        pages = {5280-5301},
          doi = {10.1093/mnras/stab3738},
archivePrefix = {arXiv},
       eprint = {2112.11272},
 primaryClass = {astro-ph.HE},
       adsurl = {https://ui.adsabs.harvard.edu/abs/2022MNRAS.510.5280M}
}

@ARTICLE{2023ApJ...955..121D,
       author = {{Dinesh}, Adithiya and {Bhatta}, Gopal and {Adhikari}, Tek P. and {Mohorian}, Maksym and {Dhital}, Niraj and {Chaudhary}, Suvas C. and {P{\'a}nis}, Radim and {G{\'o}ra}, Dariusz},
        title = "{Constraining X-Ray Variability of the Blazar 3C 273 Using XMM-Newton Observations over Two Decades}",
      journal = {\apj},
         year = 2023,
        month = oct,
       volume = {955},
       number = {2},
          eid = {121},
        pages = {121},
          doi = {10.3847/1538-4357/acf316},
archivePrefix = {arXiv},
       eprint = {2309.00406},
 primaryClass = {astro-ph.HE},
       adsurl = {https://ui.adsabs.harvard.edu/abs/2023ApJ...955..121D}
}

@ARTICLE{2025ApJ...981..118B,
       author = {{Bhatta}, Gopal and {Chaudhary}, Suvas Chandra and {Dhital}, Niraj and {Adhikari}, Tek P. and {Mohorian}, Maksym and {Dinesh}, Adithiya and {P{\'a}nis}, Radim and {Neupane}, Raghav and {Maharjan}, Yogesh Singh},
        title = "{Probing X-Ray Timing and Spectral Variability in the Blazar PKS 2155{\textendash}304 over a Decade of XMM-Newton Observations}",
      journal = {\apj},
         year = 2025,
        month = mar,
       volume = {981},
       number = {2},
          eid = {118},
        pages = {118},
          doi = {10.3847/1538-4357/adb0c9},
archivePrefix = {arXiv},
       eprint = {2410.01278},
 primaryClass = {astro-ph.HE},
       adsurl = {https://ui.adsabs.harvard.edu/abs/2025ApJ...981..118B}
}

@ARTICLE{Garson2010,
       author = {{Garson}, III, Alfred B. and {Baring}, Matthew G. and {Krawczynski}, Henric},
        title = "{A Suzaku X-ray Study of the Particle Acceleration Processes in the Relativistic Jet of Blazar Mrk 421}",
      journal = {\apj},
         year = 2010,
        month = oct,
       volume = {722},
       number = {1},
        pages = {358-366},
          doi = {10.1088/0004-637X/722/1/358},
archivePrefix = {arXiv},
       eprint = {1008.3377},
 primaryClass = {astro-ph.HE},
       adsurl = {https://ui.adsabs.harvard.edu/abs/2010ApJ...722..358G}
}

@ARTICLE{Gokus2024,
       author = {{Gokus}, A. and {Wilms}, J. and {Kadler}, M. and {Dorner}, D. and {Nowak}, M.~A. and {Kreikenbohm}, A. and {Leiter}, K. and {Bretz}, T. and {Schleicher}, B. and {Markowitz}, A.~G. and {Pottschmidt}, K. and {Mannheim}, K. and {Kreykenbohm}, I. and {Langejahn}, M. and {McBride}, F. and {Beuchert}, T. and {Dauser}, T. and {Kreter}, M. and {Abhir}, J. and {Baack}, D. and {Balbo}, M. and {Biland}, A. and {Brand}, K. and {Buss}, J. and {Eisenberger}, L. and {Elsaesser}, D. and {G{\"u}nther}, P. and {Hildebrand}, D. and {Linhoff}, M. and {Paravac}, A. and {Rhode}, W. and {Sliusar}, V. and {Hasan}, S. and {Walter}, R.},
        title = "{Rapid variability of Markarian 421 during extreme flaring as seen through the eyes of XMM-Newton}",
      journal = {\mnras},
         year = 2024,
        month = apr,
       volume = {529},
       number = {2},
        pages = {1450-1462},
          doi = {10.1093/mnras/stae643},
archivePrefix = {arXiv},
       eprint = {2402.19377},
 primaryClass = {astro-ph.HE},
       adsurl = {https://ui.adsabs.harvard.edu/abs/2024MNRAS.529.1450G}
}

@ARTICLE{Fossati1998,
       author = {{Fossati}, G. and {Maraschi}, L. and {Celotti}, A. and {Comastri}, A. and {Ghisellini}, G.},
        title = "{A unifying view of the spectral energy distributions of blazars}",
      journal = {\mnras},
         year = 1998,
        month = sep,
       volume = {299},
       number = {2},
        pages = {433-448},
          doi = {10.1046/j.1365-8711.1998.01828.x},
archivePrefix = {arXiv},
       eprint = {astro-ph/9804103},
 primaryClass = {astro-ph},
       adsurl = {https://ui.adsabs.harvard.edu/abs/1998MNRAS.299..433F}
}

@ARTICLE{Yan2018,
       author = {{Yan}, Dahai and {Yang}, Shenbang and {Zhang}, Pengfei and
         {Dai}, Benzhong and {Wang}, Jiancheng and {Zhang}, Li},
        title = "{Statistical Analysis on XMM-Newton X-Ray Flares of Mrk 421: Distributions of Peak Flux and Flaring Time Duration}",
      journal = {\apj},
         year = 2018,
        month = sep,
       volume = {864},
       number = {2},
          eid = {164},
        pages = {164},
          doi = {10.3847/1538-4357/aadd01},
archivePrefix = {arXiv},
       eprint = {1808.06029},
 primaryClass = {astro-ph.HE},
       adsurl = {https://ui.adsabs.harvard.edu/abs/2018ApJ...864..164Y}
}

@ARTICLE{Tramacere2007,
       author = {{Tramacere}, A. and {Massaro}, F. and {Cavaliere}, A.},
        title = "{Signatures of synchrotron emission and of electron acceleration in the X-ray spectra of Mrk 421}",
      journal = {\aap},
         year = 2007,
        month = may,
       volume = {466},
       number = {2},
        pages = {521-529},
          doi = {10.1051/0004-6361:20066723},
archivePrefix = {arXiv},
       eprint = {astro-ph/0702151},
 primaryClass = {astro-ph},
       adsurl = {https://ui.adsabs.harvard.edu/abs/2007A&A...466..521T}
}

@ARTICLE{Kalberla2005,
       author = {{Kalberla}, P.~M.~W. and {Burton}, W.~B. and {Hartmann}, Dap and {Arnal}, E.~M. and {Bajaja}, E. and {Morras}, R. and {P{\"o}ppel}, W.~G.~L.},
        title = "{The Leiden/Argentine/Bonn (LAB) Survey of Galactic HI. Final data release of the combined LDS and IAR surveys with improved stray-radiation corrections}",
      journal = {\aap},
         year = 2005,
        month = sep,
       volume = {440},
       number = {2},
        pages = {775-782},
          doi = {10.1051/0004-6361:20041864},
archivePrefix = {arXiv},
       eprint = {astro-ph/0504140},
 primaryClass = {astro-ph},
       adsurl = {https://ui.adsabs.harvard.edu/abs/2005A&A...440..775K}
}

@ARTICLE{Tramacere2009,
       author = {{Tramacere}, A. and {Giommi}, P. and {Perri}, M. and {Verrecchia}, F. and {Tosti}, G.},
        title = "{Swift observations of the very intense flaring activity of Mrk 421 during 2006. I. Phenomenological picture of electron acceleration and predictions for MeV/GeV emission}",
      journal = {\aap},
         year = 2009,
        month = jul,
       volume = {501},
       number = {3},
        pages = {879-898},
          doi = {10.1051/0004-6361/200810865},
archivePrefix = {arXiv},
       eprint = {0901.4124},
 primaryClass = {astro-ph.HE},
       adsurl = {https://ui.adsabs.harvard.edu/abs/2009A&A...501..879T}
}

@ARTICLE{Massaro2004,
       author = {{Massaro}, E. and {Perri}, M. and {Giommi}, P. and {Nesci}, R. and {Verrecchia}, F.},
        title = "{Log-parabolic spectra and particle acceleration in blazars.  II. The BeppoSAX wide band X-ray spectra of Mkn 501}",
      journal = {\aap},
         year = 2004,
        month = jul,
       volume = {422},
        pages = {103-111},
          doi = {10.1051/0004-6361:20047148},
archivePrefix = {arXiv},
       eprint = {astro-ph/0405152},
 primaryClass = {astro-ph},
       adsurl = {https://ui.adsabs.harvard.edu/abs/2004A&A...422..103M}
}

@ARTICLE{Massaro2004Mrk421,
       author = {{Massaro}, E. and {Perri}, M. and {Giommi}, P. and {Nesci}, R.},
        title = "{Log-parabolic spectra and particle acceleration in the BL Lac object Mkn 421: Spectral analysis of the complete BeppoSAX wide band X-ray data set}",
      journal = {\aap},
         year = 2004,
        month = jan,
       volume = {413},
        pages = {489-503},
          doi = {10.1051/0004-6361:20031558},
archivePrefix = {arXiv},
       eprint = {astro-ph/0312260},
 primaryClass = {astro-ph},
       adsurl = {https://ui.adsabs.harvard.edu/abs/2004A&A...413..489M}
}

@ARTICLE{Acciari2020,
       author = {{Acciari}, V.~A. and {Ansoldi}, S. and {Antonelli}, L.~A. and
         {Arbet Engels}, A. and {Baack}, D. and {Babi{\'c}}, A. and
         {Banerjee}, B. and {Barres de Almeida}, U. and {Barrio}, J.~A. and
         {Becerra Gonz{\'a}lez}, J. and {Bednarek}, W. and {Bellizzi}, L.~K. and
         {Bernardini}, E. and {Berti}, A. and {Besenrieder}, J. and
         {Bhattacharyya}, W. and {Bigongiari}, C. and {Biland}, A. and
         {Blanch}, O. and {Bonnoli}, G. and {Bo{\v{s}}njak}, {\v{Z}}. and
         {Busetto}, G. and {Carosi}, R. and {Ceribella}, G. and {Cerruti}, M. and
         {Chai}, Y. and {Chilingarian}, A. and {Cikota}, S. and {Colak}, S.~M. and
         {Colin}, U. and {Colombo}, E. and {Contreras}, J.~L. and {Cortina}, J. and
         {Covino}, S. and {D'Elia}, V. and {da Vela}, P. and {Dazzi}, F. and
         {de Angelis}, A. and {de Lotto}, B. and {Del Puppo}, F. and
         {Delfino}, M. and {Delgado}, J. and {Depaoli}, D. and {di Pierro}, F. and
         {di Venere}, L. and {Do Souto Espi{\~n}eira}, E. and
         {Prester}, D. Dominis and {Donini}, A. and {Dorner}, D. and {Doro}, M. and
         {Elsaesser}, D. and {Ramazani}, V. Fallah and {Fattorini}, A. and
         {Ferrara}, G. and {Foffano}, L. and {Fonseca}, M.~V. and {Font}, L. and
         {Fruck}, C. and {Fukami}, S. and {Garc{\'\i}a L{\'o}pez}, R.~J. and
         {Garczarczyk}, M. and {Gasparyan}, S. and {Gaug}, M. and
         {Giglietto}, N. and {Giordano}, F. and {Gliwny}, P. and
         {Godinovi{\'c}}, N. and {Green}, D. and {Hadasch}, D. and {Hahn}, A. and
         {Hassan}, T. and {Herrera}, J. and {Hoang}, J. and {Hrupec}, D. and
         {H{\"u}tten}, M. and {Inada}, T. and {Inoue}, S. and {Ishio}, K. and
         {Iwamura}, Y. and {Jouvin}, L. and {Kajiwara}, Y. and {Kerszberg}, D. and
         {Kobayashi}, Y. and {Kubo}, H. and {Kushida}, J. and {Lamastra}, A. and
         {Lelas}, D. and {Leone}, F. and {Lindfors}, E. and {Lombardi}, S. and
         {Longo}, F. and {L{\'o}pez}, M. and {L{\'o}pez-Coto}, R. and
         {L{\'o}pez-Oramas}, A. and {Loporchio}, S. and
         {Machado de Oliveira Fraga}, B. and {Maggio}, C. and {Majumdar}, P. and
         {Makariev}, M. and {Mallamaci}, M. and {Maneva}, G. and
         {Manganaro}, M. and {Mannheim}, K. and {Maraschi}, L. and
         {Mariotti}, M. and {Mart{\'\i}nez}, M. and {Mazin}, D. and
         {Mender}, S. and {Mi{\'c}anovi{\'c}}, S. and {Miceli}, D. and
         {Miener}, T. and {Minev}, M. and {Miranda}, J.~M. and {Mirzoyan}, R. and
         {Molina}, E. and {Moralejo}, A. and {Morcuende}, D. and {Moreno}, V. and
         {Moretti}, E. and {Munar-Adrover}, P. and {Neustroev}, V. and
         {Nigro}, C. and {Nilsson}, K. and {Ninci}, D. and {Nishijima}, K. and
         {Noda}, K. and {Nogu{\'e}s}, L. and {Nozaki}, S. and {Ohtani}, Y. and
         {Oka}, T. and {Otero-Santos}, J. and {Palatiello}, M. and
         {Paneque}, D. and {Paoletti}, R. and {Paredes}, J.~M. and
         {Pavleti{\'c}}, L. and {Pe{\~n}il}, P. and {Peresano}, M. and
         {Persic}, M. and {Moroni}, P.~G. Prada and {Prandini}, E. and
         {Puljak}, I. and {Rhode}, W. and {Rib{\'o}}, M. and {Rico}, J. and
         {Righi}, C. and {Rugliancich}, A. and {Saha}, L. and {Sahakyan}, N. and
         {Saito}, T. and {Sakurai}, S. and {Satalecka}, K. and {Schleicher}, B. and
         {Schmidt}, K. and {Schweizer}, T. and {Sitarek}, J. and
         {{\v{S}}nidari{\'c}}, I. and {Sobczynska}, D. and {Spolon}, A. and
         {Stamerra}, A. and {Strom}, D. and {Strzys}, M. and {Suda}, Y. and
         {Suri{\'c}}, T. and {Takahashi}, M. and {Tavecchio}, F. and
         {Temnikov}, P. and {Terzi{\'c}}, T. and {Teshima}, M. and
         {Torres-Alb{\`a}}, N. and {Tosti}, L. and {van Scherpenberg}, J. and
         {Vanzo}, G. and {Vazquez Acosta}, M. and {Ventura}, S. and
         {Verguilov}, V. and {Vigorito}, C.~F. and {Vitale}, V. and {Vovk}, I. and
         {Will}, M. and {Zari{\'c}}, D. and {MAGIC Collaboration} and
         {Finke}, J. and {D'Ammando}, F. and {Balokovi{\'c}}, M. and
         {Madejski}, G. and {Mori}, K. and {Puccetti}, Simonetta and {Leto}, C. and
         {Perri}, M. and {Verrecchia}, F. and {Villata}, M. and
         {Raiteri}, C.~M. and {Agudo}, I. and {Bachev}, R. and {Berdyugin}, A. and
         {Blinov}, D.~A. and {Chanishvili}, R. and {Chen}, W.~P. and
         {Chigladze}, R. and {Damljanovic}, G. and {Eswaraiah}, C. and
         {Grishina}, T.~S. and {Ibryamov}, S. and {Jordan}, B. and
         {Jorstad}, S.~G. and {Joshi}, M. and {Kopatskaya}, E.~N. and
         {Kurtanidze}, O.~M. and {Kurtanidze}, S.~O. and {Larionova}, E.~G. and
         {Larionova}, L.~V. and {Larionov}, V.~M. and {Latev}, G. and
         {Lin}, H.~C. and {Marscher}, A.~P. and {Mokrushina}, A.~A. and
         {Morozova}, D.~A. and {Nikolashvili}, M.~G. and {Semkov}, E. and
         {Smith}, P.~S. and {Strigachev}, A. and {Troitskaya}, Yu. V. and
         {Troitsky}, I.~S. and {Vince}, O. and {Barnes}, J. and {G{\"u}ver}, T. and
         {Moody}, J.~W. and {Sadun}, A.~C. and {Hovatta}, T. and
         {Richards}, J.~L. and {Max-Moerbeck}, W. and {Readhead}, A.~C.~S. and
         {L{\"a}hteenm{\"a}ki}, A. and {Tornikoski}, M. and {Tammi}, J. and
         {Ramakrishnan}, V. and {Reinthal}, R.},
        title = "{Unraveling the Complex Behavior of Mrk 421 with Simultaneous X-Ray and VHE Observations during an Extreme Flaring Activity in 2013 April}",
      journal = {\apjs},
         year = 2020,
        month = jun,
       volume = {248},
       number = {2},
          eid = {29},
        pages = {29},
          doi = {10.3847/1538-4365/ab89b5},
archivePrefix = {arXiv},
       eprint = {2001.08678},
 primaryClass = {astro-ph.HE},
       adsurl = {https://ui.adsabs.harvard.edu/abs/2020ApJS..248...29A}
}

@article{urry1995unified,
  title={Unified schemes for radio-loud active galactic nuclei},
  author={Urry, C Megan and Padovani, Paolo},
  journal={Publications of the Astronomical Society of the Pacific},
  volume={107},
  number={715},
  pages={803},
  year={1995},
  publisher={IOP Publishing}
}

@ARTICLE{Wilms2000,
       author = {{Wilms}, J. and {Allen}, A. and {McCray}, R.},
        title = "{On the Absorption of X-Rays in the Interstellar Medium}",
      journal = {\apj},
         year = 2000,
        month = oct,
       volume = {542},
       number = {2},
        pages = {914-924},
          doi = {10.1086/317016},
archivePrefix = {arXiv},
       eprint = {astro-ph/0008425},
 primaryClass = {astro-ph},
       adsurl = {https://ui.adsabs.harvard.edu/abs/2000ApJ...542..914W}
}

@ARTICLE{Maraschi1992,
       author = {{Maraschi}, L. and {Ghisellini}, G. and {Celotti}, A.},
        title = "{A Jet Model for the Gamma-Ray--emitting Blazar 3C 279}",
      journal = {\apjl},
         year = 1992,
        month = sep,
       volume = {397},
        pages = {L5},
          doi = {10.1086/186531},
       adsurl = {https://ui.adsabs.harvard.edu/abs/1992ApJ...397L...5M}
}

@ARTICLE{Dermer1993,
       author = {{Dermer}, Charles D. and {Schlickeiser}, Reinhard},
        title = "{Model for the High-Energy Emission from Blazars}",
      journal = {\apj},
         year = 1993,
        month = oct,
       volume = {416},
        pages = {458},
          doi = {10.1086/173251},
       adsurl = {https://ui.adsabs.harvard.edu/abs/1993ApJ...416..458D}
}

@ARTICLE{Sikora1994,
       author = {{Sikora}, Marek and {Begelman}, Mitchell C. and {Rees}, Martin J.},
        title = "{Comptonization of Diffuse Ambient Radiation by a Relativistic Jet: The Source of Gamma Rays from Blazars?}",
      journal = {\apj},
         year = 1994,
        month = jan,
       volume = {421},
        pages = {153},
          doi = {10.1086/173633},
       adsurl = {https://ui.adsabs.harvard.edu/abs/1994ApJ...421..153S}
}

@ARTICLE{Warwick1986,
       author = {{Warwick}, R.~S. and {George}, I.~M. and {McHardy}, I. and {Pounds}, K.~A.},
        title = "{The X-ray spectrum and variability of Markarian 421.}",
      journal = {\mnras},
         year = 1986,
        month = mar,
       volume = {219},
        pages = {39-45},
          doi = {10.1093/mnras/219.1.39},
       adsurl = {https://ui.adsabs.harvard.edu/abs/1986MNRAS.219...39W}
}

@ARTICLE{Cui2004,
       author = {{Cui}, Wei},
        title = "{X-Ray Flaring Activity of Markarian 421}",
      journal = {\apj},
         year = 2004,
        month = apr,
       volume = {605},
       number = {2},
        pages = {662-669},
          doi = {10.1086/382587},
archivePrefix = {arXiv},
       eprint = {astro-ph/0401222},
 primaryClass = {astro-ph},
       adsurl = {https://ui.adsabs.harvard.edu/abs/2004ApJ...605..662C}
}

@ARTICLE{Mondal2022,
       author = {{Mondal}, S. and {Rani}, P. and {Stalin}, C.~S. and {Chakrabarti}, S.~K. and {Rakshit}, S.},
        title = "{Flux and spectral variability of Mrk 421 during its moderate activity state using NuSTAR: Possible accretion disc contribution?}",
      journal = {\aap},
         year = 2022,
        month = jul,
       volume = {663},
          eid = {A178},
        pages = {A178},
          doi = {10.1051/0004-6361/202141990},
archivePrefix = {arXiv},
       eprint = {2204.02132},
 primaryClass = {astro-ph.HE},
       adsurl = {https://ui.adsabs.harvard.edu/abs/2022A&A...663A.178M}
}

@ARTICLE{Acciari2011,
       author = {{Acciari}, V.~A. and {Aliu}, E. and {Arlen}, T. and {Aune}, T. and {Beilicke}, M. and {Benbow}, W. and {Boltuch}, D. and {Bradbury}, S.~M. and {Buckley}, J.~H. and {Bugaev}, V. and {Byrum}, K. and {Cannon}, A. and {Cesarini}, A. and {Ciupik}, L. and {Cui}, W. and {Dickherber}, R. and {Duke}, C. and {Falcone}, A. and {Finley}, J.~P. and {Finnegan}, G. and {Fortson}, L. and {Furniss}, A. and {Galante}, N. and {Gall}, D. and {Gillanders}, G.~H. and {Godambe}, S. and {Grube}, J. and {Guenette}, R. and {Gyuk}, G. and {Hanna}, D. and {Holder}, J. and {Hui}, C.~M. and {Humensky}, T.~B. and {Imran}, A. and {Kaaret}, P. and {Karlsson}, N. and {Kertzman}, M. and {Kieda}, D. and {Konopelko}, A. and {Krawczynski}, H. and {Krennrich}, F. and {Lang}, M.~J. and {Maier}, G. and {McArthur}, S. and {McCutcheon}, M. and {Moriarty}, P. and {Ong}, R.~A. and {Otte}, A.~N. and {Ouellette}, M. and {Pandel}, D. and {Perkins}, J.~S. and {Pichel}, A. and {Pohl}, M. and {Quinn}, J. and {Ragan}, K. and {Reyes}, L.~C. and {Reynolds}, P.~T. and {Roache}, E. and {Rose}, H.~J. and {Rovero}, A.~C. and {Schroedter}, M. and {Sembroski}, G.~H. and {Senturk}, G. Demet and {Steele}, D. and {Swordy}, S.~P. and {Theiling}, M. and {Thibadeau}, S. and {Varlotta}, A. and {Vassiliev}, V.~V. and {Vincent}, S. and {Wagner}, R.~G. and {Wakely}, S.~P. and {Ward}, J.~E. and {Weekes}, T.~C. and {Weinstein}, A. and {Weisgarber}, T. and {Williams}, D.~A. and {Wissel}, S. and {Wood}, M. and {Zitzer}, B. and {Garson}, III, A. and {Lee}, K. and {Sadun}, A.~C. and {Carini}, M. and {Barnaby}, D. and {Cook}, K. and {Maune}, J. and {Pease}, A. and {Smith}, S. and {Walters}, R. and {Berdyugin}, A. and {Lindfors}, E. and {Nilsson}, K. and {Pasanen}, M. and {Sainio}, J. and {Sillanpaa}, A. and {Takalo}, L.~O. and {Villforth}, C. and {Montaruli}, T. and {Baker}, M. and {Lahteenmaki}, A. and {Tornikoski}, M. and {Hovatta}, T. and {Nieppola}, E. and {Aller}, H.~D. and {Aller}, M.~F.},
        title = "{TeV and Multi-wavelength Observations of Mrk 421 in 2006-2008}",
      journal = {\apj},
         year = 2011,
        month = sep,
       volume = {738},
       number = {1},
          eid = {25},
        pages = {25},
          doi = {10.1088/0004-637X/738/1/25},
archivePrefix = {arXiv},
       eprint = {1106.1210},
 primaryClass = {astro-ph.HE},
       adsurl = {https://ui.adsabs.harvard.edu/abs/2011ApJ...738...25A}
}

@ARTICLE{DiGesu2022,
       author = {{Di Gesu}, Laura and {Donnarumma}, Immacolata and {Tavecchio}, Fabrizio and {Agudo}, Iv{\'a}n and {Barnounin}, Thibault and {Cibrario}, Nicol{\`o} and {Di Lalla}, Niccol{\`o} and {Di Marco}, Alessandro and {Escudero}, Juan and {Errando}, Manel and {Jorstad}, Svetlana G. and {Kim}, Dawoon E. and {Kouch}, Pouya M. and {Liodakis}, Ioannis and {Lindfors}, Elina and {Madejski}, Grzegorz and {Marshall}, Herman L. and {Marscher}, Alan P. and {Middei}, Riccardo and {Muleri}, Fabio and {Myserlis}, Ioannis and {Negro}, Michela and {Omodei}, Nicola and {Pacciani}, Luigi and {Paggi}, Alessandro and {Perri}, Matteo and {Puccetti}, Simonetta and {Antonelli}, Lucio A. and {Bachetti}, Matteo and {Baldini}, Luca and {Baumgartner}, Wayne H. and {Bellazzini}, Ronaldo and {Bianchi}, Stefano and {Bongiorno}, Stephen D. and {Bonino}, Raffaella and {Brez}, Alessandro and {Bucciantini}, Niccol{\`o} and {Capitanio}, Fiamma and {Castellano}, Simone and {Cavazzuti}, Elisabetta and {Ciprini}, Stefano and {Costa}, Enrico and {De Rosa}, Alessandra and {Del Monte}, Ettore and {Doroshenko}, Victor and {Dov{\v{c}}iak}, Michal and {Ehlert}, Steven R. and {Enoto}, Teruaki and {Evangelista}, Yuri and {Fabiani}, Sergio and {Ferrazzoli}, Riccardo and {Garcia}, Javier A. and {Gunji}, Shuichi and {Hayashida}, Kiyoshi and {Heyl}, Jeremy and {Iwakiri}, Wataru and {Karas}, Vladimir and {Kitaguchi}, Takao and {Kolodziejczak}, Jeffery J. and {Krawczynski}, Henric and {La Monaca}, Fabio and {Latronico}, Luca and {Maldera}, Simone and {Manfreda}, Alberto and {Marin}, Fr{\'e}d{\'e}ric and {Marinucci}, Andrea and {Massaro}, Francesco and {Matt}, Giorgio and {Mitsuishi}, Ikuyuki and {Mizuno}, Tsunefumi and {Ng}, C. -Y. and {O'Dell}, Stephen L. and {Oppedisano}, Chiara and {Papitto}, Alessandro and {Pavlov}, George G. and {Peirson}, Abel L. and {Pesce-Rollins}, Melissa and {Petrucci}, Pierre-Olivier and {Pilia}, Maura and {Possenti}, Andrea and {Poutanen}, Juri and {Ramsey}, Brian D. and {Rankin}, John and {Ratheesh}, Ajay and {Romani}, Roger W. and {Sgr{\`o}}, Carmelo and {Slane}, Patrick and {Soffitta}, Paolo and {Spandre}, Gloria and {Tamagawa}, Toru and {Taverna}, Roberto and {Tawara}, Yuzuru and {Tennant}, Allyn F. and {Thomas}, Nicolas E. and {Tombesi}, Francesco and {Trois}, Alessio and {Tsygankov}, Sergey and {Turolla}, Roberto and {Vink}, Jacco and {Weisskopf}, Martin C. and {Wu}, Kinwah and {Xie}, Fei and {Zane}, Silvia},
        title = "{The X-Ray Polarization View of Mrk 421 in an Average Flux State as Observed by the Imaging X-Ray Polarimetry Explorer}",
      journal = {\apjl},
         year = 2022,
        month = oct,
       volume = {938},
       number = {1},
          eid = {L7},
        pages = {L7},
          doi = {10.3847/2041-8213/ac913a},
archivePrefix = {arXiv},
       eprint = {2209.07184},
 primaryClass = {astro-ph.HE},
       adsurl = {https://ui.adsabs.harvard.edu/abs/2022ApJ...938L...7D}
}

@ARTICLE{bhattacharyya2020blazar,
       author = {{Bhattacharyya}, Joy and {Ghosh}, Ritesh and {Chatterjee}, Ritaban and {Das}, Nabanita},
        title = "{Blazar Variability: A Study of Nonstationarity and the Flux-Rms Relation}",
      journal = {\apj},
         year = 2020,
        month = jul,
       volume = {897},
       number = {1},
          eid = {25},
        pages = {25},
          doi = {10.3847/1538-4357/ab91a8},
archivePrefix = {arXiv},
       eprint = {2005.05230},
 primaryClass = {astro-ph.HE},
       adsurl = {https://ui.adsabs.harvard.edu/abs/2020ApJ...897...25B}
}

@ARTICLE{bhatta2020nature,
       author = {{Bhatta}, Gopal and {Dhital}, Niraj},
        title = "{The Nature of {\ensuremath{\gamma}}-Ray Variability in Blazars}",
      journal = {\apj},
         year = 2020,
        month = mar,
       volume = {891},
       number = {2},
          eid = {120},
        pages = {120},
          doi = {10.3847/1538-4357/ab7455},
archivePrefix = {arXiv},
       eprint = {1911.08198},
 primaryClass = {astro-ph.HE},
       adsurl = {https://ui.adsabs.harvard.edu/abs/2020ApJ...891..120B}
}

@ARTICLE{kim2024,
       author = {{Kim}, Dawoon E. and {Di Gesu}, Laura and {Liodakis}, Ioannis and {Marscher}, Alan P. and {Jorstad}, Svetlana G. and {Middei}, Riccardo and {Marshall}, Herman L. and {Pacciani}, Luigi and {Agudo}, Iv{\'a}n and {Tavecchio}, Fabrizio and {Cibrario}, Nicol{\'o} and {Tugliani}, Stefano and {Bonino}, Raffaella and {Negro}, Michela and {Puccetti}, Simonetta and {Tombesi}, Francesco and {Costa}, Enrico and {Donnarumma}, Immacolata and {Soffitta}, Paolo and {Mizuno}, Tsunefumi and {Fukazawa}, Yasushi and {Kawabata}, Koji S. and {Nakaoka}, Tatsuya and {Uemura}, Makoto and {Imazawa}, Ryo and {Sasada}, Mahito and {Akitaya}, Hiroshi and {Jos{\`e} Aceituno}, Francisco and {Bonnoli}, Giacomo and {Casanova}, V{\`\i}ctor and {Myserlis}, Ioannis and {Sievers}, Albrecht and {Angelakis}, Emmanouil and {Kraus}, Alexander and {Yeon Cheong}, Whee and {Jeong}, Hyeon-Woo and {Kang}, Sincheol and {Kim}, Sang-Hyun and {Lee}, Sang-Sung and {Ag{\`\i}s-Gonz{\`a}lez}, Beatriz and {Sota}, Alfredo and {Escudero}, Juan and {Gurwell}, Mark and {Keating}, Garrett K. and {Rao}, Ramprasad and {Kouch}, Pouya M. and {Lindfors}, Elina and {Bourbah}, Ioakeim G. and {Kiehlmann}, Sebastian and {Kontopodis}, Evangelos and {Mandarakas}, Nikos and {Romanopoulos}, Stylianos and {Skalidis}, Raphael and {Vervelaki}, Anna and {Savchenko}, Sergey S. and {Antonelli}, Lucio A. and {Bachetti}, Matteo and {Baldini}, Luca and {Baumgartner}, Wayne H. and {Bellazzini}, Ronaldo and {Bianchi}, Stefano and {Bongiorno}, Stephen D. and {Brez}, Alessandro and {Bucciantini}, Niccol{\'o} and {Capitanio}, Fiamma and {Castellano}, Simone and {Cavazzuti}, Elisabetta and {Chen}, Chien-Ting and {Ciprini}, Stefano and {De Rosa}, Alessandra and {Del Monte}, Ettore and {Di Lalla}, Niccol{\'o} and {Di Marco}, Alessandro and {Doroshenko}, Victor and {Dov{\v{c}}iak}, Michal and {Ehlert}, Steven R. and {Enoto}, Teruaki and {Evangelista}, Yuri and {Fabiani}, Sergio and {Ferrazzoli}, Riccardo and {Garcia}, Javier A. and {Gunji}, Shuichi and {Hayashida}, Kiyoshi and {Heyl}, Jeremy and {Iwakiri}, Wataru and {Kaaret}, Philip and {Karas}, Vladimir and {Kislat}, Fabian and {Kitaguchi}, Takao and {Kolodziejczak}, Jeffery J. and {Krawczynski}, Henric and {La Monaca}, Fabio and {Latronico}, Luca and {Maldera}, Simone and {Manfreda}, Alberto and {Marin}, Fr{\'e}d{\'e}ric and {Marinucci}, Andrea and {Massaro}, Francesco and {Matt}, Giorgio and {Mitsuishi}, Ikuyuki and {Muleri}, Fabio and {Ng}, C. -Y. and {O'Dell}, Stephen L. and {Omodei}, Nicola and {Oppedisano}, Chiara and {Papitto}, Alessandro and {Pavlov}, George G. and {Peirson}, Abel L. and {Perri}, Matteo and {Pesce-Rollins}, Melissa and {Petrucci}, Pierre-Olivier and {Pilia}, Maura and {Possenti}, Andrea and {Poutanen}, Juri and {Ramsey}, Brian D. and {Rankin}, John and {Ratheesh}, Ajay and {Roberts}, Oliver and {Romani}, Roger W. and {Sgr{\'o}}, Carmelo and {Slane}, Patrick and {Spandre}, Gloria and {Swartz}, Doug and {Tamagawa}, Toru and {Taverna}, Roberto and {Tawara}, Yuzuru and {Tennant}, Allyn F. and {Thomas}, Nicholas E. and {Trois}, Alessio and {Tsygankov}, Sergey S. and {Turolla}, Roberto and {Vink}, Jacco and {Weisskopf}, Martin C. and {Wu}, Kinwah and {Xie}, Fei and {Zane}, Silvia},
        title = "{Magnetic field properties inside the jet of Mrk 421. Multiwavelength polarimetry, including the Imaging X-ray Polarimetry Explorer}",
      journal = {\aap},
         year = 2024,
        month = jan,
       volume = {681},
          eid = {A12},
        pages = {A12},
          doi = {10.1051/0004-6361/202347408},
archivePrefix = {arXiv},
       eprint = {2310.06097},
 primaryClass = {astro-ph.HE},
       adsurl = {https://ui.adsabs.harvard.edu/abs/2024A&A...681A..12K}
}

@ARTICLE{Kizhakkekalam2026,
       author = {{Kizhakkekalam}, Sangeetha and {Bhatta}, Gopal and {Navaneeth}, P.~K. and {Adhikari}, Tek P.},
        title = "{NICER Perspective on TeV Blazar Mrk 421: X-Ray Variability and Particle Acceleration}",
      journal = {\apj},
         year = 2026,
        month = jul,
       volume = {1005},
       number = {2},
          eid = {216},
        pages = {216},
          doi = {10.3847/1538-4357/ae7a51},
archivePrefix = {arXiv},
       eprint = {2512.08531},
 primaryClass = {astro-ph.HE},
       adsurl = {https://ui.adsabs.harvard.edu/abs/2026ApJ..1005..216K}
}

@ARTICLE{Blazejowski2000,
       author = {{B{\l}a{\.z}ejowski}, M. and {Sikora}, M. and {Moderski}, R. and {Madejski}, G.~M.},
        title = "{Comptonization of Infrared Radiation from Hot Dust by Relativistic Jets in Quasars}",
      journal = {\apj},
         year = 2000,
        month = dec,
       volume = {545},
       number = {1},
        pages = {107-116},
          doi = {10.1086/317791},
archivePrefix = {arXiv},
       eprint = {astro-ph/0008154},
 primaryClass = {astro-ph},
       adsurl = {https://ui.adsabs.harvard.edu/abs/2000ApJ...545..107B}
}

@ARTICLE{Balokovi2016,
       author = {{Balokovi{\'c}}, M. and {Paneque}, D. and {Madejski}, G. and {Furniss}, A. and {Chiang}, J. and {Ajello}, M. and {Alexander}, D.~M. and {Barret}, D. and {Blandford}, R.~D. and {Boggs}, S.~E. and {Christensen}, F.~E. and {Craig}, W.~W. and {Forster}, K. and {Giommi}, P. and {Grefenstette}, B. and {Hailey}, C. and {Harrison}, F.~A. and {Hornstrup}, A. and {Kitaguchi}, T. and {Koglin}, J.~E. and {Madsen}, K.~K. and {Mao}, P.~H. and {Miyasaka}, H. and {Mori}, K. and {Perri}, M. and {Pivovaroff}, M.~J. and {Puccetti}, S. and {Rana}, V. and {Stern}, D. and {Tagliaferri}, G. and {Urry}, C.~M. and {Westergaard}, N.~J. and {Zhang}, W.~W. and {Zoglauer}, A. and {NuSTAR Team} and {Archambault}, S. and {Archer}, A. and {Barnacka}, A. and {Benbow}, W. and {Bird}, R. and {Buckley}, J.~H. and {Bugaev}, V. and {Cerruti}, M. and {Chen}, X. and {Ciupik}, L. and {Connolly}, M.~P. and {Cui}, W. and {Dickinson}, H.~J. and {Dumm}, J. and {Eisch}, J.~D. and {Falcone}, A. and {Feng}, Q. and {Finley}, J.~P. and {Fleischhack}, H. and {Fortson}, L. and {Griffin}, S. and {Griffiths}, S.~T. and {Grube}, J. and {Gyuk}, G. and {Huetten}, M. and {H{\r{a}}kansson}, N. and {Holder}, J. and {Humensky}, T.~B. and {Johnson}, C.~A. and {Kaaret}, P. and {Kertzman}, M. and {Khassen}, Y. and {Kieda}, D. and {Krause}, M. and {Krennrich}, F. and {Lang}, M.~J. and {Maier}, G. and {McArthur}, S. and {Meagher}, K. and {Moriarty}, P. and {Nelson}, T. and {Nieto}, D. and {Ong}, R.~A. and {Park}, N. and {Pohl}, M. and {Popkow}, A. and {Pueschel}, E. and {Reynolds}, P.~T. and {Richards}, G.~T. and {Roache}, E. and {Santander}, M. and {Sembroski}, G.~H. and {Shahinyan}, K. and {Smith}, A.~W. and {Staszak}, D. and {Telezhinsky}, I. and {Todd}, N.~W. and {Tucci}, J.~V. and {Tyler}, J. and {Vincent}, S. and {Weinstein}, A. and {Wilhelm}, A. and {Williams}, D.~A. and {Zitzer}, B. and {VERITAS Collaboration} and {Ahnen}, M.~L. and {Ansoldi}, S. and {Antonelli}, L.~A. and {Antoranz}, P. and {Babic}, A. and {Banerjee}, B. and {Bangale}, P. and {Barres de Almeida}, U. and {Barrio}, J.~A. and {Becerra Gonz{\'a}lez}, J. and {Bednarek}, W. and {Bernardini}, E. and {Biasuzzi}, B. and {Biland}, A. and {Blanch}, O. and {Bonnefoy}, S. and {Bonnoli}, G. and {Borracci}, F. and {Bretz}, T. and {Carmona}, E. and {Carosi}, A. and {Chatterjee}, A. and {Clavero}, R. and {Colin}, P. and {Colombo}, E. and {Contreras}, J.~L. and {Cortina}, J. and {Covino}, S. and {Da Vela}, P. and {Dazzi}, F. and {De Angelis}, A. and {De Lotto}, B. and {de O{\~n}a Wilhelmi}, E. and {Delgado Mendez}, C. and {Di Pierro}, F. and {Dominis Prester}, D. and {Dorner}, D. and {Doro}, M. and {Einecke}, S. and {Elsaesser}, D. and {Fern{\'a}ndez-Barral}, A. and {Fidalgo}, D. and {Fonseca}, M.~V. and {Font}, L. and {Frantzen}, K. and {Fruck}, C. and {Galindo}, D. and {Garc{\'\i}a L{\'o}pez}, R.~J. and {Garczarczyk}, M. and {Garrido Terrats}, D. and {Gaug}, M. and {Giammaria}, P. and {Glawion (Eisenacher}, D. and {Godinovi{\'c}}, N. and {Gonz{\'a}lez Mu{\~n}oz}, A. and {Guberman}, D. and {Hahn}, A. and {Hanabata}, Y. and {Hayashida}, M. and {Herrera}, J. and {Hose}, J. and {Hrupec}, D. and {Hughes}, G. and {Idec}, W. and {Kodani}, K. and {Konno}, Y. and {Kubo}, H. and {Kushida}, J. and {La Barbera}, A. and {Lelas}, D. and {Lindfors}, E. and {Lombardi}, S. and {Longo}, F. and {L{\'o}pez}, M. and {L{\'o}pez-Coto}, R. and {L{\'o}pez-Oramas}, A. and {Lorenz}, E. and {Majumdar}, P. and {Makariev}, M. and {Mallot}, K. and {Maneva}, G. and {Manganaro}, M. and {Mannheim}, K. and {Maraschi}, L. and {Marcote}, B. and {Mariotti}, M. and {Mart{\'\i}nez}, M. and {Mazin}, D. and {Menzel}, U. and {Miranda}, J.~M. and {Mirzoyan}, R. and {Moralejo}, A. and {Moretti}, E. and {Nakajima}, D. and {Neustroev}, V. and {Niedzwiecki}, A. and {Nievas Rosillo}, M. and {Nilsson}, K. and {Nishijima}, K. and {Noda}, K.},
        title = "{Multiwavelength Study of Quiescent States of Mrk 421 with Unprecedented Hard X-Ray Coverage Provided by NuSTAR in 2013}",
      journal = {\apj},
         year = 2016,
        month = mar,
       volume = {819},
       number = {2},
          eid = {156},
        pages = {156},
          doi = {10.3847/0004-637X/819/2/156},
archivePrefix = {arXiv},
       eprint = {1512.02235},
 primaryClass = {astro-ph.HE},
       adsurl = {https://ui.adsabs.harvard.edu/abs/2016ApJ...819..156B}
}

@ARTICLE{Markowitz2022,
       author = {{Markowitz}, Alex G. and {Nalewajko}, Krzysztof and {Bhatta}, Gopal and {Dewangan}, Gulab C. and {Chandra}, Sunil and {Dorner}, Daniela and {Schleicher}, Bernd and {Pajdosz-{\'S}mierciak}, Urszula and {Stawarz}, {\L}ukasz and {Zola}, Staszek and {Ostrowski}, Micha{\l} and {Carosati}, Daniele and {Krishnan}, Saikruba and {Bachev}, Rumen and {Ben{\'\i}tez}, Erika and {Gazeas}, Kosmas and {Hiriart}, David and {Hu}, Shao-Ming and {Larionov}, Valeri and {Marchini}, Alessandro and {Matsumoto}, Katsura and {Nikiforova}, A.~A. and {Pursimo}, Tapio and {Raiteri}, Claudia M. and {Reichart}, Daniel E. and {Rodriguez}, Diego and {Semkov}, Evgeni and {Strigachev}, Anton and {Sugiura}, Yuki and {Villata}, Massimo and {Webb}, James R. and {Arbet-Engels}, Axel and {Baack}, Dominik and {Balbo}, Matteo and {Biland}, Adrian and {Bretz}, Thomas and {Buss}, Jens and {Eisenberger}, Laura and {Elsaesser}, Dominik and {Hildebrand}, Dorothee and {Iotov}, Roman and {Kalenski}, Adelina and {Mannheim}, Karl and {Mitchell}, Alison and {Neise}, Dominik and {Noethe}, Maximilian and {Paravac}, Aleksander and {Rhode}, Wolfgang and {Sliusar}, Vitalii and {Walter}, Roland},
        title = "{Rapid X-ray variability in Mkn 421 during a multiwavelength campaign}",
      journal = {\mnras},
         year = 2022,
        month = jun,
       volume = {513},
       number = {2},
        pages = {1662-1679},
          doi = {10.1093/mnras/stac917},
archivePrefix = {arXiv},
       eprint = {2206.03874},
 primaryClass = {astro-ph.HE},
       adsurl = {https://ui.adsabs.harvard.edu/abs/2022MNRAS.513.1662M}
}

@ARTICLE{Aleksic2015,
       author = {{Aleksi{\'c}}, J. and {Ansoldi}, S. and {Antonelli}, L.~A. and {Antoranz}, P. and {Babic}, A. and {Bangale}, P. and {Barres de Almeida}, U. and {Barrio}, J.~A. and {Becerra Gonz{\'a}lez}, J. and {Bednarek}, W. and {Berger}, K. and {Bernardini}, E. and {Biland}, A. and {Blanch}, O. and {Bock}, R.~K. and {Bonnefoy}, S. and {Bonnoli}, G. and {Borracci}, F. and {Bretz}, T. and {Carmona}, E. and {Carosi}, A. and {Carreto Fidalgo}, D. and {Colin}, P. and {Colombo}, E. and {Contreras}, J.~L. and {Cortina}, J. and {Covino}, S. and {Da Vela}, P. and {Dazzi}, F. and {De Angelis}, A. and {De Caneva}, G. and {De Lotto}, B. and {Delgado Mendez}, C. and {Doert}, M. and {Dom{\'\i}nguez}, A. and {Dominis Prester}, D. and {Dorner}, D. and {Doro}, M. and {Einecke}, S. and {Eisenacher}, D. and {Elsaesser}, D. and {Farina}, E. and {Ferenc}, D. and {Fonseca}, M.~V. and {Font}, L. and {Frantzen}, K. and {Fruck}, C. and {Garc{\'\i}a L{\'o}pez}, R.~J. and {Garczarczyk}, M. and {Garrido Terrats}, D. and {Gaug}, M. and {Giavitto}, G. and {Godinovi{\'c}}, N. and {Gonz{\'a}lez Mu{\~n}oz}, A. and {Gozzini}, S.~R. and {Hadamek}, A. and {Hadasch}, D. and {Herrero}, A. and {Hildebrand}, D. and {Hose}, J. and {Hrupec}, D. and {Idec}, W. and {Kadenius}, V. and {Kellermann}, H. and {Knoetig}, M.~L. and {Krause}, J. and {Kushida}, J. and {La Barbera}, A. and {Lelas}, D. and {Lewandowska}, N. and {Lindfors}, E. and {Longo}, F. and {Lombardi}, S. and {L{\'o}pez}, M. and {L{\'o}pez-Coto}, R. and {L{\'o}pez-Oramas}, A. and {Lorenz}, E. and {Lozano}, I. and {Makariev}, M. and {Mallot}, K. and {Maneva}, G. and {Mankuzhiyil}, N. and {Mannheim}, K. and {Maraschi}, L. and {Marcote}, B. and {Mariotti}, M. and {Mart{\'\i}nez}, M. and {Mazin}, D. and {Menzel}, U. and {Meucci}, M. and {Miranda}, J.~M. and {Mirzoyan}, R. and {Moralejo}, A. and {Munar-Adrover}, P. and {Nakajima}, D. and {Niedzwiecki}, A. and {Nilsson}, K. and {Nowak}, N. and {Orito}, R. and {Overkemping}, A. and {Paiano}, S. and {Palatiello}, M. and {Paneque}, D. and {Paoletti}, R. and {Paredes}, J.~M. and {Paredes-Fortuny}, X. and {Partini}, S. and {Persic}, M. and {Prada}, F. and {Prada Moroni}, P.~G. and {Prandini}, E. and {Preziuso}, S. and {Puljak}, I. and {Reinthal}, R. and {Rhode}, W. and {Rib{\'o}}, M. and {Rico}, J. and {RodriguezGarcia}, J. and {R{\"u}gamer}, S. and {Saggion}, A. and {Saito}, K. and {Salvati}, M. and {Satalecka}, K. and {Scalzotto}, V. and {Scapin}, V. and {Schultz}, C. and {Schweizer}, T. and {Shore}, S.~N. and {Sillanp{\"a}{\"a}}, A. and {Sitarek}, J. and {Snidaric}, I. and {Sobczynska}, D. and {Spanier}, F. and {Stamatescu}, V. and {Stamerra}, A. and {Steinbring}, T. and {Storz}, J. and {Sun}, S. and {Suri{\'c}}, T. and {Takalo}, L. and {Tavecchio}, F. and {Temnikov}, P. and {Terzi{\'c}}, T. and {Tescaro}, D. and {Teshima}, M. and {Thaele}, J. and {Tibolla}, O. and {Torres}, D.~F. and {Toyama}, T. and {Treves}, A. and {Uellenbeck}, M. and {Vogler}, P. and {Wagner}, R.~M. and {Zandanel}, F. and {Zanin}, R. and {MAGIC Collaboration} and {Archambault}, S. and {Behera}, B. and {Beilicke}, M. and {Benbow}, W. and {Bird}, R. and {Buckley}, J.~H. and {Bugaev}, V. and {Cerruti}, M. and {Chen}, X. and {Ciupik}, L. and {Collins-Hughes}, E. and {Cui}, W. and {Dumm}, J. and {Eisch}, J.~D. and {Falcone}, A. and {Federici}, S. and {Feng}, Q. and {Finley}, J.~P. and {Fleischhack}, H. and {Fortin}, P. and {Fortson}, L. and {Furniss}, A. and {Griffin}, S. and {Griffiths}, S.~T. and {Grube}, J. and {Gyuk}, G. and {Hanna}, D. and {Holder}, J. and {Hughes}, G. and {Humensky}, T.~B. and {Johnson}, C.~A. and {Kaaret}, P. and {Kertzman}, M. and {Khassen}, Y. and {Kieda}, D. and {Krawczynski}, H. and {Krennrich}, F. and {Kumar}, S. and {Lang}, M.~J. and {Maier}, G. and {McArthur}, S. and {Meagher}, K. and {Moriarty}, P. and {Mukherjee}, R.},
        title = "{The 2009 multiwavelength campaign on Mrk 421: Variability and correlation studies}",
      journal = {\aap},
         year = 2015,
        month = apr,
       volume = {576},
          eid = {A126},
        pages = {A126},
          doi = {10.1051/0004-6361/201424216},
archivePrefix = {arXiv},
       eprint = {1502.02650},
 primaryClass = {astro-ph.HE},
       adsurl = {https://ui.adsabs.harvard.edu/abs/2015A&A...576A.126A}
}

@ARTICLE{Blazejowski2005,
       author = {{B{\l}a{\.z}ejowski}, M. and {Blaylock}, G. and {Bond}, I.~H. and {Bradbury}, S.~M. and {Buckley}, J.~H. and {Carter-Lewis}, D.~A. and {Celik}, O. and {Cogan}, P. and {Cui}, W. and {Daniel}, M. and {Duke}, C. and {Falcone}, A. and {Fegan}, D.~J. and {Fegan}, S.~J. and {Finley}, J.~P. and {Fortson}, L. and {Gammell}, S. and {Gibbs}, K. and {Gillanders}, G.~G. and {Grube}, J. and {Gutierrez}, K. and {Hall}, J. and {Hanna}, D. and {Holder}, J. and {Horan}, D. and {Humensky}, B. and {Kenny}, G. and {Kertzman}, M. and {Kieda}, D. and {Kildea}, J. and {Knapp}, J. and {Kosack}, K. and {Krawczynski}, H. and {Krennrich}, F. and {Lang}, M. and {LeBohec}, S. and {Linton}, E. and {Lloyd-Evans}, J. and {Maier}, G. and {Mendoza}, D. and {Milovanovic}, A. and {Moriarty}, P. and {Nagai}, T.~N. and {Ong}, R.~A. and {Power-Mooney}, B. and {Quinn}, J. and {Quinn}, M. and {Ragan}, K. and {Reynolds}, P.~T. and {Rebillot}, P. and {Rose}, H.~J. and {Schroedter}, M. and {Sembroski}, G.~H. and {Swordy}, S.~P. and {Syson}, A. and {Valcarel}, L. and {Vassiliev}, V.~V. and {Wakely}, S.~P. and {Walker}, G. and {Weekes}, T.~C. and {White}, R. and {Zweerink}, J. and {VERITAS Collaboration} and {Mochejska}, B. and {Smith}, B. and {Aller}, M. and {Aller}, H. and {Ter{\"a}sranta}, H. and {Boltwood}, P. and {Sadun}, A. and {Stanek}, K. and {Adams}, E. and {Foster}, J. and {Hartman}, J. and {Lai}, K. and {B{\"o}ttcher}, M. and {Reimer}, A. and {Jung}, I.},
        title = "{A Multiwavelength View of the TeV Blazar Markarian 421: Correlated Variability, Flaring, and Spectral Evolution}",
      journal = {\apj},
         year = 2005,
        month = sep,
       volume = {630},
       number = {1},
        pages = {130-141},
          doi = {10.1086/431925},
archivePrefix = {arXiv},
       eprint = {astro-ph/0505325},
 primaryClass = {astro-ph},
       adsurl = {https://ui.adsabs.harvard.edu/abs/2005ApJ...630..130B}
}

@ARTICLE{fossati2000x,
       author = {{Fossati}, G. and {Celotti}, A. and {Chiaberge}, M. and {Zhang}, Y.~H. and {Chiappetti}, L. and {Ghisellini}, G. and {Maraschi}, L. and {Tavecchio}, F. and {Pian}, E. and {Treves}, A.},
        title = "{X-Ray Emission of Markarian 421: New Clues from Its Spectral Evolution. I. Temporal Analysis}",
      journal = {\apj},
         year = 2000,
        month = sep,
       volume = {541},
       number = {1},
        pages = {153-165},
          doi = {10.1086/309422},
archivePrefix = {arXiv},
       eprint = {astro-ph/0005066},
 primaryClass = {astro-ph},
       adsurl = {https://ui.adsabs.harvard.edu/abs/2000ApJ...541..153F}
}

@ARTICLE{Hota2021,
       author = {{Hota}, Jyotishree and {Shah}, Zahir and {Khatoon}, Rukaiya and {Misra}, Ranjeev and {Pradhan}, Ananta C. and {Gogoi}, Rupjyoti},
        title = "{Understanding the X-ray spectral curvature of Mkn 421 using broad-band AstroSat observations}",
      journal = {\mnras},
         year = 2021,
        month = dec,
       volume = {508},
       number = {4},
        pages = {5921-5934},
          doi = {10.1093/mnras/stab2903},
archivePrefix = {arXiv},
       eprint = {2110.03344},
 primaryClass = {astro-ph.GA},
       adsurl = {https://ui.adsabs.harvard.edu/abs/2021MNRAS.508.5921H}
}

@ARTICLE{Ravasio2004,
       author = {{Ravasio}, M. and {Tagliaferri}, G. and {Ghisellini}, G. and {Tavecchio}, F.},
        title = "{Observing Mkn 421 with XMM-Newton: The EPIC-PN point of view}",
      journal = {\aap},
         year = 2004,
        month = sep,
       volume = {424},
        pages = {841-855},
          doi = {10.1051/0004-6361:20034545},
archivePrefix = {arXiv},
       eprint = {astro-ph/0406481},
 primaryClass = {astro-ph},
       adsurl = {https://ui.adsabs.harvard.edu/abs/2004A&A...424..841R}
}

@ARTICLE{Sobolewska2014,
       author = {{Sobolewska}, M.~A. and {Siemiginowska}, A. and {Kelly}, B.~C. and {Nalewajko}, K.},
        title = "{Stochastic Modeling of the Fermi/LAT {\ensuremath{\gamma}}-Ray Blazar Variability}",
      journal = {\apj},
         year = 2014,
        month = may,
       volume = {786},
       number = {2},
          eid = {143},
        pages = {143},
          doi = {10.1088/0004-637X/786/2/143},
archivePrefix = {arXiv},
       eprint = {1403.5276},
 primaryClass = {astro-ph.HE},
       adsurl = {https://ui.adsabs.harvard.edu/abs/2014ApJ...786..143S}
}

@ARTICLE{Brinkmann2001,
       author = {{Brinkmann}, W. and {Sembay}, S. and {Griffiths}, R.~G. and {Branduardi-Raymont}, G. and {Gliozzi}, M. and {Boller}, Th. and {Tiengo}, A. and {Molendi}, S. and {Zane}, S.},
        title = "{XMM-Newton observations of Markarian 421}",
      journal = {\aap},
         year = 2001,
        month = jan,
       volume = {365},
        pages = {L162-L167},
          doi = {10.1051/0004-6361:20000084},
archivePrefix = {arXiv},
       eprint = {astro-ph/0011193},
 primaryClass = {astro-ph},
       adsurl = {https://ui.adsabs.harvard.edu/abs/2001A&A...365L.162B}
}

@ARTICLE{Brinkmann2005,
       author = {{Brinkmann}, W. and {Papadakis}, I.~E. and {Raeth}, C. and {Mimica}, P. and {Haberl}, F.},
        title = "{XMM-Newton timing mode observations of Mrk 421}",
      journal = {\aap},
         year = 2005,
        month = nov,
       volume = {443},
       number = {2},
        pages = {397-411},
          doi = {10.1051/0004-6361:20052767},
archivePrefix = {arXiv},
       eprint = {astro-ph/0508433},
 primaryClass = {astro-ph},
       adsurl = {https://ui.adsabs.harvard.edu/abs/2005A&A...443..397B}
}

@ARTICLE{Panis2025,
       author = {{P{\'a}nis}, Radim and {Bhatta}, Gopal and {Adhikari}, Tek P. and {Mohorian}, Maksym and {Chaudhary}, Suvas Chandra and {Dinesh}, Adithiya and {Bachchan}, Rajesh K. and {Dhital}, Niraj and {Stuchl{\'\i}k}, Zden{\v{e}}k},
        title = "{Deterministic and Stochastic Study of the X-Ray Emission from the TeV Blazar Mrk 421}",
      journal = {\apj},
         year = 2025,
        month = aug,
       volume = {989},
       number = {1},
          eid = {106},
        pages = {106},
          doi = {10.3847/1538-4357/ade9b1},
archivePrefix = {arXiv},
       eprint = {2506.21346},
 primaryClass = {astro-ph.HE},
       adsurl = {https://ui.adsabs.harvard.edu/abs/2025ApJ...989..106P}
}

@BOOK{Bevington1992,
       author = {{Bevington}, Philip R. and {Robinson}, D. Keith},
        title = "{Data reduction and error analysis for the physical sciences}",
         year = 1992,
       adsurl = {https://ui.adsabs.harvard.edu/abs/1992drea.book.....B}
}

@ARTICLE{Akaike1974,
       author = {{Akaike}, H.},
        title = "{A New Look at the Statistical Model Identification}",
      journal = {IEEE Transactions on Automatic Control},
         year = 1974,
        month = jan,
       volume = {19},
        pages = {716-723},
          doi = {10.1109/TAC.1974.1100705},
       adsurl = {https://ui.adsabs.harvard.edu/abs/1974ITAC...19..716A}
}

@ARTICLE{Liddle2007,
       author = {{Liddle}, Andrew R.},
        title = "{Information criteria for astrophysical model selection}",
      journal = {\mnras},
         year = 2007,
        month = may,
       volume = {377},
       number = {1},
        pages = {L74-L78},
          doi = {10.1111/j.1745-3933.2007.00306.x},
archivePrefix = {arXiv},
       eprint = {astro-ph/0701113},
 primaryClass = {astro-ph},
       adsurl = {https://ui.adsabs.harvard.edu/abs/2007MNRAS.377L..74L}
}

@ARTICLE{1978AnSta...6..461S,
       author = {{Schwarz}, Gideon},
        title = "{Estimating the Dimension of a Model}",
      journal = {Annals of Statistics},
         year = 1978,
        month = jul,
       volume = {6},
       number = {2},
        pages = {461-464},
       adsurl = {https://ui.adsabs.harvard.edu/abs/1978AnSta...6..461S}
}

@article{Sugiura1978,
        author = {Nariaki Sugiura},
        title = {Further analysis of the data by Akaike's information criterion and the finite corrections},
        journal = {Communications in Statistics - Theory and Methods},
        volume = {7},
        number = {1},
        pages = {13--26},
        year = {1978},
        publisher = {Taylor \& Francis},
        doi = {10.1080/03610927808827599},
}

@ARTICLE{Kardashev1962,
       author = {{Kardashev}, N.~S.},
        title = "{Nonstationarity of Spectra of Young Sources of Nonthermal Radio Emission}",
      journal = {\sovast},
         year = 1962,
        month = dec,
       volume = {6},
        pages = {317},
       adsurl = {https://ui.adsabs.harvard.edu/abs/1962SvA.....6..317K}
}

@BOOK{Rybicki1979,
       author = {{Rybicki}, George B. and {Lightman}, Alan P.},
        title = "{Radiative processes in astrophysics}",
         year = 1979,
       adsurl = {https://ui.adsabs.harvard.edu/abs/1979rpa..book.....R}
}

@BOOK{Longair2011,
       author = {{Longair}, Malcolm S.},
        title = "{High Energy Astrophysics}",
         year = 2011,
       adsurl = {https://ui.adsabs.harvard.edu/abs/2011hea..book.....L}
}

@ARTICLE{Kirk1998,
       author = {{Kirk}, J.~G. and {Rieger}, F.~M. and {Mastichiadis}, A.},
        title = "{Particle acceleration and synchrotron emission in blazar jets}",
      journal = {\aap},
         year = 1998,
        month = may,
       volume = {333},
        pages = {452-458},
          doi = {10.48550/arXiv.astro-ph/9801265},
archivePrefix = {arXiv},
       eprint = {astro-ph/9801265},
 primaryClass = {astro-ph},
       adsurl = {https://ui.adsabs.harvard.edu/abs/1998A&A...333..452K}
}

@ARTICLE{Ghisellini2005,
       author = {{Ghisellini}, G. and {Tavecchio}, F. and {Chiaberge}, M.},
        title = "{Structured jets in TeV BL Lac objects and radiogalaxies.  Implications for the observed properties}",
      journal = {\aap},
         year = 2005,
        month = mar,
       volume = {432},
       number = {2},
        pages = {401-410},
          doi = {10.1051/0004-6361:20041404},
archivePrefix = {arXiv},
       eprint = {astro-ph/0406093},
 primaryClass = {astro-ph},
       adsurl = {https://ui.adsabs.harvard.edu/abs/2005A&A...432..401G}
}

@ARTICLE{Tavecchio1998,
       author = {{Tavecchio}, Fabrizio and {Maraschi}, Laura and {Ghisellini}, Gabriele},
        title = "{Constraints on the Physical Parameters of TeV Blazars}",
      journal = {\apj},
         year = 1998,
        month = dec,
       volume = {509},
       number = {2},
        pages = {608-619},
          doi = {10.1086/306526},
archivePrefix = {arXiv},
       eprint = {astro-ph/9809051},
 primaryClass = {astro-ph},
       adsurl = {https://ui.adsabs.harvard.edu/abs/1998ApJ...509..608T}
}

@ARTICLE{Bottcher2002,
       author = {{B{\"o}ttcher}, Markus and {Chiang}, James},
        title = "{X-Ray Spectral Variability Signatures of Flares in BL Lacertae Objects}",
      journal = {\apj},
         year = 2002,
        month = dec,
       volume = {581},
       number = {1},
        pages = {127-142},
          doi = {10.1086/344155},
archivePrefix = {arXiv},
       eprint = {astro-ph/0208238},
 primaryClass = {astro-ph},
       adsurl = {https://ui.adsabs.harvard.edu/abs/2002ApJ...581..127B}
}

@ARTICLE{Zheng2019,
       author = {{Zheng}, Y.~G. and {Kang}, S.~J. and {Yang}, C.~Y. and {Bai}, J.~M.},
        title = "{Particle Acceleration and Synchrotron Self-Compton Emission in Blazar Jets. I. An Application to Quiescent Emission}",
      journal = {\apj},
         year = 2019,
        month = mar,
       volume = {873},
       number = {1},
          eid = {7},
        pages = {7},
          doi = {10.3847/1538-4357/ab02fb},
archivePrefix = {arXiv},
       eprint = {1902.08205},
 primaryClass = {astro-ph.HE},
       adsurl = {https://ui.adsabs.harvard.edu/abs/2019ApJ...873....7Z}
}

@ARTICLE{Takahashi1996,
       author = {{Takahashi}, T. and {Tashiro}, M. and {Madejski}, G. and {Kubo}, H. and {Kamae}, T. and {Kataoka}, J. and {Kii}, T. and {Makino}, F. and {Makishima}, K. and {Yamasaki}, N.},
        title = "{ASCA Observation of an X-Ray/TeV Flare from the BL Lacertae Object Markarian 421}",
      journal = {\apjl},
         year = 1996,
        month = oct,
       volume = {470},
        pages = {L89},
          doi = {10.1086/310302},
       adsurl = {https://ui.adsabs.harvard.edu/abs/1996ApJ...470L..89T}
}

@ARTICLE{Zhang2002,
       author = {{Zhang}, Y.~H. and {Treves}, A. and {Celotti}, A. and {Chiappetti}, L. and {Fossati}, G. and {Ghisellini}, G. and {Maraschi}, L. and {Pian}, E. and {Tagliaferri}, G. and {Tavecchio}, F.},
        title = "{Four Years of Monitoring Blazar PKS 2155-304 with BeppoSAX: Probing the Dynamics of the Jet}",
      journal = {\apj},
         year = 2002,
        month = jun,
       volume = {572},
       number = {2},
        pages = {762-785},
          doi = {10.1086/340349},
archivePrefix = {arXiv},
       eprint = {astro-ph/0202378},
 primaryClass = {astro-ph},
       adsurl = {https://ui.adsabs.harvard.edu/abs/2002ApJ...572..762Z}
}

@ARTICLE{vaughan2003characterizing,
       author = {{Vaughan}, S. and {Edelson}, R. and {Warwick}, R.~S. and {Uttley}, P.},
        title = "{On characterizing the variability properties of X-ray light curves from active galaxies}",
      journal = {\mnras},
         year = 2003,
        month = nov,
       volume = {345},
       number = {4},
        pages = {1271-1284},
          doi = {10.1046/j.1365-2966.2003.07042.x},
archivePrefix = {arXiv},
       eprint = {astro-ph/0307420},
 primaryClass = {astro-ph},
       adsurl = {https://ui.adsabs.harvard.edu/abs/2003MNRAS.345.1271V}
}

@ARTICLE{edelson1990broad,
       author = {{Edelson}, R.~A. and {Krolik}, J.~H. and {Pike}, G.~F.},
        title = "{Broad-Band Properties of the CfA Seyfert Galaxies. III. Ultraviolet Variability}",
      journal = {\apj},
         year = 1990,
        month = aug,
       volume = {359},
        pages = {86},
          doi = {10.1086/169036},
       adsurl = {https://ui.adsabs.harvard.edu/abs/1990ApJ...359...86E}
}

@ARTICLE{bhatta2018microvariability,
       author = {{Bhatta}, Gopal and {Webb}, James},
        title = "{Microvariability in BL Lacertae: ``Zooming'' into the Innermost Blazar Regions}",
      journal = {Galaxies},
         year = 2018,
        month = jan,
       volume = {6},
       number = {1},
        pages = {2},
          doi = {10.3390/galaxies6010002},
archivePrefix = {arXiv},
       eprint = {1711.08698},
 primaryClass = {astro-ph.HE},
       adsurl = {https://ui.adsabs.harvard.edu/abs/2018Galax...6....2B}
}

@ARTICLE{bottcher2003coordinated,
       author = {{B{\"o}ttcher}, M. and {Marscher}, A.~P. and {Ravasio}, M. and {Villata}, M. and {Raiteri}, C.~M. and {Aller}, H.~D. and {Aller}, M.~F. and {Ter{\"a}sranta}, H. and {Mang}, O. and {Tagliaferri}, G. and {Aharonian}, F. and {Krawczynski}, H. and {Kurtanidze}, O.~M. and {Nikolashvili}, M.~G. and {Ibrahimov}, M.~A. and {Papadakis}, I.~E. and {Tsinganos}, K. and {Sadakane}, K. and {Okada}, N. and {Takalo}, L.~O. and {Sillanp{\"a}{\"a}}, A. and {Tosti}, G. and {Ciprini}, S. and {Frasca}, A. and {Marilli}, E. and {Robb}, R. and {Noble}, J.~C. and {Jorstad}, S.~G. and {Hagen-Thorn}, V.~A. and {Larionov}, V.~M. and {Nesci}, R. and {Maesano}, M. and {Schwartz}, R.~D. and {Basler}, J. and {Gorham}, P.~W. and {Iwamatsu}, H. and {Kato}, T. and {Pullen}, C. and {Ben{\'\i}tez}, E. and {de Diego}, J.~A. and {Moilanen}, M. and {Oksanen}, A. and {Rodriguez}, D. and {Sadun}, A.~C. and {Kelly}, M. and {Carini}, M.~T. and {Miller}, H.~R. and {Catalano}, S. and {Dultzin-Hacyan}, D. and {Fan}, J.~H. and {Ghisellini}, G. and {Ishioka}, R. and {Karttunen}, H. and {Kein{\"a}nen}, P. and {Kudryavtseva}, N.~A. and {Lainela}, M. and {Lanteri}, L. and {Larionova}, E.~G. and {Matsumoto}, K. and {Mattox}, J.~R. and {McHardy}, I. and {Montagni}, F. and {Nucciarelli}, G. and {Ostorero}, L. and {Papamastorakis}, J. and {Pasanen}, M. and {Sobrito}, G. and {Uemura}, M.},
        title = "{Coordinated Multiwavelength Observations of BL Lacertae in 2000}",
      journal = {\apj},
         year = 2003,
        month = oct,
       volume = {596},
       number = {2},
        pages = {847-859},
          doi = {10.1086/378156},
archivePrefix = {arXiv},
       eprint = {astro-ph/0307022},
 primaryClass = {astro-ph},
       adsurl = {https://ui.adsabs.harvard.edu/abs/2003ApJ...596..847B}
}

@ARTICLE{bottcher2019,
       author = {{B{\"o}ttcher}, Markus},
        title = "{Progress in Multi-Wavelength and Multi-Messenger Observations of Blazars and Theoretical Challenges}",
      journal = {Galaxies},
         year = 2019,
        month = jan,
       volume = {7},
       number = {1},
          eid = {20},
        pages = {20},
          doi = {10.3390/galaxies7010020},
archivePrefix = {arXiv},
       eprint = {1901.04178},
 primaryClass = {astro-ph.HE},
       adsurl = {https://ui.adsabs.harvard.edu/abs/2019Galax...7...20B}
}

@ARTICLE{Mastichiadis2002,
       author = {{Mastichiadis}, Apostolos and {Kirk}, John G.},
        title = "{Models of Variability in Blazar Jets}",
      journal = {\pasa},
         year = 2002,
        month = jan,
       volume = {19},
       number = {1},
        pages = {138-142},
          doi = {10.1071/AS01108},
       adsurl = {https://ui.adsabs.harvard.edu/abs/2002PASA...19..138M}
}

@article{ghisellini1998,
  author       = {Ghisellini, G. and Celotti, A. and Fossati, G. and Maraschi, L. and Comastri, A.},
  title        = {A theoretical unifying scheme for gamma-ray bright blazars},
  journal      = {Monthly Notices of the Royal Astronomical Society},
  volume       = {301},
  number       = {2},
  pages        = {451--468},
  year         = {1998},
  doi          = {10.1046/j.1365-8711.1998.02032.x},
  adsurl       = {https://ui.adsabs.harvard.edu/abs/1998MNRAS.301..451G}
}

@ARTICLE{falomo2014optical,
       author = {{Falomo}, Renato and {Pian}, Elena and {Treves}, Aldo},
        title = "{An optical view of BL Lacertae objects}",
      journal = {\aapr},
         year = 2014,
        month = sep,
       volume = {22},
          eid = {73},
        pages = {73},
          doi = {10.1007/s00159-014-0073-z},
archivePrefix = {arXiv},
       eprint = {1407.7615},
 primaryClass = {astro-ph.HE},
       adsurl = {https://ui.adsabs.harvard.edu/abs/2014A&ARv..22...73F}
}

@ARTICLE{abdo2010spectral,
       author = {{Abdo}, A.~A. and {Ackermann}, M. and {Agudo}, I. and {Ajello}, M. and {Aller}, H.~D. and {Aller}, M.~F. and {Angelakis}, E. and {Arkharov}, A.~A. and {Axelsson}, M. and {Bach}, U. and {Baldini}, L. and {Ballet}, J. and {Barbiellini}, G. and {Bastieri}, D. and {Baughman}, B.~M. and {Bechtol}, K. and {Bellazzini}, R. and {Benitez}, E. and {Berdyugin}, A. and {Berenji}, B. and {Blandford}, R.~D. and {Bloom}, E.~D. and {Boettcher}, M. and {Bonamente}, E. and {Borgland}, A.~W. and {Bregeon}, J. and {Brez}, A. and {Brigida}, M. and {Bruel}, P. and {Burnett}, T.~H. and {Burrows}, D. and {Buson}, S. and {Caliandro}, G.~A. and {Calzoletti}, L. and {Cameron}, R.~A. and {Capalbi}, M. and {Caraveo}, P.~A. and {Carosati}, D. and {Casandjian}, J.~M. and {Cavazzuti}, E. and {Cecchi}, C. and {{\c{C}}elik}, {\"O}. and {Charles}, E. and {Chaty}, S. and {Chekhtman}, A. and {Chen}, W.~P. and {Chiang}, J. and {Chincarini}, G. and {Ciprini}, S. and {Claus}, R. and {Cohen-Tanugi}, J. and {Colafrancesco}, S. and {Cominsky}, L.~R. and {Conrad}, J. and {Costamante}, L. and {Cutini}, S. and {D'ammando}, F. and {Deitrick}, R. and {D'Elia}, V. and {Dermer}, C.~D. and {de Angelis}, A. and {de Palma}, F. and {Digel}, S.~W. and {Donnarumma}, I. and {Silva}, E. do Couto e. and {Drell}, P.~S. and {Dubois}, R. and {Dultzin}, D. and {Dumora}, D. and {Falcone}, A. and {Farnier}, C. and {Favuzzi}, C. and {Fegan}, S.~J. and {Focke}, W.~B. and {Forn{\'e}}, E. and {Fortin}, P. and {Frailis}, M. and {Fuhrmann}, L. and {Fukazawa}, Y. and {Funk}, S. and {Fusco}, P. and {G{\'o}mez}, J.~L. and {Gargano}, F. and {Gasparrini}, D. and {Gehrels}, N. and {Germani}, S. and {Giebels}, B. and {Giglietto}, N. and {Giommi}, P. and {Giordano}, F. and {Giuliani}, A. and {Glanzman}, T. and {Godfrey}, G. and {Grenier}, I.~A. and {Gronwall}, C. and {Grove}, J.~E. and {Guillemot}, L. and {Guiriec}, S. and {Gurwell}, M.~A. and {Hadasch}, D. and {Hanabata}, Y. and {Harding}, A.~K. and {Hayashida}, M. and {Hays}, E. and {Healey}, S.~E. and {Heidt}, J. and {Hiriart}, D. and {Horan}, D. and {Hoversten}, E.~A. and {Hughes}, R.~E. and {Itoh}, R. and {Jackson}, M.~S. and {J{\'o}hannesson}, G. and {Johnson}, A.~S. and {Johnson}, W.~N. and {Jorstad}, S.~G. and {Kadler}, M. and {Kamae}, T. and {Katagiri}, H. and {Kataoka}, J. and {Kawai}, N. and {Kennea}, J. and {Kerr}, M. and {Kimeridze}, G. and {Kn{\"o}dlseder}, J. and {Kocian}, M.~L. and {Kopatskaya}, E.~N. and {Koptelova}, E. and {Konstantinova}, T.~S. and {Kovalev}, Y.~Y. and {Kovalev}, Yu. A. and {Kurtanidze}, O.~M. and {Kuss}, M. and {Lande}, J. and {Larionov}, V.~M. and {Latronico}, L. and {Leto}, P. and {Lindfors}, E. and {Longo}, F. and {Loparco}, F. and {Lott}, B. and {Lovellette}, M.~N. and {Lubrano}, P. and {Madejski}, G.~M. and {Makeev}, A. and {Marchegiani}, P. and {Marscher}, A.~P. and {Marshall}, F. and {Max-Moerbeck}, W. and {Mazziotta}, M.~N. and {McConville}, W. and {McEnery}, J.~E. and {Meurer}, C. and {Michelson}, P.~F. and {Mitthumsiri}, W. and {Mizuno}, T. and {Moiseev}, A.~A. and {Monte}, C. and {Monzani}, M.~E. and {Morselli}, A. and {Moskalenko}, I.~V. and {Murgia}, S. and {Nestoras}, I. and {Nilsson}, K. and {Nizhelsky}, N.~A. and {Nolan}, P.~L. and {Norris}, J.~P. and {Nuss}, E. and {Ohsugi}, T. and {Ojha}, R. and {Omodei}, N. and {Orlando}, E. and {Ormes}, J.~F. and {Osborne}, J. and {Ozaki}, M. and {Pacciani}, L. and {Padovani}, P. and {Pagani}, C. and {Page}, K. and {Paneque}, D. and {Panetta}, J.~H. and {Parent}, D. and {Pasanen}, M. and {Pavlidou}, V. and {Pelassa}, V. and {Pepe}, M. and {Perri}, M. and {Pesce-Rollins}, M. and {Piranomonte}, S. and {Piron}, F. and {Pittori}, C. and {Porter}, T.~A. and {Puccetti}, S. and {Rahoui}, F. and {Rain{\`o}}, S. and {Raiteri}, C. and {Rando}, R. and {Razzano}, M. and {Reimer}, A. and {Reimer}, O.},
        title = "{The Spectral Energy Distribution of Fermi Bright Blazars}",
      journal = {\apj},
         year = 2010,
        month = jun,
       volume = {716},
       number = {1},
        pages = {30-70},
          doi = {10.1088/0004-637X/716/1/30},
archivePrefix = {arXiv},
       eprint = {0912.2040},
 primaryClass = {astro-ph.CO},
       adsurl = {https://ui.adsabs.harvard.edu/abs/2010ApJ...716...30A}
}

@ARTICLE{punch1992detection,
       author = {{Punch}, M. and {Akerlof}, C.~W. and {Cawley}, M.~F. and {Chantell}, M. and {Fegan}, D.~J. and {Fennell}, S. and {Gaidos}, J.~A. and {Hagan}, J. and {Hillas}, A.~M. and {Jiang}, Y. and {Kerrick}, A.~D. and {Lamb}, R.~C. and {Lawrence}, M.~A. and {Lewis}, D.~A. and {Meyer}, D.~I. and {Mohanty}, G. and {O'Flaherty}, K.~S. and {Reynolds}, P.~T. and {Rovero}, A.~C. and {Schubnell}, M.~S. and {Sembroski}, G. and {Weekes}, T.~C. and {Whitaker}, T. and {Wilson}, C.},
        title = "{Detection of TeV photons from the active galaxy Markarian 421}",
      journal = {\nat},
         year = 1992,
        month = aug,
       volume = {358},
       number = {6386},
        pages = {477-478},
          doi = {10.1038/358477a0},
       adsurl = {https://ui.adsabs.harvard.edu/abs/1992Natur.358..477P}
}

@ARTICLE{Aharonian2006,
       author = {{Aharonian}, F. and {Akhperjanian}, A.~G. and {Bazer-Bachi}, A.~R. and {Beilicke}, M. and {Benbow}, W. and {Berge}, D. and {Bernl{\"o}hr}, K. and {Boisson}, C. and {Bolz}, O. and {Borrel}, V. and {Braun}, I. and {Breitling}, F. and {Brown}, A.~M. and {Chadwick}, P.~M. and {Chounet}, L.-M. and {Cornils}, R. and {Costamante}, L. and {Degrange}, B. and {Dickinson}, H.~J. and {Djannati-Ata{\"\i}}, A. and {Drury}, L. O'c. and {Dubus}, G. and {Emmanoulopoulos}, D. and {Espigat}, P. and {Feinstein}, F. and {Fontaine}, G. and {Fuchs}, Y. and {Funk}, S. and {Gallant}, Y.~A. and {Giebels}, B. and {Gillessen}, S. and {Glicenstein}, J.~F. and {Goret}, P. and {Hadjichristidis}, C. and {Hauser}, D. and {Hauser}, M. and {Heinzelmann}, G. and {Henri}, G. and {Hermann}, G. and {Hinton}, J.~A. and {Hofmann}, W. and {Holleran}, M. and {Horns}, D. and {Jacholkowska}, A. and {de Jager}, O.~C. and {Kh{\'e}lifi}, B. and {Klages}, S. and {Komin}, Nu. and {Konopelko}, A. and {Latham}, I.~J. and {Le Gallou}, R. and {Lemi{\`e}re}, A. and {Lemoine-Goumard}, M. and {Leroy}, N. and {Lohse}, T. and {Martin}, J.~M. and {Martineau-Huynh}, O. and {Marcowith}, A. and {Masterson}, C. and {McComb}, T.~J.~L. and {de Naurois}, M. and {Nolan}, S.~J. and {Noutsos}, A. and {Orford}, K.~J. and {Osborne}, J.~L. and {Ouchrif}, M. and {Panter}, M. and {Pelletier}, G. and {Pita}, S. and {P{\"u}hlhofer}, G. and {Punch}, M. and {Raubenheimer}, B.~C. and {Raue}, M. and {Raux}, J. and {Rayner}, S.~M. and {Reimer}, A. and {Reimer}, O. and {Ripken}, J. and {Rob}, L. and {Rolland}, L. and {Rowell}, G. and {Sahakian}, V. and {Saug{\'e}}, L. and {Schlenker}, S. and {Schlickeiser}, R. and {Schuster}, C. and {Schwanke}, U. and {Siewert}, M. and {Sol}, H. and {Spangler}, D. and {Steenkamp}, R. and {Stegmann}, C. and {Tavernet}, J.-P. and {Terrier}, R. and {Th{\'e}oret}, C.~G. and {Tluczykont}, M. and {van Eldik}, C. and {Vasileiadis}, G. and {Venter}, C. and {Vincent}, P. and {V{\"o}lk}, H.~J. and {Wagner}, S.~J.},
        title = "{A low level of extragalactic background light as revealed by {\ensuremath{\gamma}}-rays from blazars}",
      journal = {\nat},
         year = 2006,
        month = apr,
       volume = {440},
       number = {7087},
        pages = {1018-1021},
          doi = {10.1038/nature04680},
archivePrefix = {arXiv},
       eprint = {astro-ph/0508073},
 primaryClass = {astro-ph},
       adsurl = {https://ui.adsabs.harvard.edu/abs/2006Natur.440.1018A}
}

@ARTICLE{Noel2022,
       author = {{Noel}, A. Priyana and {Gaur}, Haritma and {Gupta}, Alok C. and {Wierzcholska}, Alicja and {Ostrowski}, Micha{\l} and {Dhiman}, Vinit and {Bhatta}, Gopal},
        title = "{X-Ray Intraday Variability of the TeV Blazar Markarian 421 with XMM-Newton}",
      journal = {\apjs},
         year = 2022,
        month = sep,
       volume = {262},
       number = {1},
          eid = {4},
        pages = {4},
          doi = {10.3847/1538-4365/ac7799},
archivePrefix = {arXiv},
       eprint = {2206.02159},
 primaryClass = {astro-ph.HE},
       adsurl = {https://ui.adsabs.harvard.edu/abs/2022ApJS..262....4N}
}

@ARTICLE{Brinkmann2003,
       author = {{Brinkmann}, W. and {Papadakis}, I.~E. and {den Herder}, J.~W.~A. and {Haberl}, F.},
        title = "{Temporal variability of Mrk 421 from XMM-Newton observations}",
      journal = {\aap},
         year = 2003,
        month = may,
       volume = {402},
        pages = {929-947},
          doi = {10.1051/0004-6361:20030264},
       adsurl = {https://ui.adsabs.harvard.edu/abs/2003A&A...402..929B}
}

@ARTICLE{Li2000,
       author = {{Li}, Hui and {Kusunose}, Masaaki},
        title = "{Temporal and Spectral Variabilities of High-Energy Emission from Blazars Using Synchrotron Self-Compton Models}",
      journal = {\apj},
         year = 2000,
        month = jun,
       volume = {536},
       number = {2},
        pages = {729-741},
          doi = {10.1086/308960},
archivePrefix = {arXiv},
       eprint = {astro-ph/0002134},
 primaryClass = {astro-ph},
       adsurl = {https://ui.adsabs.harvard.edu/abs/2000ApJ...536..729L}
}

@ARTICLE{Kataoka2000,
       author = {{Kataoka}, J. and {Takahashi}, T. and {Makino}, F. and {Inoue}, S. and {Madejski}, G.~M. and {Tashiro}, M. and {Urry}, C.~M. and {Kubo}, H.},
        title = "{Variability Pattern and the Spectral Evolution of the BL Lacertae Object PKS 2155-304}",
      journal = {\apj},
         year = 2000,
        month = jan,
       volume = {528},
       number = {1},
        pages = {243-253},
          doi = {10.1086/308154},
archivePrefix = {arXiv},
       eprint = {astro-ph/9908090},
 primaryClass = {astro-ph},
       adsurl = {https://ui.adsabs.harvard.edu/abs/2000ApJ...528..243K}
}

@ARTICLE{Falcone2004,
       author = {{Falcone}, A.~D. and {Cui}, W. and {Finley}, J.~P.},
        title = "{X-Ray Spectral Variability of Extreme BL Lacertae Active Galactic Nucleus H1426+428}",
      journal = {\apj},
         year = 2004,
        month = jan,
       volume = {601},
       number = {1},
        pages = {165-172},
          doi = {10.1086/380447},
archivePrefix = {arXiv},
       eprint = {astro-ph/0310226},
 primaryClass = {astro-ph},
       adsurl = {https://ui.adsabs.harvard.edu/abs/2004ApJ...601..165F}
}
\bibliographystyle{aasjournal}

\end{document}